\documentclass[preprint,nopreprintline,3p,times]{elsarticle}

\usepackage{import}  
\usepackage{lineno,hyperref}
\modulolinenumbers[5]

\usepackage[Export]{adjustbox}  
\usepackage[
 ruled,
 vlined,
 linesnumbered,
 algosection,  
]{algorithm2e}
\usepackage{amsthm}  
\usepackage{amssymb}  
\biboptions{longnamesfirst} 
\usepackage{booktabs} 
\usepackage[
 labelfont=bf,  
 margin=3em,  
]{caption}  
\usepackage{colortbl} 
\usepackage{enumitem}
\usepackage{etoolbox} 
\usepackage{float}
\usepackage[T1]{fontenc} 
\usepackage{hyphenat} 
\usepackage[utf8]{inputenc}
\usepackage{longtable}
\usepackage[tbtags]{mathtools}
\usepackage{multicol} 
\usepackage{multirow}
\usepackage{placeins} 
\usepackage{rotating} 
\usepackage[
 labelformat=simple,  
]{subcaption}  
\usepackage[dvipsnames]{xcolor}

\usepackage[
 capitalise,
 nameinlink,
 noabbrev]{cleveref}  

\newcommand{\bkt}[1]{\bktR{#1}}
\newcommand{\Bkt}[1]{\BktR{#1}}

\newcommand{\DBkt}[1]{\Bkt{\Bkt{#1}}}

\newcommand{\bktR}[1]{\setWithBracketsBase{(}{#1}{)}}
\newcommand{\BktR}[1]{\SetWithBracketsBase{(}{#1}{)}}

\newcommand{\bktC}[1]{\setWithBrackets{\{}{#1}{\}}}
\newcommand{\BktC}[1]{\SetWithBrackets{\{}{#1}{\}}}

\newcommand{\bktU}[1]{\setWithBrackets{\lceil}{#1}{\rceil}}
\newcommand{\BktU}[1]{\SetWithBrackets{\lceil}{#1}{\rceil}}

\newcommand{\bktD}[1]{\setWithBrackets{\lfloor}{#1}{\rfloor}}
\newcommand{\BktD}[1]{\SetWithBrackets{\lfloor}{#1}{\rfloor}}

\newcommand{\fixedbkt}[1]{\csname #1\endcsname}

\newcommand{\bktRfixed}[2]{
\setWithBracketsBase{\fixedbkt{#1}(}{#2}{\fixedbkt{#1})}}

\newcommand{\bktCfixed}[2]{
\setWithBracketsBase{\fixedbkt{#1}\{}{#2}{\fixedbkt{#1}\}}}

\newcommand{\bktUfixed}[2]{
\setWithBracketsBase{\fixedbkt{#1}\lceil}{#2}{\fixedbkt{#1}\rceil}}

\newcommand{\bktDfixed}[2]{
\setWithBracketsBase{\fixedbkt{#1}\lfloor}{#2}{\fixedbkt{#1}\rfloor}}

\newcommand{\innerBracketSpace}{2mu}

\newcommand{\setWithBracketsBase}[3]{#1{#2}#3}
\newcommand{\SetWithBracketsBase}[3]{\setWithBracketsBase{\left#1}{#2}{\right#3}}
\DeclareRobustCommand{\setWithBrackets}[3]{%
 \ifInnermost{\setWithBrackets}{#2}{%
  \setWithBracketsBase{#1}{\mspace{\innerBracketSpace}#2\mspace{\innerBracketSpace}}{#3}%
 }{%
  \setWithBracketsBase{#1}{#2}{#3}%
 }%
}
\DeclareRobustCommand{\SetWithBrackets}[3]{%
 \setWithBrackets{\left#1}{#2}{\right#3}%
}

\makeatletter

\newcommand{\ifInnermost}[4]{%
  \let\ifInnermost@oldBraces#1%
  \gdef\ifInnermost@foundBraces{0}
  \renewcommand{#1}[3]{%
    \gdef\ifInnermost@foundBraces{1}%
    \renewcommand{#1}[3]{}%
  }%
  \sbox0{$#2$}%
  \let#1\ifInnermost@oldBraces%
  \ifnum\numexpr\ifInnermost@foundBraces\relax=0%
    #3%
  \else%
    #4%
  \fi%
}

\makeatother

\definecolor{magenta}{RGB}{255,0,255}
\definecolor{gray83}{RGB}{212,212,212}
\definecolor{gray67}{RGB}{171,171,171}
\definecolor{gray50}{RGB}{128,128,128}
\definecolor{gray33}{RGB}{86,86,86}
\definecolor{gray17}{RGB}{43,43,43}

\definecolor{reddish_brown}{RGB}{149,  82,  81}
\definecolor{brown}{RGB}{199, 141, 107}
\definecolor{purple}{RGB}{154,  51, 143}
\definecolor{violet}{RGB}{225,  15, 125}
\definecolor{pink}{RGB}{255,  62, 181}
\definecolor{flieder}{RGB}{236, 166, 192}
\definecolor{red}{RGB}{217,  38,  41}
\definecolor{mid_red}{rgb}{1,0,0}
\definecolor{orange}{RGB}{232, 121,  40}
\definecolor{mid_orange}{RGB}{250, 121,  40}
\definecolor{yellow}{RGB}{244, 236,  39}
\definecolor{mid_yellow}{rgb}{1,1,0}
\definecolor{yellowish_green}{RGB}{151, 215,   0}
\definecolor{light_green}{RGB}{111, 184,  76}
\definecolor{mid_green}{rgb}{0, 0.8, 0.3}
\definecolor{dark_green}{RGB}{ 13, 149,  75}
\definecolor{turkoise}{RGB}{125, 207, 182}
\definecolor{default_blue}{RGB}{  0, 181, 226}
\definecolor{light_blue}{RGB}{ 36, 156, 216}
\definecolor{dark_blue}{RGB}{ 29,  78, 137}
\definecolor{dark_orange}{RGB}{189, 100, 33}

\definecolor{emphcolor}{rgb}{0.175,0.175,0.64}

\newlength{\boxwidth}
\newcommand{\linebox}[1]{\begin{center}\framebox[\textwidth][c]{\parbox[t]{\boxwidth}{#1}}\end{center}}

\newtheorem{theorem}{Theorem}
\numberwithin{theorem}{section}
\newtheorem{proposition}[theorem]{Proposition}
\newtheorem{lemma}[theorem]{Lemma}
\newtheorem{corollary}[theorem]{Corollary}

\theoremstyle{definition}
\newtheorem{definition}[theorem]{Definition}
\newtheorem{observation}[theorem]{Observation}
\newtheorem{notation}[theorem]{Notation}

\newtheorem{remark}[theorem]{Remark}
\newtheorem*{remark_no_num}{Remark}

\theoremstyle{remark}

\newcounter{claimcount}
\newenvironment{claim}
   {\par\addvspace{0.5\baselineskip}
   \noindent \refstepcounter{claimcount}\textit{Claim \arabic{claimcount}.}}
   {\par\addvspace{0.5\baselineskip}}
\makeatletter
\AtBeginEnvironment{proof}{\setcounter{claimcount}{0}}
\makeatother

\newtheorem*{claim_no_num}{Claim}
\newcommand{\claimqed}{\scriptsize \qed}
\newenvironment{proof_of_claim}{\par\noindent\textit{Proof of claim.}}{\claimqed}
\newenvironment{proof_of_claim_no_qed}{\par\noindent\textit{Proof of claim.}}{}

\Crefname{observation}{Observation}{Observations}
\Crefname{conjecture}{Conjecture}{Conjectures}
\crefname{claimcount}{Claim}{Claims}
\Crefname{claimcount}{Claim}{Claims}

\newenvironment{toshowequation}{\equation}
   {\endequation}

\newenvironment{qedtheorem}
 {\pushQED{\qed}\theorem}
 {\popQED\endtheorem}

\newenvironment{qedproposition}
 {\pushQED{\qed}\proposition}
 {\popQED\endproposition}

 \newenvironment{qedlemma}
 {\pushQED{\qed}\lemma}
 {\popQED\endlemma}

\newenvironment{qedcorollary}
 {\pushQED{\qed}\corollary}
 {\popQED\endcorollary}

\newcommand{\boxedDescription}[5]{%
 {%
  \setlength{\parskip}{\topsep}
   {%
    \setlength{\fboxsep}{5pt}
     {%
      \setlength{\parindent}{0pt}
       \framebox[\textwidth][c]{%
        \begin{tabular}{%
          @{\hspace{\fboxsep}}
          l@{}
          @{\hspace{\fboxsep}}
          p{\textwidth-3\fboxsep-\maxof{\widthof{#1}}{\widthof{#2}}}@{\hspace{\fboxsep}}
        }%
         \multicolumn{2}{p{\textwidth-2\fboxsep}}{%
          {\Large{\textsc{#3}}}} \\ %
         \textit{#1:} & #4 \\ %
         \textit{#2:} & #5 \\ %
        \end{tabular}%
       }%
     }%
   }%
  \vspace{\topsep}
 }%
}

\newcommand{\problemBase}[3]{%
 \boxedDescription{Instance}{Task}{#1}{#2}{#3}%
}

\newcommand{\problemIndex}[4]{
 \index{#2|(}%
 \problemBase{#1}{#3}{#4}%
 \index{#2|)}%
}

\newcommand{\problem}[4]{
 \problemIndex{#1}{#1@\textsc{#1}}{#2}{#3}
}

\subimport{./}{shortcref.tex}

\newcommand{\N}{\mathbb N}
\newcommand{\R}{\mathbb R}
\newcommand{\Z}{\mathbb Z}
\newcommand{\I}{\set{0, 1}}
\newcommand{\cupdot}{\mathbin{\mathaccent\cdot\cup}} 
\newcommand{\Cupdot}{\mathbin{\dot{\bigcup}}} 
\newcommand{\set}{\BktC}
\DeclareMathOperator{\IH}{\text{(IH)}}

\newcommand{\blank}{\_}
\newcommand{\where}{:}
\newcommand{\mdoubleplus}{\mathbin{+\mkern-8mu+}}
\newcommand{\const}{\text{const}}
\newcommand{\indhyp}{\text{(IH)}}

\DeclareMathOperator{\flodd}{flodd}

\newcommand{\tupleconcat}{\mdoubleplus}

\newcommand{\AND}{\textsc{And}}
\newcommand{\OR}{\textsc{Or}}

\DeclareMathOperator{\sym}{\AND}
\DeclareMathOperator{\symdual}{\OR}

\newcommand{\xor}{\oplus}

\newcommand{\aop}{\textsc{And}-\textsc{Or} path}
\newcommand{\AOP}{\textsc{And}-\textsc{Or} Path}
\newcommand{\praopdepthopt}{\textsc{And-Or Path Optimization Problem}}

\newcommand{\prsymdelayopt}{\textsc{Symmetric Function Delay Optimization Problem}}
\newcommand{\prmultiand}{\textsc{Parallel \AND{}-Prefix Problem}}

\newcommand{\pradderopt}{\textsc{Adder Optimization Problem}}

\newcommand{\prprefix}{\textsc{Parallel Prefix Problem}}

\newcommand{\prefix}[2]{#1'}
\newcommand{\suffix}[2]{#1''}
\newcommand{\everysecond}[1]{\widehat{#1}}

\newcommand{\dmin}{d_{\min}}

\DeclareMathOperator{\pgate}{\circ_p}

\DeclareMathOperator{\cktins}{\mathcal I}
\DeclareMathOperator{\cktouts}{\mathcal O}
\DeclareMathOperator{\cktgates}{\mathcal G}
\DeclareMathOperator{\cktnodes}{\mathcal V}
\DeclareMathOperator{\cktedges}{\mathcal E}
\DeclareMathOperator{\basis}{\Omega}

\DeclareMathOperator{\depthckt}{C}
\DeclareMathOperator{\cktout}{out}

\newcommand{\ckt}[3]{{#1}^{(#2)}_{#3}}
\newcommand{\inpart}[1]{P^{(#1)}}
\newcommand{\andprefixckt}{\AND{}-prefix circuit}
\newcommand{\nachfahre}{descendant}

\DeclareMathOperator{\depth}{d}
\DeclareMathOperator{\delay}{delay}
\DeclareMathOperator{\size}{s}
\DeclareMathOperator{\fanout}{fanout}

\newcommand{\totalref}[1]{\cref{#1} (\cpageref{#1})}
\newcount\myloopcounter

\newcommand{\define}[1]{\emph{#1}}

\newcommand{\printktimes}[2]{%
  \myloopcounter0
  \loop\ifnum\myloopcounter < #1 
  #2%
  \advance\myloopcounter by 1 %
  \repeat 
}

\newcommand{\mw}[1]{\mathmakebox[\widthof{#1}]}

\newcommand{\ca}[2]{&\mw{\printktimes{#1}{=}}{#2}}

\newcommand{\rmDecorate}[2][]{%
 \ifthenelse{\equal{#1}{p}}{%
   \IfDecimal{#2}{\SI[retain-explicit-plus, group-minimum-digits=3]{#2}{\percent}}{#2}%
 }{%
   \IfDecimal{#2}{\num[retain-explicit-plus, group-minimum-digits=3]{#2}}{#2}%
 }
}

\newcommand{\dashrule}[1][black]{%
  \color{#1}\rule[\dimexpr.5ex-.2pt]{4pt}{.4pt}\xleaders\hbox{\rule{4pt}{0pt}\rule[\dimexpr.5ex-.2pt]{4pt}{.4pt}}\hfill\kern0pt%
}

\newcommand{\spacingbetweenappproofs}{\smallskip}

\usepackage{tikz}
\usetikzlibrary{arrows}
\usetikzlibrary{calc}
\usetikzlibrary{shapes}

\usetikzlibrary{positioning}
\usetikzlibrary{shapes.misc}
\usetikzlibrary{fit}
\usetikzlibrary{quotes}
\usetikzlibrary{angles}
\usetikzlibrary{through}
\usetikzlibrary{decorations.pathmorphing}
\usetikzlibrary{shapes.gates.logic.US,shapes.gates.logic.IEC}
\usepackage{pgf}
\usepackage{pgfplots}
\usepgfplotslibrary{colormaps}

\pgfplotsset{compat=1.14}

\colorlet{atcolor}{blue}

\tikzset{and-gate/.style={fill=mid_red,outer sep=0pt, thick, and gate US, draw, rotate=270, text=atcolor}}
\tikzset{or-gate/.style={fill=mid_green,outer sep=0pt, thick, or gate US, draw, rotate=270, text=atcolor}}
\tikzset{nand-gate/.style={fill=mid_red,outer sep=0pt, thick, nand gate US, draw, rotate=270, text=atcolor}}
\tikzset{nor-gate/.style={fill=mid_green,outer sep=0pt, thick, nor gate US, draw, rotate=270, text=atcolor}}
\tikzset{xor-gate/.style={fill=cyan,outer sep=0pt, thick, xor gate US, draw, rotate=270, text=atcolor}}
\tikzset{xnor-gate/.style={fill=cyan,outer sep=0pt, thick, xnor gate US, draw, rotate=270, text=atcolor}}
\tikzset{prefix-gate/.style={outer sep=0pt, circle, scale=2, thick, draw, text=atcolor}}
\tikzset{and3-gate/.style={and-gate, logic gate inputs=nnn}}
\tikzset{and4-gate/.style={and-gate, logic gate inputs=nnnn}}
\tikzset{or3-gate/.style={or-gate, logic gate inputs=nnn}}
\tikzset{or4-gate/.style={or-gate, logic gate inputs=nnnn}}
\tikzset{and5-gate/.style={and-gate, logic gate inputs=nnnnn}}
\tikzset{inv-gate/.style={fill=mid_yellow,outer sep=0pt, thick, not gate US, draw, rotate=270, text=atcolor}}
\tikzset{buf-gate/.style={fill=mid_yellow,outer sep=0pt, thick, buffer gate US, draw, rotate=270, text=atcolor}}
\tikzset{sym-and-gate/.style={fill=mid_yellow,outer sep=0pt, thick, and gate US, draw, rotate=270, text=atcolor}}
\tikzset{sym-or-gate/.style={fill=mid_yellow,outer sep=0pt, thick, or gate US, draw, rotate=270, text=atcolor}}
\tikzset{large-node/.style={scale=1.6}}
\tikzset{uncolored-and-gate/.style={outer sep=0pt, thick, and gate US, draw, rotate=270, text=atcolor}}
\tikzset{uncolored-or-gate/.style={outer sep=0pt, thick, or gate US, draw, rotate=270, text=atcolor}}
\tikzset{concat-and-gate/.style={fill=cyan,outer sep=0pt, thick, and gate US, draw, rotate=270, text=atcolor}}
\tikzset{concat-or-gate/.style={fill=cyan,outer sep=0pt, thick, or gate US, draw, rotate=270, text=atcolor}}
\tikzset{input/.style={scale=1.6}}
\tikzset{smallinput/.style={}}
\tikzset{input-at/.style={atcolor}}
\tikzset{output-at/.style={atcolor}}
\tikzset{output/.style={scale=1.6}}
\tikzset{input/.style={scale=1.6}}
\tikzset{every path/.style={thick, -}}
\tikzset{output-edge/.style={thick, ->}}
\tikzset{marked-edge/.style={very thick, cyan}}
\tikzset{marked-prop/.style={mid_red, draw, thick, scale = 0.9}}
\tikzset{marked-gen/.style={mid_green, draw, thick, scale = 0.9}}

 \definecolor{bn_no_color}{RGB}{0,0,0}
 \definecolor{bn_black}{RGB}{0,0,0}
 \definecolor{bn_cyan}{RGB}{0,255,255}
 \definecolor{bn_magenta}{RGB}{255,0,255}
 \definecolor{bn_yellow}{RGB}{255,255,0}
 \definecolor{bn_blue}{RGB}{0,0,255}
 \definecolor{bn_orange}{RGB}{255,179,0}
 \definecolor{bn_white}{RGB}{255,255,255}
 \definecolor{bn_gray83}{RGB}{212,212,212}
 \definecolor{bn_gray67}{RGB}{171,171,171}
 \definecolor{bn_gray50}{RGB}{128,128,128}
 \definecolor{bn_gray33}{RGB}{86,86,86}
 \definecolor{bn_gray17}{RGB}{43,43,43}
 \definecolor{bn_brown}{RGB}{140,70,20}
 \definecolor{bn_light_cyan}{RGB}{128,255,255}
 \definecolor{bn_light_magenta}{RGB}{255,128,255}
 \definecolor{bn_light_yellow}{RGB}{255,255,128}
 \definecolor{bn_light_blue}{RGB}{128,128,255}
 \definecolor{bn_light_red}{RGB}{255,128,128}
 \definecolor{bn_light_green}{RGB}{128,255,128}
 \definecolor{bn_light_orange}{RGB}{255,210,0}
 \definecolor{bn_light_brown}{RGB}{190,160,138}
 \definecolor{bn_dark_cyan}{RGB}{0,128,128}
 \definecolor{bn_dark_magenta}{RGB}{128,0,128}
 \definecolor{bn_dark_yellow}{RGB}{128,128,0}
 \definecolor{bn_dark_blue}{RGB}{0,0,128}
 \definecolor{bn_dark_red}{RGB}{128,0,0}
 \definecolor{bn_dark_green}{RGB}{0,128,0}
 \definecolor{bn_dark_orange}{RGB}{204,143,0}
 \definecolor{bn_dark_brown}{RGB}{70,35,10}

\makeatletter
\def\ps@pprintTitle{%
}
\makeatother

\begin{document}

\begin{frontmatter}

\title{Faster Linear-Size \AOP{} and Adder Circuits}
\author{Ulrich Brenner\footnote[1]{Corresponding author}}
\ead{brenner@dm.uni-bonn.de}
\author{Anna Silvanus\footnote[2]{Now with Synopsys GmbH, Germany.}}
\address{University of Bonn  \\ Research Institute for Discrete Mathematics, University of Bonn \\ Lennéstr. 2, 53113 Bonn, Germany}
\ead{anna.silvanus@synopsys.com}
\date{\today}

\begin{abstract}
We consider the fundamental problem of
constructing fast and small circuits for binary addition.
We propose a new algorithm with running time $\mathcal O(n \log_2 n)$
for constructing linear-size $n$-bit adder circuits with a
significantly better depth guarantee compared to previous approaches:
Our circuits have a depth of at most
$\log_2 n + \log_2 \log_2 n + \log_2 \log_2 \log_2 n + \const$,
improving upon the previously best circuits by \citet{HSAdders}
with a depth of at most
$\log_2 n + 8 \sqrt{\log_2 n} + 6 \log_2 \log_2 n + \const$.
Hence,
we decrease the gap to the lower bound
of $\log_2 n + \log_2 \log_2 n + \const$ by \citet{Commentz-Walter-mon}
significantly
from $\mathcal O \Bkt{\sqrt{\log_2 n}}$
to $\mathcal O(\log_2 \log_2 \log_2 n)$.

Our core routine is a new algorithm
for the construction of a circuit for a single carry bit,
or, more generally, for an \aop{},
i.e., a Boolean function of type
$t_0 \lor \bkt{ t_1 \land \bkt{t_2 \lor \bkt{ \dots t_{m-1}} \dots }}$.
We compute linear-size \aop{} circuits
with a depth of at most $\log_2 m + \log_2 \log_2 m + 0.65$
in time $\mathcal O(m \log_2 m)$.
These are the first \aop{} circuits known that, up to an additive constant,
match the lower bound by \citet{Commentz-Walter-mon}
and at the same time have a linear size.
The previously fastest \aop{} circuits 
are only by an additive constant worse in depth,
but have a much higher size in the order of $\mathcal O (m \log_2 m)$.
\end{abstract}

\thispagestyle{empty}



\end{frontmatter}

\section{Introduction} \label{sec::intro}


Constructing efficient binary adder circuits belongs to the classical
problems in Boolean circuit design and has been studied for many decades.
Assume that we want to compute the sum of two $n$-bit binary numbers
$a = \sum_{i=0}^{n-1}a_i 2^i$ and $b = \sum_{i=0}^{n-1}b_i 2^i$.
A circuit computing the binary sum can be constructed via carry bits:
For $i \in \set{0, \dotsc, n-1}$, we call $g_i := a_i \land b_i \in \I$
the $i$-th \define{generate signal} and
$p_i := a_i \xor b_i \in \I$ the $i$-th \define{propagate signal} for $a$ and $b$.
Recursively, we define the \define{carry bits} $c_0, \dotsc, c_n \in \I$ by $c_0 = 0$ and
\begin{align}
 c_{i+1} & = g_i \lor \Bkt{p_i \land c_i} \text{ for } 0 \leq i \leq n-1\,. \label{carry-recursive}
\end{align}
From the carry bits, we can easily read off the sum
$\sum_{i=0}^{n}s_i 2^i$ via
$s_i = c_i \xor p_i$
for $i \in \set{0, \dotsc, n-1}$
and
$s_n = c_n$. 
Thus, each carry bit $c_{i+1}$ can be computed from 
$g_i, p_i, g_{i-1}, p_{i-1}, \dotsc, g_1, p_1, g_0$ with the following formula:
\begin{align}
\begin{split}
 c_{i+1} = g_i \lor \Bkt{p_i \land c_i}
         &= g_i \lor \Bkt{p_i \land \Bkt{g_{i-1} \lor \Bkt{p_{i-1} \land c_{i-1}}}} \\
         &= g_i \lor \Bkt{p_i \land \Bkt{g_{i-1} \lor \Bkt{p_{i-1} \land \Bkt{g_{i-2} \lor \Bkt{p_{i-2} \land \dotsc \Bkt{p_1 \land g_0}}}}}} \label{carry-aop}
\end{split}
\end{align}
We call this kind of a Boolean formula an \aop{}.

Since, on the other hand, it is easy to compute the carry bits from the 
input an the bits $(s_0,\dots,s_n)$, 
constructing a circuit that computes all carry bits is equivalent
to constructing an adder circuit.

Before we formalize the problem that we consider, we introduce some notation.
We call a function
$f \colon \I^m \to \I$ for some $m \in \N$
a \define{Boolean function}.
We often write $t = \Bkt{t_0, \dotsc, t_{m-1}}$ for
the \define{input variables}, short \define{inputs}, of $f$.
In this paper, a \define{circuit} $C$ is a connected acyclic digraph
with vertex set $\cktnodes(C)$\label{cktnodes-page} and
edge set $\cktedges(C)$\label{cktedges-page}
such that $\cktnodes(C)$ can be partitioned into two sets:
\define{inputs} $\cktins(C)$\label{cktins-page}
with no incoming edges that each represent a Boolean variable
or a constant $0$ or $1$, and
\define{gates} $\cktgates(C)$\label{cktgates-page} with exactly $2$ incoming edges
that each represent a Boolean function
(usually, an elementary Boolean function, e.g.,
$\AND{}2$ or $\OR{}2$, i.e.,
logical \AND{} and logical \OR{} on two inputs).
There is a subset $\cktouts(C) \subseteq \cktnodes(C)$
called \define{outputs}.
All vertices with no outgoing edges are outputs, but there might be others.
In the special case that $C$ contains only a single output,
we denote this output by $\cktout(C)$.

The \define{Boolean function computed at a vertex} of a circuit $C$
can be read off recursively by combining the logical functions
represented by the gates.
If $C$ has a single output which computes a Boolean function $f$,
we also say that $C$ \define{computes} or \define{realizes} $f$
and write $f = f(C)$ for the function computing $C$.
Two circuits that realize the same Boolean function are
called \define{equivalent}.
Given a set $\Omega$ of Boolean functions,
$C$ is a circuit \define{over the basis} $\Omega$
if for all gates $g$ in $C$ the Boolean function
represented by $g$ is in $\Omega$.
A circuit where each gate has only one successor is called a \define{formula circuit}.

Given a Boolean function $f$, there are numerous circuits computing $f$.
In \cref{ex-aop-delay},
we see two circuits realizing the function
$t_0 \land \Bkt{t_1 \lor \Bkt{t_2 \land \Bkt{t_3 \lor t_4}}}$.
Here, \AND{} gates are shown in red and \OR{} gates in green;
the edges are implicitely directed from top to bottom.
The circuit outputs are marked by an arrow.

In order to evaluate the quality of different circuits, we introduce
the following measures motivated by VLSI design:
By $\size(C)$,\label{size-page} we denote the \define{size} of a
circuit $C$, i.e., its number of gates.
As an estimation for the time needed to excecute the computation of a circuit $C$,
we use the \define{circuit depth}, denoted by $\depth(C)$,\label{depth-page}
i.e., the maximum number of gates on any directed path $P$ in $C$.

We formalize the concept of \aop{}s, a special type of Boolean functions
examined in this work, as follows.

\begin{figure}[t]
 \hfill
 \begin{subfigure}[t]{0.3\textwidth}
 \begin{center}
 \adjustbox{max width=0.65\textwidth}{
  \centering{\begin{tikzpicture}

\node[or-gate] at (5,2) (or1){};
\node[and-gate] at (4,1) (and1){};

\node[or-gate] at (3,0) (or4){};
\node[and-gate] at (2,-1) (and2){};

\node[input] (i5) at (1.5, 3.2){$t_0$};
\node[input] (i4) at (2.5, 3.2){$t_1$};
\node[input] (i3) at (3.5, 3.2){$t_2$};
\node[input] (i2) at (4.5, 3.2){$t_3$};
\node[input] (i1) at (5.5, 3.2){$t_4$};

\draw (or1.output) -- (and1.input 1);
\draw (i2) -- (or1.input 2);
\draw (i3) -- (and1.input 2);
\draw (i1) -- (or1.input 1);

\draw (or4.output) -- (and2.input 1);

\draw (i5) -- (and2.input 2);

\draw (i4) -- (or4.input 2);
\draw (and1.output) -- (or4.input 1);

\node[output-at] (output) at (2, -2) {};
\draw[output-edge] (and2.output) -- (output) {};

\end{tikzpicture}}
  }
  \end{center}
  \caption{Circuit $C_1$ with $\depth(C_1) = 4$ and $\size(C_1) = 4$.}
  \label{ex-aop-delay-1}
 \end{subfigure}
\hspace*{0.03\textwidth}
 \begin{subfigure}[t]{0.3\textwidth}
 \begin{center}
  \adjustbox{max width=0.65\textwidth}{
  \centering{\begin{tikzpicture}

\node[input] (i5) at (1.5, 3.2){$t_0$};
\node[input] (i4) at (2.5, 3.2){$t_1$};
\node[input] (i3) at (3.5, 3.2){$t_2$};
\node[input] (i2) at (4.5, 3.2){$t_3$};
\node[input] (i1) at (5.5, 3.2){$t_4$};

\node[or-gate] at (3,2) (or1){};
\draw (i3) -- (or1.input 1);
\draw (i4) -- (or1.input 2);

\node[or-gate] at (4.2,2) (or2){};
\draw (i2) -- (or2.input 1);
\draw (i4) -- (or2.input 2);

\node[or-gate] at (4.5,1) (or3){};
\draw (or2.output)   --  (or3.input 2);
\draw (i1) -- (or3.input 1);

\node[and-gate] at (2.5,1) (and1){};
\draw (or1.output) -- (and1.input 1);
\draw (i5) -- (and1.input 2);

\node[and-gate] at (3.5,0) (and2){};
\draw (or3.output) -- (and2.input 1);
\draw (and1.output)   -- (and2.input 2);

\node[output-at] (output) at (3.5, -1) {};
\draw[output-edge] (and2.output) -- (output) {};

\phantom{\node[output-at] (output) at (2, -2) {};}
\end{tikzpicture}}
  }
 \end{center}
\caption{Circuit $C_2$ with $\depth(C_2) = 3$ and $\size(C_2) = 5$.}
  \label{ex-aop-delay-2}
 \end{subfigure}
\hspace*{0.03\textwidth}
 \begin{subfigure}[t]{0.3\textwidth}
 \begin{center}
  \adjustbox{max width=0.65\textwidth}{
  \centering{\begin{tikzpicture}

\node[input] (i5) at (1.5, 3.2){$t_0$};
\node[input] (i4) at (2.5, 3.2){$t_1$};
\node[input] (i3) at (3.5, 3.2){$t_2$};
\node[input] (i2) at (4.5, 3.2){$t_3$};
\node[input] (i1) at (5.5, 3.2){$t_4$};

\node[or-gate] at (5,2) (or1){};
\draw (i1) -- (or1.input 1);
\draw (i2) -- (or1.input 2);

\node[and-gate] at (3.5,2) (and2){};
\draw (i3) -- (and2.input 1);
\draw (i5) -- (and2.input 2);

\node[and-gate] at (2,2) (and3){};
\draw (i4) -- (and3.input 1);
\draw (i5) -- (and3.input 2);

\node[and-gate] at (4.25,1) (and4){};
\draw (or1.output) -- (and4.input 1);
\draw (and2.output) -- (and4.input 2);

\node[or-gate] at (3,0) (or5){};
\draw (and4.output) -- (or5.input 1);
\draw (and3.output) -- (or5.input 2);

\draw[output-edge] (or5.output) -- ($(or5.output) - (0, .5)$) {};

\phantom{\node[output-at] (output) at (2, -2) {};}

\end{tikzpicture}}
  }
 \end{center}
\caption{Circuit $C_3$ with $\depth(C_3) = 3$ and $\size(C_2) = 5$.}
  \label{ex-aop-delay-3}
 \end{subfigure}
\hfill
 \caption{Three circuits realizing
 $g\DBkt{t_0, \dotsc, t_4} = t_0 \land \Bkt{t_1 \lor \Bkt{t_2 \land \Bkt{t_3 \lor t_4}}}$.}
 \label{ex-aop-delay}
\end{figure}

\begin{definition}\label{def-aop}
   Let inputs $t = (t_0, \dotsc, t_{m-1})$ for some $m \in \N_{> 0}$ be given.
   We call each of the recursively defined functions
   \[g(t) = \begin{cases}
                    t_0 & m = 1 \\
                    t_0 \land g^*((t_1, \dotsc, t_{m-1})) & m > 1
                 \end{cases} \quad \text{ and } \quad
    g^*(t) = \begin{cases}
                    t_0 & m = 1 \\
                    t_0 \lor g((t_1, \dotsc, t_{m-1})) & m > 1
                 \end{cases} \]
   an \emph{\aop} on $m$ inputs.
\end{definition}

The circuits arising from this \lcnamecref{def-aop} are called
\define{standard \aop{} circuits}.
In \cref{ex-aop-delay-1}, we see a standard circuit
for $g\DBkt{t_0, \dotsc, t_4}$,
while \cref{ex-aop-delay-2} displays
an equivalent circuit with a better depth.
Due to the duality principle of Boolean algebra
(see also \citet{Crama}, Theorem 1.3),
any circuit for $g(t)$ over the basis $\{\land, \lor\}$
can be transformed into a circuit for $g^*(t)$ with the same depth and vice
versa by switching all \AND~and \OR~gates.
We also say that $g(t)$ and $g^*(t)$ are \define{dual} Boolean functions.
By duality, the following problem is well-defined.

\problem{\praopdepthopt}
        {$m \in \N$.}
        {Compute a circuit over the basis $\set{\AND2, \OR2}$ realizing an \aop{} on $m$ inputs with minimum depth.}
        {praopdepthopt}

Since, as noted above, computing the propagate and generate signals
from $a$ and $b$
and computing the sum from the propagate signals and carry bits
requires only a constant depth and linear size,
most researchers including ourselves
define the \pradderopt{} as follows.

\problem{\pradderopt}
        {$n \in \N$.}
        {Construct a circuit over the basis $\set{\AND2, \OR2}$
         on $n$ input pairs $p_0, g_0, \dotsc, p_{n-1}, g_{n-1}$
         computing all the carry bits $c_1, \dotsc, c_{n}$.}
        {pradderopt}

\begin{definition} \label{def-adder}
 A circuit solving the \pradderopt{} for some $n \in \N$
 is called an \define{adder circuit} or, short, \define{adder}.
 A family of circuits $\Bkt{A_n}_{n \in \N_{\geq 1}}$
 where circuit $A_n$ solves the \pradderopt{} on $n$ input pairs $p_0, g_0, \dotsc, p_{n-1}, g_{n-1}$
 is called a \define{family of adder circuits}.
 Given $n \in \N_{\geq 1}$,
 $i \in \set{1, \dotsc, n}$, and an adder circuit $A_n$,
 we denote the output of $A_n$ computing the carry bit $c_i$ by $\cktout_i(A_n)$.
\end{definition}

By \cref{carry-recursive},
an adder circuit on $n$ input pairs $p_0, g_0, \dotsc, p_{n-1}, g_{n-1}$
does not depend on $p_0$.
However, to simplify notation, we mention $p_0$ as an input in \cref{def-adder}.


\begin{remark} \label{all aops sep}
 Given $n$ \aop{} circuits $\Bkt{AOP_i}_{i = 1, \dotsc, n}$,
 where $AOP_i$ is an \aop{} circuit on $i$ input pairs,
 we can construct an adder circuit $C$ on $n$ input pairs
 with $\depth(C) = \max_{i \in \set{1, \dotsc, n}} d(AOP_i)$
 and $\size(C) = \sum_{i=1}^n s(AOP_i)$.
\end{remark}

Hence, when the only objective function is depth,
the fastest adder circuits
can be obtained by applying
the best possible \aop{} optimization algorithms to compute all the carry bits separately.
However,
this leads to a size at least quadratic in $n$ as the \aop{}s have at least a size linear in $n$.

Thus, the two problems examined in this work are as follows:
First, find a good algorithm for the
\aop{} optimization problem constructing fast circuits
with a linear size (see \cref{sec-aop-opt}),
secondly, apply this algorithm for adder optimization
while maintaining a linear size (see \cref{sec-adder}).

\subsection{Previous Work} \label{sec::previous_work}

For general circuits using only two-input gates, there is a well-known lower bound on the optimum depth.

\begin{observation}[\citet{Winograd}] \label{general-circuit-lb}
Given $n \in \N$ and any circuit $C$ that depends on $n$ inputs,
a depth-optimum circuit for $C$ has depth at least $\bktU{\log_2 n}$.
\end{observation}

Most adder circuits known so far are prefix adders,
i.e., adder circuits derived via the \prprefix{}.

\problem{\prprefix}
        {An associative operator $\circ \colon \I \times \I \to \I$, $n \in \N$.}
        {Construct a circuit over the basis $\set{\circ}$
         on inputs $z_0, \dotsc z_{n-1}$
         computing the prefixes $Z_i := z_i \circ z_{i-1} \circ \dotsc \circ z_0$
         for all $i = 0, \dotsc, n-1$.}
        {prprefix}

A solution to the \prprefix{} is also called a
\define{parallel prefix graph},
and $\circ$ is called a \define{prefix operator}.

In the easiest special case of the \prprefix{}, the operator $\circ$
is not only associative, but also symmetric.
For instance, we define two \define{symmetric functions}
$\sym\DBkt{x_0, \dotsc, x_{n-1}} := x_0 \land \dotsc \land x_{n-1}$
and $\symdual\DBkt{x_0, \dotsc, x_{n-1}} := x_0 \lor \dotsc \lor x_{n-1}$,
which are dual, and the following sub-problem of the \prprefix{}:

\problem{\prmultiand}
        {$n \in \N$.}
        {Construct a circuit over the basis $\set{\AND2}$
         on inputs $z_0, \dotsc z_{n-1}$
         computing the function $\sym(z_i, \dotsc, z_{0}) = z_i \land z_{i-1} \land \dotsc \land z_0$
         for all $i = 0, \dotsc, n-1$.}
        {prmultiand}

We call a circuit solving the \prmultiand{} an \define{\andprefixckt{}}.

Using a slightly more complicated operator $\circ$,
we obtain the classical way to construct adder circuits,
see, e.g., \citet{Knowles}.
Assume that we compute an adder circuit on $n$ input pairs $p_0, g_0, \dotsc, p_{n-1}, g_{n-1}$,
and define auxiliary variables $z_i = (g_i, p_i) \in \I^2$ for all $i = 0, \dotsc, n-1$.
The \define{adder prefix operator} is given by
\[
\blank \pgate \blank \colon \I^2 \times \I^2 \to \I^2 \,,
{y_1 \choose x_1}
\pgate {y_0 \choose x_0} = {y_1 \lor (x_1 \land y_0) \choose x_1 \land x_0}
\]
for input pairs $(x_0, y_0), (x_1, y_1) \in \I^2$.
It is easy to see that the adder prefix operator is associative.
By \cref{carry-aop}, we have
\begin{equation*} 
 {c_{i+1} \choose p_i \land p_{i-1} \land \dotsc \land p_0}
 = {g_i \choose p_i} \pgate {g_{i-1} \choose p_{i-1}} \pgate \dotsc \pgate {g_0 \choose p_0} 
 = z_i \pgate z_{i-1} \pgate \dotsc \pgate z_0
\end{equation*}
for every $i = 0, \dotsc, n-1$.
Thus, any parallel prefix graph for the prefix operator $\pgate$
yields both an \andprefixckt{} and an adder circuit on $n$ input pairs,
and adders that can be obtained this way are called \define{prefix adders}.
Note that transforming a parallel prefix graph into an adder circuit
means replacing each prefix gate by $3$ logic gates, i.e., \AND{}2 or \OR{}2 gates.
Hence, a prefix-gate size of $n$ yields a logic-gate size of $3n$,
and a prefix-gate depth of $d$ yields a logic-gate depth between $d$ and $2d$.

Parallel prefix graphs with optimum depth $\log_2 n$
in terms of prefix gates can be found, e.g, in
\citet{Sklansky60}, 
\citet{KS73}, or 
\citet{LF80}. 
In practice, the construction by \citet{KS73} is widely used
as it has a fanout of $2$,\label{fanout-page}
i.e., at most two successors per node.
See \citet{Zim98} for an overview on these and other constructions for parallel prefix graphs.
We especially note the method by \citet{LF80}
which we will use as a sub-routine in \cref{sec-adder}.
\citet{LF80} propose several families of parallel prefix graphs
in order to optimize different tradeoffs of depth and size.

\begin{qedtheorem}[\citet{LF80}] \label{Ladner Fischer}
Let $\circ$ be a binary associative operator.
For any $n \in \N_{> 0}$ and any $0 \leq f \leq \lceil \log_2 n \rceil$,
the \prprefix{} can be solved with depth at most $\lceil \log_2 n \rceil + f$
and size at most $2\left(1 + 2^{-f}\right)n$.

In particular, for any $0 \leq f \leq \lceil \log_2 n \rceil$,
there is a circuit $S\!^f_n$ solving the \prmultiand{}
with $\depth\Bkt{S\!^f_n} \leq \lceil \log_2 n \rceil + f$
and $\size\Bkt{S\!^f_n} \leq 2\left(1 + 2^{-f}\right)n$.

Moreover, for any $0 \leq f \leq \lceil \log_2 n \rceil$,
there is a combined circuit $L_n^f$ solving the \pradderopt{}
and the \prmultiand{} on $n$ input pairs at the same time.
For the adder circuit, the resulting depth is at most $2(\lceil \log_2 n \rceil + f)$,
while for the \andprefixckt{}, it is at most $\lceil \log_2 n \rceil + f$.
The total size of the circuit is bounded by $6\left(1 + 2^{-f}\right)n$.
\end{qedtheorem}

Ladner-Fischer circuits for the parallel prefix problem
are constructed by induction on $n$.
Following the detailed size analysis by \citet{Wegener},
it is not hard to show that the number of steps for the construction
of the circuit for $n$ inputs and $0 \leq f \leq \bktU{\log_2 n}$
is bounded by $2\left(1 + 2^{-f}\right)n$.
This implies the following statement.

\begin{qedproposition}\label{LF-runtime}
 Given $n \in \N_{> 0}$ and $0 \leq f \leq \bktU{\log_2 n}$,
 the parallel prefix circuits $S\!^f_n$ by \citet{LF80} can be computed in time $\mathcal O(n)$.
\end{qedproposition}

However, if no care is taken, a depth-optimum parallel prefix graph
with a depth of $\log_2 n$ in terms of prefix gates
leads to a prefix adder with a high depth of $2 \log_2 n$
in terms of $\AND{}2$ and $\OR{}2$ gates.
Some prefix adders optimize \AND{}2-\OR{}2 depth directly
in order to overcome this problem.
\citet{RWS07} provide a prefix adder with \AND{}2-\OR{}2 depth
$2 \log_2 n + 6 \log_2 \log_2 n + \const$
using at most $\mathcal O(n \log_2 \log_2 n)$ logic gates.
\citet{Held-Spirkl-AOPs} improved this result to a depth of
$1.441 \log_2 n + 5 \log_2 \log_2 n + \const$.
Regarding \aop{} computation,
based on \citet{Werber-Fib-AOPs},
\citet{Held-Spirkl-AOPs}
obtain a size of $3n - 3$ and a depth of 
\begin{equation}\label{eq::sophie_aop_depth}
1.441 \log_2 m + 2.674\,.
\end{equation}
\citet{Held-Spirkl-AOPs} give a lower bound of
$\log_\varphi n - 1 \approx 1.44 \log_2 n - 1$
on the depth of \aop{} circuits on $n$ input pairs constructed via prefix graphs
(thus also of prefix adders),
where $\varphi = \frac{1+\sqrt{5}}{2} \approx 1.618$ is the golden ratio.
Hence, at least for \aop{} optimization,
the circuits by \citet{Held-Spirkl-AOPs} are
the best possible prefix adders up to a small additive constant;
and using prefix adders, no substantial improvement is possible.

Thus, we now consider non-prefix adders.
A non-prefix adder with depth $\log_2 n + 7 \sqrt{2 \log_2 n} + 14$ and size $9n$
has been proposed by \citet{Khrapchenko-construction}.
\citet{gashkov07} provided improvements for several small values of $n$.
For arbitrarily large $n$, but still only for $n$ that are a power of $2$,
the construction has been improved by \citet{HSAdders} to a
depth of
\[\log_2 n + 8 \BktU{\sqrt{\log_2 n}} + 6 \BktU{\log_2 \BktU{\sqrt{\log_2 n}}} + 2\]
and a size of at most $13.5 n$ (even at most $9.5 n$ if $n \geq 4096$).
This is the previously best known upper bound on the depth of a linear-size adder circuit with $n$ input pairs.

For \aop{}s, when ignoring circuit size,
the depth optimization problem is actually solved from an asymptotic point of view:
\citet{Grinchuk} showed that given $m \in \N_{\geq 2}$,
an \aop{} on $m+1$ inputs can be realized by a circuit with depth at most
\begin{equation} \label{grin-depth}
\log_2 m + \log_2 \log_2 m + 3\,. 
\end{equation}

Note that the depth bound claimed in \cite{Grinchuk} is actually by $1$ better than
the bound stated in \cref{grin-depth}
but there appears to be a gap in the proof of Lemma 7 in \cite{Grinchuk}.
The upper bound stated in \cref{grin-depth} follows easily from the existing part
of the proof of Grinchuk's Lemma 7.

Using the same methods but a more sophisticated analysis,
\citet{Ley2022} improved Grinchuk's depth bound of
$\log_2 m + \log_2 \log_2 m + 3$ to  
\begin{equation}\label{ley-depth}
\log_2 m + \log_2 \log_2 m + 1.5\,. 
\end{equation}
So far, this was the best bound on the depth of a \aop{} circuit with
$m+1$ inputs. The work by \citet{Ley2022} also contains a detailed
description of the gap in the analysis by \cite{Grinchuk}.

\citet{Commentz-Walter-mon} proved that there is a constant $c$
and some $M \in \N$ such that for any $m \geq M$,
the optimum depth of an \aop{} circuit
is at least
$\log_2 m + \log_2 \log_2 m + c$.
For $M = 2^{2^{18}}$, \citet{Hitzschke2018} showed $c = -5.02$.
Hence, Grinchuk's circuits are depth-optimum up to an additive
constant of roughly $8$.
Grinchuk does not analyze the size or fanout of the circuits arising from his proof.
But using that they are formula circuits
with a special structure, it is not hard to see
that the fanout is bounded by the depth, so logarithmic in $m$,
and that the size is thus at most $\mathcal O(m \log_2 m)$.

\begin{sloppypar}
As a depth-optimum $n$-bit adder circuit can be constructed from
depth-optimum \aop{} circuits on $1, 3, \dotsc, 2n - 1$ inputs,
the lower bound of $\log_2 n + \log_2 \log_2  n + \const$ on the depth
by \citet{Commentz-Walter-mon} also holds for adder circuits.
Note that (for $n$ large enough) the additive constant is roughly 
$-4.01$ in this case.
\end{sloppypar}

\subsection{Our Contributions}

In \cref{sec-aop-opt},
we develop an algorithm
with running time $\mathcal O(m \log_2 m)$
that constructs the fastest known circuits for \aop{}s
with respect to depth.
The depth $\depth(C)$ of the \aop{} circuit $C$ constructed by our algorithm
for $m$ inputs can be bounded by
\begin{equation} \label{our-aop-depth}
\depth(C) \leq \log_2 m + \log_2 \log_2 m + 0.65\,,
\end{equation}
i.e., we improve the additive constant compared to \cref{grin-depth}
and  \cref{ley-depth} to $0.65$.
However, the main advantage of our approach over the previous ones
is that we can bound the size of $C$ by $\size(C) \leq 3.67 m - 2$.
Thus, we present the first \aop{} circuits known to
reach the best possible
asymptotic depth bound by \citet{Commentz-Walter-mon}
and, simultaneously, a linear size.
Our result is based on our novel constructions for symmetric trees,
where symmetric trees are circuits whose underlying undirected graph 
is a tree and that compute a symmetric function. We construct them in 
such a way that different trees may share the same subtrees, which
enables us to save gates. We present these circuits 
in \cref{sec-sym-share}.

In \cref{sec-adder},
we consider the \pradderopt{}.
We construct several families $\Bkt{A_{n}}_{n \in \N}$
of adder circuits with a linear size and a very good depth.
Our core idea for adder optimization
is to apply our \aop{} optimization algorithm from \cref{sec-aop-opt}
in a careful way such that the total size of the circuit
remains linear instead of becoming quadratic.
To this end, we proceed in several steps:
In \cref{sec: step 1}, we construct a family of adder circuits $\Bkt{A^1_n}_{n \in \N}$
with $d(A^1_n) \leq \log_2 n + \log_2 \log_2 n + 2.65$
and sub-quadratic size $s(A^1_n) \leq 6.2 n \log_2 n$.
In \cref{sec: linearization framework},
we develop a general framework for linearizing the size of adder circuits,
increasing the depth only by an additive term of $\log_2 \log_2 \log_2 n + \const$.
In \cref{sec: step 2}, we apply our linearization
to the adder family $\Bkt{A^1_n}_{n \in \N}$,
obtaining an adder family $\Bkt{A^2_n}_{n \in \N}$ with
\[d(A^2_n) \leq \log_2 n + \log_2 \log_2 n + \log_2 \log_2 \log_2 n + 6.6\]
and $s(A^2_n) \leq 21.6 n$.
If we allow the additive constant in the depth bound to increase from $6.6$ to $7.6$,
then, applying our linearization framework with slightly different parameters,
we can construct adder circuits $A^3_n$ with $s(A^3_n) \leq 16.7 n$.
The running time needed to construct $A^2_n$ is $\mathcal O(n \log_2 n)$,
while it is $\mathcal O(n \log_2 \log_2 n)$ for $A^3_n$.

Hence, the depth of both $A^2_n$ and $A^3_n$
is by an additive term of $\log_2 \log_2 \log_2 n + \const$
away from the lower bound
of $\log_2 n + \log_2 \log_2 n + \const$ on depth by \citet{Commentz-Walter-mon}.
This significantly improves the gap to the lower bound in comparison
with the previously best depth of
$\log_2 n + 8 \BktU{\sqrt{\log_2 n}} + 6 \BktU{\log_2 \BktU{\sqrt{\log_2 n}}} + 2$
achieved by the adder circuits of \citet{HSAdders}
(here, $n$ needs to be a power of $2$),
for which the gap is in the order of $\Theta\Bkt{\sqrt{\log_2 n}}$.
The size of the circuits by Held and Spirkl is at most $13.5n$,
and even at most $9.5n$ if $n \geq 4096$.
As for their analysis, Held and Spirkl assume that $n$ is a power of two,
for arbitrary $n$, their depth bound increases by a constant,
and their size bound increases up to a factor of $2$.
Thus, there are arbitrarily large instances where our circuits $A^3_n$
have a better size than the circuits by \citet{HSAdders}.

\begin{table}
\begin{center}
\begin{tabular}{|l|l|l|l|}\hline
Author & Year & Upper bound on depth & Upper bound on size \\ \hline \hline
\citet{Sklansky60}$^*$ & 1960
       & $2 \lceil \log_2 n \rceil$         & $3 n \lceil \log_2 n \rceil$\\ \hline
\citet{Ofman63}$^*$ & 1962
       & $4 \lceil \log_2 n \rceil - 2$         & $18n - 3\lceil \log_2 n \rceil - 6$\\ \hline
\citet{Khrapchenko-construction} & 1967
       & $\lceil\log_2 n\rceil + 7 \sqrt{2 \lceil\log_2 n\rceil} + 14$ 
                              & $9n$ \\ \hline
\citet{brent1970addition} & 1970
       & $\lceil \log_2 n \rceil + \sqrt{2 \log_2 n} + 2$
                              & $\mathcal{O}(n^2)$ \\ \hline
\citet{KS73}$^*$ & 1973
       & $2 \lceil \log_2 n \rceil$         & $6\,  n \lceil \log_2 n \rceil - 3 n$\\ \hline 
\citet{LF80}$^*$ & 1980
       & $2 \lceil \log_2 n \rceil + 2f$
                              & $6(1+ 2^{-f})n$ \\ \hline
\citet{BrentKung82}$^*$ & 1982
       & $4 \lceil \log_2 n \rceil - 2$     & $15n - 1.5 \lceil \log_2 n \rceil - 7.5$\\ \hline
\citet{RWS07} & 2007
       & $2 \log_2 n + 6 \log_2 \log_2 n + \const$
                              & $6\, n \log_2 \log_2 n$\\  \hline
\citet{Held-Spirkl-AOPs} & 2017
       & $1.441 \log_2 n + 5 \log_2 \log_2 n + \const$
                              & $6\, n \log_2 \log_2 n$ \\  \hline
\citet{HSAdders} & 2017
       & $ \log_2 n + 8 \BktU{\sqrt{\log_2 n}} + 6 \BktU{\log_2 \BktU{\sqrt{\log_2 n}}} + \const$
                              & $27\, n$\\ \hline \hline
Here (\cref{theorem::step_1_red_version}) &
       & $\log_2 n + \log_2 \log_2 n + 2.65$
                              & $6.2\, n \log_2 n$ \\ \hline
Here (\cref{n log log logn red}) &
       & $\log_2 n + \log_2 \log_2 n + \log_2 \log_2 \log_2 n + 6.6$
                              & $21.6\, n$ \\ \hline
Here (\cref{better-size-n log log logn red}) & 
       & $\log_2 n + \log_2 \log_2 n + \log_2 \log_2 \log_2 n + 7.6$
                              & $16.7\, n$ \\ \hline
\end{tabular}
\end{center}
\caption{Upper bounds on depths and sizes of different
adder circuits for two $n$-bit numbers including our new results
(in chronological order).
Note that all depths and sizes are given with respect to 
\AND{}2 and \OR{}2 gates, which means that depths and sizes that were achieved with
respect to prefix gates (marked by $^*$, i.e.\ \citet{Sklansky60}, \citet{Ofman63},
\citet{KS73}, \citet{BrentKung82},
and \citet{LF80}) had to be multiplied by~$2$ or $3$, respectively.
For most of the older results (\citet{Sklansky60}, 
\citet{Khrapchenko-construction}, \citet{brent1970addition}, \citet{KS73}, and
\citet{BrentKung82}) we follow the depth and size analysis in 
\citet{SpirklMaster} since the authors of the papers do not provide tight bounds. 
For some circuits better bounds on the size are possible if $n$ is a power of $2$.
The circuit proposed by \citet{brent1970addition} in fact computes only
one \aop{} with linear size, so for computing all carry bits, its size had to be multiplied by $n$.
In the adder circuit by \citet{LF80}, $f$ may be chosen between $0$ and 
$\lceil \log_2 n \rceil$.}
\label{table::overview}
\end{table}

\cref{table::overview} summarizes results from literature and our results on
adder circuits. 


Before we describe our new circuits and algorithms in detail,
in \cref{sec::aop_and_adders},
we summarize well-known techniques for constructing
efficient adder and \aop{} circuits that have been used in the most successful
approaches.
Technical proofs are outsourced to \cref{app-proofs}. Finally, \cref{sec::glossary}
contains a list of symbols used in this paper.

\section{\AOP{}s and Adders: Basic Ideas and Recursive Circuit Construction}\label{sec::aop_and_adders}

By \cref{carry-recursive}, the carry-bit function is monotone.
Thus, 
there is always a monotone formula for each carry bit,
and hence a circuit over $\set{\AND, \OR}$ solving the \pradderopt{}.
In contrast, the summation function $s_n$ is non-monotone:
E.g., for $n = 1$, $a = (1)$, $b_0 = (0)$ and $b_1 = (1)$
with $b_0 < b_1$,
we have $s_1\DBkt{a, b_0} = (0, 1) \nless (1, 0) = s_1\DBkt{a, b_1}$.
Hence, a circuit computing the sum of $a$ and $b$ is always non-monotone.
However, it is an open question whether
-- beyond the computation of the propagate signals and the final sum --
inverters can help to construct
adder circuits with, say, a good depth.
Hence, as all previous approaches, we restrict ourselves to \AND{} and
\OR{} gates when stating the {\sc Adder Optimization Problem}.

\begin{figure}[ht!]
  \begin{center}
   \newcommand{\localwidth}{0.42\textwidth}
   \hfill{}
   \begin{subfigure}[t]{\localwidth}
    \begin{center}
     \adjustbox{max width=.7\textwidth}{
      \centering{\begin{tikzpicture}

\node[and-gate] at (4.6,2) (and1){};
\node[or-gate] at (3.6,1) (or1){};

\node[and-gate] at (2.6,0) (and2){};
\node[or-gate] at (1.6,-1) (or2){};

\node[input] at (6.5, 3.5){$p_0$};
\node[input] at (5.5, 3.5){$g_0$};
\node[input] at (4.5, 3.5){$p_1$};
\node[input] at (3.5, 3.5){$g_1$};
\node[input] at (2.5, 3.5){$p_2$};
\node[input] at (1.5, 3.5){$g_2$};

\node[output] (c_1) at (5.5, -2.5){$c_1$};
\node[output] (c_2) at (3.6, -2.5){$c_2$};
\node[output] (c_3) at (1.6, -2.5){$c_3$};

\draw[thick] (and1.output) -- (or1.input 1);
\draw[thick] (and2.output) -- (or2.input 1);
\draw[thick] (or1.output) -- (and2.input 1);

\draw[thick] (1.5, 3) -- (or2.input 2);
\draw[thick] (2.5, 3) -- (and2.input 2);
\draw[thick] (3.5, 3) -- (or1.input 2);
\draw[thick] (4.5, 3) -- (and1.input 2);
\draw[thick] (5.5, 3) -- (and1.input 1);

\draw[thick, ->] (or2.output) -- (c_3);
\draw[thick, ->] (or1.output) -- (c_2);
\draw[thick, ->] (5.5, 3) -- (c_1);

\end{tikzpicture}
     }
     }
     \caption{A ripple-carry adder, see \cref{rem-ripple}.}
     \label{fig-simple-adders-1}
    \end{center}
   \end{subfigure}
   \hfill{}
   \begin{subfigure}[t]{\localwidth}
    \begin{center}
      \adjustbox{max width=.7\textwidth}{
      \centering{\begin{tikzpicture}

\node[input] (i5) at (1.5, 3.2){$g_2$};
\node[input] (i4) at (2.5, 3.2){$p_2$};
\node[input] (i3) at (3.5, 3.2){$g_1$};
\node[input] (i2) at (4.5, 3.2){$p_1$};
\node[input] (i1) at (5.5, 3.2){$g_0$};
\node[input] (i0) at (6.5, 3.2){$p_0$};

\node[and-gate] at (3,2) (and1){};
\node[and-gate] at (3.8,2) (and2){};
\node[or-gate] at (2.5,1) (or1){};
\node[and-gate] at (4.2,1) (and3){};
\node[or-gate] at (3.5,0) (or2){};

\draw (i3) -- (and1.input 1);
\draw (i4) -- (and1.input 2);
\draw (i2) -- (and2.input 1);
\draw (i4) -- (and2.input 2);
\draw (and1.output) -- (or1.input 1);
\draw (i5) -- (or1.input 2);
\draw (and2.output)   --  (and3.input 2);
\draw (i1) -- (and3.input 1);
\draw (and3.output) -- (or2.input 1);
\draw (or1.output)   -- (or2.input 2);

\node[and-gate] at (6.1,2) (and4){};
\draw (i1) -- (and4.input 1);
\draw (i2) -- (and4.input 2);
\node[or-gate] at (5.1,1) (or3){};
\draw (and4.output) -- (or3.input 1);
\draw (i3) -- (or3.input 2);

\node[output] at ($(or2.output) - (0, 0.7)$) (c_3) {$c_3$};
\node[output] at ($(c_3) + (2, 0)$) (c_1) {$c_1$};
\node[output] at ($(c_1)!0.5!(c_3)$) (c_2) {$c_2$};
\draw[->] (or2.output) -- (c_3);
\draw[->] (i1) -- (c_1);
\draw[->] (or3.output) -- (c_2);

\phantom{\node[output] (c_2) at (3.6, -2.8){$c_2$};}

\end{tikzpicture}
     }
     }
     \caption{An adder circuit arising from applying \cref{all aops sep}
        to depth-optimum \aop{} circuits.}
     \label{fig-simple-adders-2}
    \end{center}
   \end{subfigure}
   \hfill{}
   \caption{Two adder circuits on $n = 3$ input pairs.}
   \label{fig-simple-adders}
  \end{center}
 \end{figure}
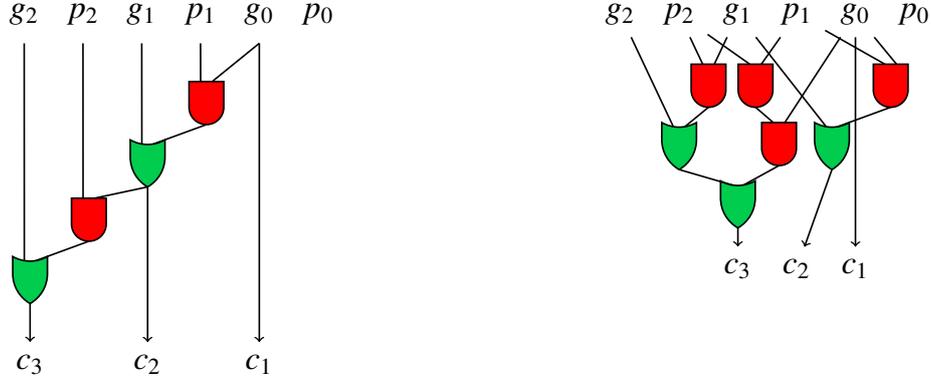

The circuit that directly emerges from \cref{carry-aop} applied for computing all carry bits
is called a \define{ripple-carry adder}.
A ripple-carry adder on $n = 3$ input pairs is shown in
\cref{fig-simple-adders-1}.

\begin{observation} \label{rem-ripple}
 The ripple carry adder for $n$ input pairs has depth and size $2n - 2$.
\end{observation}

Instead of computing each carry bit by a standard \aop{} circuit as in the ripple carry adder,
one might use fast \aop{} circuits to compute each carry bit separately.
\cref{fig-simple-adders-2} gives an example with $n = 3$ input pairs
where each carry bit is realized by a depth-optimum \aop{} circuit.


For \aop{}s with a very short length, depth-optimum circuits are easy to find.
For instance, for $m = 5$,
\totalref{ex-aop-delay-2} shows a realization for $g(t)$ with depth $3$,
which is optimum by \cref{general-circuit-lb}.

For general $m$, a common strategy to find \aop{} realizations with good depth is
to reduce the problem to the construction of \aop{}s of strictly smaller lengths.
In the rest of this \lcnamecref{sec::aop_and_adders},
we shall see \define{recursion formulas} that help improving the upper bound on the optimum depth.
All these recursion formulas can be found in \citet{Grinchuk},
and in different form and for certain special cases also in earlier works.
In order to describe the recursion formulas,
we need the following generalization of \aop{}s.

\begin{definition}\label{def-extended-aop}
 Let $n, m \in \N$ and
 inputs ${s = \Bkt{s_0, \dotsc, s_{n-1}}}$ and ${t = \Bkt{t_0, \dotsc, t_{m-1}}}$ be given.
 We call each of the functions
 \newcommand{\longestterm}{$t_0 \land g^*\DBkt{t_1, \dotsc, t_{m-1}}$}
 \begin{align*}
  f\bkt{s, t} = \begin{cases}
   \sym(s) & \text{if } m = 0 \\
   \sym(s) \land g\bkt{t} & \text{if } m \geq 1
                \end{cases}
  \quad \text{and} \quad
  f^*\bkt{s, t} =
  \begin{cases}
   \symdual(s) & \text{if } m = 0 \\
   \symdual(s) \lor g^*\bkt{t} & \text{if } m \geq 1
                \end{cases}
 \end{align*}
 an \define{extended \aop} on $n+m$ inputs.
 We call $t$ the \define{alternating inputs}
 and $s$ the \define{symmetric inputs} of the extended \aop{}s $f(s, t)$ and $f^*(s, t)$.
\end{definition}

\begin{figure}[t]
  \begin{center}
  \begin{subfigure}[t]{0.47\textwidth}
  \adjustbox{max width=\textwidth}{
  \centering{\begin{tikzpicture}
 \node[input] (x0)  at (-1.5, 6.2) {$s_0$};
\node[input] (x1)  at (-0.5, 6.2) {$s_1$};
\node[input] (x2)  at (0.5,  6.2) {$t_0$};
\node[input] (x3)  at (1.5,  6.2) {$t_1$};
\node[input] (x4)  at (2.5,  6.2) {$t_2$};
\node[input] (x5)  at (3.5,  6.2) {$t_3$};
\node[input] (x6)  at (4.5,  6.2) {$t_4$};
\node[input] (x7)  at (5.5,  6.2) {$t_5$};
\node[input] (x8)  at (6.5,  6.2) {$t_6$};
\node[input] (x9)  at (7.5,  6.2) {$t_7$};
\node[input] (x10) at (8.5,  6.2) {$t_{8}$};
\node[input] (x11) at (9.5,  6.2) {$t_{9}$};

\draw[thick, decorate,decoration={brace, amplitude=5pt, raise=10pt}] (x0.west) -- (x1.east) node [midway, above, sloped, yshift=15pt, scale = 1.6] {$s$};
\draw[thick, decorate,decoration={brace, amplitude=5pt, raise=10pt}] (x2.west) -- (x6.east) node [midway, above, sloped, yshift=15pt, scale = 1.6] {$t'$};
\draw[thick, decorate,decoration={brace, amplitude=5pt, raise=10pt}] (x7.west) -- (x11.east) node [midway, above, sloped, yshift=15pt, scale = 1.6] {$t''$};

\node[or-gate] at (7,3) (or1){};
\node[and-gate] at (6,2) (and1){};

\node[or-gate] at (5,1) (or4){};
\node[and-gate] at (4,0) (and2){};

\node[or-gate] at (3,-1) (or5){};

\node[and-gate] at (8,4) (and7){};
\node[or-gate] at (9,5) (or8){};

\node[and-gate] at (2,-2) (and6){};
\node[or-gate] at (1,-3) (or9){};

\node[sym-or-gate] at (0,-4) (and10){};
\node[sym-or-gate] at (-1,-5) (or11){};

\draw (or1.output) -- (and1.input 1);
\draw (x8) -- (or1.input 2);
\draw (x7) -- (and1.input 2);
\draw (x9) -- (and7.input 2);

\draw (or4.output) -- (and2.input 1);
\draw (or8.output) -- (and7.input 1);

\draw (x5) -- (and2.input 2);
\draw (x10) -- (or8.input 2);
\draw (x11) -- (or8.input 1);

\draw (x6) -- (or4.input 2);
\draw (and1.output) -- (or4.input 1);

\draw (and2.output) -- (or5.input 1);
\draw (or5.output) -- (and6.input 1);

\draw (x4) -- (or5.input 2);
\draw (x3) -- (and6.input 2);

\draw (x2) -- (or9.input 2);
\draw (and6.output) -- (or9.input 1);
\draw (and7.output) -- (or1.input 1);

\draw (or9.output) -- (and10.input 1);
\draw (x1) -- (and10.input 2);
\draw (and10.output) -- (or11.input 1);
\draw (x0) -- (or11.input 2);

\draw[->] (or11.output) -- ($(or11.output) - (0, 0.5)$);

\end{tikzpicture}}
  }
  \caption{The standard \aop{} circuit for the extended \aop{} $f^*\Bkt{\Bkt{s_0, s_1}, \Bkt{t_0, \dotsc, t_{9}}}$.}
  \label{fig-aop-split-1}
 \end{subfigure}
 \hfill
 \begin{subfigure}[t]{0.47\textwidth}
   \adjustbox{max width=\textwidth}{
  \centering{\begin{tikzpicture}

\node[input] (x0) at (-3.5, 6.2){$s_0$};
\node[input] (x1) at (-2.5, 6.2){$s_1$};
\node[input] (x2) at (-1.5, 6.2){$t_0$};
\node[input] (x3) at (-0.5, 6.2){$t_1$};
\node[input] (x4) at (0.5, 6.2){$t_2$};
\node[input] (x5) at (1.5, 6.2){$t_3$};
\node[input] (x6) at (2.5, 6.2){$t_4$};
\node[input] (x7) at (3.5, 6.2){$t_5$};
\node[input] (x8) at (4.5, 6.2){$t_6$};
\node[input] (x9) at (5.5, 6.2){$t_7$};
\node[input] (x10) at (6.5, 6.2){$t_{8}$};
\node[input] (x11) at (7.5, 6.2){$t_{9}$};

\draw[thick, decorate,decoration={brace, amplitude=5pt, raise=10pt}] (x0.west) -- (x1.east) node [midway, above, sloped, yshift=15pt, scale = 1.6] {$s$};
\draw[thick, decorate,decoration={brace, amplitude=5pt, raise=10pt}] (x2.west) -- (x6.east) node [midway, above, sloped, yshift=15pt, scale = 1.6] {$t'$};
\draw[thick, decorate,decoration={brace, amplitude=5pt, raise=10pt}] (x7.west) -- (x11.east) node [midway, above, sloped, yshift=15pt, scale = 1.6] {$t''$};

\node[and-gate] at (2,5) (and1){};
\draw[thick] (x6) -- (and1.input 1);
\draw[thick] (x5) -- (and1.input 2);

\node[or-gate] at (1.5,4) (or1){};
\draw[thick] (and1.output) -- (or1.input 1);
\draw[thick] (x4) -- (or1.input 2);

\node[and-gate] at (1,3) (and2){};
\draw[thick] (or1.output) -- (and2.input 1);
\draw[thick] (x3) -- (and2.input 2);

\node[or-gate] at (.5,2) (or9){};
\draw[thick] (and2.output) -- (or9.input 1);
\draw[thick] (x2) -- (or9.input 2);

\node[sym-or-gate] at (0,1) (and10){};
\draw[thick] (or9.output) -- (and10.input 1);
\draw[thick] (x1) -- (and10.input 2);

\node[sym-or-gate] at (-.5,0) (or11){};
\draw[thick] (and10.output) -- (or11.input 1);
\draw[thick] (x0) -- (or11.input 2);

\node[or-gate] at (7,5) (or6){};
\draw[thick] (x11) -- (or6.input 1);
\draw[thick] (x10) -- (or6.input 2);

\node[and-gate] at (6.5,4) (and2){};
\draw[thick] (or6.output) -- (and2.input 1);
\draw[thick] (x9) -- (and2.input 2);

\node[or-gate] at (6,3) (or3){};
\draw[thick] (and2.output) -- (or3.input 1);
\draw[thick] (x8) -- (or3.input 2);

\node[and-gate] at (5.5,2) (and9){};
\draw[thick] (or3.output) -- (and9.input 1);
\draw[thick] (x7) -- (and9.input 2);

\node[sym-and-gate] at (4.9,1) (and8){};
\draw[thick]       (and9.output) -- (and8.input 1);
\draw[thick] (x5) -- (and8.input 2);

\node[sym-and-gate] at (4.3,0) (and5){};
\draw[thick] (and8.output) -- (and5.input 1);
\draw[thick] (x3) -- (and5.input 2);

\node[concat-or-gate] at (1.5,-2) (or4){};
\draw[thick] (and5.output) -- (or4.input 1);
\draw[thick] (or11.output) -- (or4.input 2);
\node[outer sep=0pt] (x10) at (2.1, -2.1){};

\draw[->] (or4.output) -- ($(or4.output)- (0, 0.5)$);

\phantom{\node[sym-or-gate] at (-1,-5) (or11){};}
\phantom{\draw[->] (or4.output) -- ($(or11.output)- (0, 0.5)$);}
\end{tikzpicture}}
  }
   \caption{A circuit realizing the function $f^*\Bkt{\Bkt{s_0, s_1}, \Bkt{t_0, \dotsc, t_{9}}}$ as indicated by \cref{aop-split} with $k = 5$.}
  \label{fig-aop-split-2}
  \end{subfigure}
  \end{center}
 \caption{Illustration of the alternating split.
 For an extended \aop{},
 the gates fed by alternating inputs are colored red (\AND) and green (\OR);
 the gates fed by symmetric inputs are colored yellow;
 the concatenation gate for the split is colored blue.}
 \label{fig-aop-split}
\end{figure}
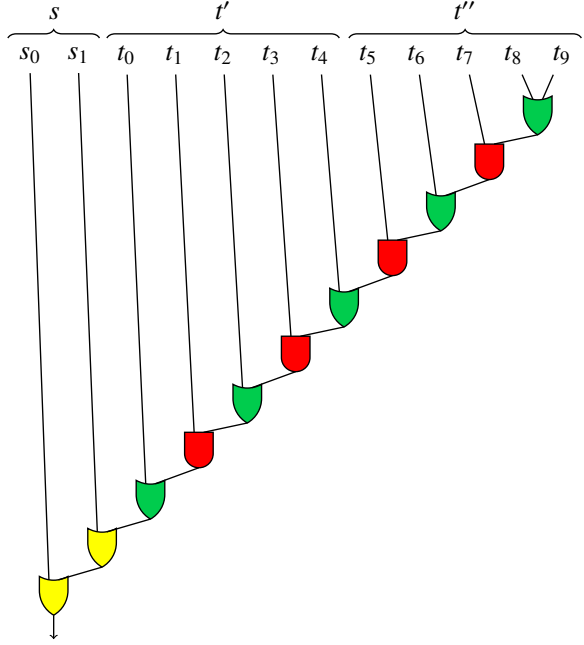
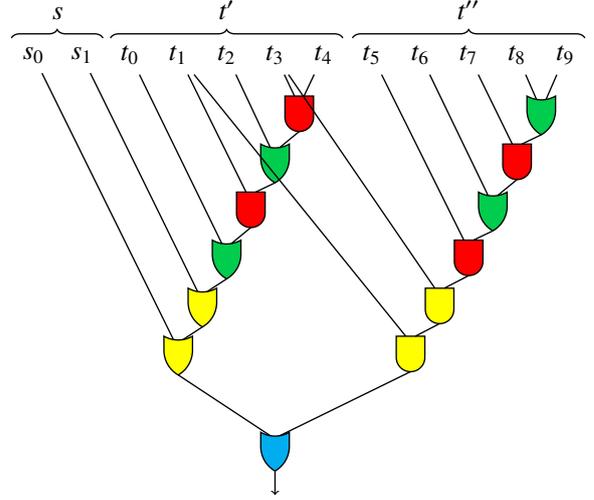

Again, we call the canonical circuits arising from this \lcnamecref{def-extended-aop}
standard circuits.
\cref{fig-aop-split-1} shows the standard circuit for an extended \aop{}
with $2$ symmetric inputs and $10$ alternating inputs computing
$f\bkt{\Bkt{s_0, s_{1}}, \Bkt{t_0, \dotsc, t_{9}}}$.

We shortly consider the special case $m \leq 1$, where
$f(s, t)$ and $f^*(s, t)$ are symmetric functions.
It is well known that a depth-optimum circuit for a symmetric function
can be built easily as part of a full binary tree.

\begin{observation} \label{sym-log-n}
Given $n \in \N$, a depth-optimum circuit for $\sym\DBkt{x_0, \dotsc, x_{n-1}}$
has depth $\bktU{\log_2 n}$
and can be constructed in linear time.
\end{observation}

Note that a circuit for $\sym\DBkt{x_0, \dotsc, x_{n-2} \lor x_{n-1}}$
can be constructed from a circuit for $\sym\DBkt{x_0, \dotsc, x_{n-1}}$
by reordering the inputs such that $x_{n-2}$ and $x_{n-1}$ are
connected by an \AND{} gate,
and then replacing $x_{n-2} \land x_{n-1}$ by $x_{n-2} \lor x_{n-1}$.

\begin{qedcorollary} \label{almost-sym-log-n}
	Given $n \in \N$, a depth-optimum circuit for $\sym\DBkt{x_0, \dotsc, x_{n-3}, x_{n-2} \lor x_{n-1}}$
has depth $\bktU{\log_2 n}$
and can be constructed in linear time.
\end{qedcorollary}

For larger $m$, we apply the following recursive strategies
for circuit construction.
In order to allow a compact description,
given input variables $t = \Bkt{t_0, \dotsc, t_{m-1}}$ with $m$ odd,
we write
$\everysecond{t} := (t_1, t_3, t_5, \dotsc, t_{m-2})$.\label{everysecond-page}

The most important recursion formula used in this work is as follows.
For a proof, see, e.g., Lemma 3 in \citet{Grinchuk}
or Lemma 2.5 in \citet{BH-Theory-AOPs}.

\begin{qedproposition}[Alternating split] \label{aop-split}
 Let inputs $s = \Bkt{s_0, \dotsc, s_{n-1}}$ and $t = \Bkt{t_0, \dotsc, t_{m-1}}$
 and an odd integer $k$ with $1 \leq k < m$ be given.
 Denote by $\prefix{t}{2k+1}$ the odd-length prefix $\prefix{t}{k} = (t_0, t_1, \dotsc, t_{k-1})$ of $t$,
 and by $\suffix{t}{k}$ the remaining inputs of $t$, i.e., $\suffix{t}{k} = (t_{k}, \dotsc, t_{m-1})$.
 Then, we have
 \begin{equation*}
 f^*(s, t) = f^*\Bkt{s, \prefix{t}{2k+1}} \lor f\Bkt{\everysecond{\prefix{t}{2k+1}}, \suffix{t}{2k+1}}
 \quad \text{and} \quad
 f(s, t) = f\Bkt{s, \prefix{t}{2k+1}} \land f^*\Bkt{\everysecond{\prefix{t}{2k+1}}, \suffix{t}{2k+1}}\,. \qedhere
 \end{equation*}
\end{qedproposition}

\cref{fig-aop-split} shows an illustration of the alternating split
with a prefix of length $5$.
We use standard circuits for $f^*\Bkt{s, \Bkt{t_0, \dotsc, t_4}}$
and $f\Bkt{\Bkt{t_1, t_3}, \Bkt{t_5, \dotsc, t_{9}}}$
for illustration purposes.
Any circuits realizing these functions could be used here.
We can see directly that the circuits in \cref{fig-aop-split-1,fig-aop-split-2}
are logically equivalent:
Consider any truth assignment of the inputs.
In the case that $t_5$ is false, clearly both circuits
compute the extended \aop{} $f^*\Bkt{s, \Bkt{t_0, \dotsc, t_4}}$.
If $t_5$ is true, the circuit in \cref{fig-aop-split-1}
computes a true output signal if and only if $f^*\Bkt{s, \Bkt{t_0, \dotsc, t_4}}$
is true or both $f\Bkt{(), \Bkt{t_6, \dotsc, t_9}}$ and
$t_3 \land t_1$ are true.
Note that exactly the same holds for the circuit in \cref{fig-aop-split-2}.
This argumententation for the example can easily be generalized to prove
\cref{aop-split}.

A special case of \cref{aop-split} is
the following statement which is used, e.g., by \citet{Werber-Fib-AOPs}.

\begin{corollary}\label{alternating-split-propagate}
Let input variables $t = \Bkt{t_0, \dotsc, t_{m-1}}$
and an odd integer $k$ with $1 \leq k < m$ be given.
Then, we have
\begin{align*}
g^*(t) &= g^*\DBkt{t_0, \dotsc, t_{k-1}} \lor \Bkt{\sym\DBkt{t_1, t_3, \dotsc, t_{k}} \land g^*\DBkt{t_{k+1}, \dotsc, t_{m-1}}}\,, \\
g(t)   &= g\DBkt{t_0, \dotsc, t_{k-1}} \land \Bkt{\symdual\DBkt{t_1, t_3, \dotsc, t_{k}} \lor g\DBkt{t_{k+1}, \dotsc, t_{m-1}}}\,.
\end{align*}
\begin{proof}
 By duality, it suffices to prove the first statement.
 From \cref{aop-split}, we deduce
 \begin{align*}
 g^*(t) &=
   g^*\DBkt{t_0, \dotsc, t_{k-1}} \lor f\Bkt{(t_1, t_3, \dotsc, t_{k-2}), (t_k, \dotsc, t_{m-1})} \\
    &= g^*\DBkt{t_0, \dotsc, t_{k-1}} \lor \Bkt{\sym\DBkt{t_1, , t_3, \dotsc, t_{k-2}} \land g\DBkt{t_k, \dotsc, t_{m-1}}} \\
    &= g^*\DBkt{t_0, \dotsc, t_{k-1}} \lor \Bkt{\sym\DBkt{t_1, , t_3, \dotsc, t_{k}} \land g^*\DBkt{t_{k+1}, \dotsc, t_{m-1}}}\,. \tag*{\qedhere}
 \end{align*}
\end{proof}
\end{corollary}

The definition of extended \aop{}s implies other ways to realize extended \aop{}s recursively.
We shall call these \define{symmetric splits}.

\begin{observation}[Symmetric splits] \label{obs-sym-splits}
Given inputs $s = \Bkt{s_0, \dotsc, s_{n-1}}$ and $t = \Bkt{t_0, \dotsc, t_{m-1}}$,
any $k \in \set{1, \dotsc, n}$
and any $k$-element sub-tuple $s' = (s_{i_0}, \dotsc, s_{i_{k-1}})$ of $s$,
we have
\begin{alignat}{5}
f\Bkt{s, t} &= \sym(s') \land f\Bkt{s \backslash s', t} \label{symmetric split subset}
\quad && \text{and} \quad
f^*\Bkt{s, t} &&= \symdual(s') \lor f^*\Bkt{s \backslash s', t} \,.
\intertext{In particular, with $k = n$, we have}
f\Bkt{s, t} &= \sym(s) \land g(t) \label{symmetric split A}
\quad && \text{and} \quad
f^*\Bkt{s, t} &&= \symdual(s) \lor  g^*(t)\,.
\end{alignat}
\end{observation}

\section{Optimization of \AOP{}s} \label{sec-aop-opt}

In this \lcnamecref{sec-aop-opt},
we present the first algorithm for \aop{} optimization
which constructs \aop{} circuits
that are depth-optimum up to an additive constant
and, at the same time, have a linear size.
In \cref{sec-grinchuk}, we describe our algorithm,
and in \cref{sec-grinchuk-depth}, we estimate the depth bound
of the arising circuits.
In \cref{sec-grinchuk-size} to \cref{sec::linear_aops},
we show that our algorithm can be implemented such that
our circuits have a linear size.
The main ingredient for this is our novel symmetric tree construction
based on so-called \emph{leftist circuits} developed in \cref{sec-sym-share}.

\subsection{Algorithm for Depth Optimization of \AOP{}s} \label{sec-grinchuk}

The idea of our algorithm is based on \citet{Grinchuk}.
The algorithm is a recursive approach using some of the recursion strategies presented 
in \cref{sec::aop_and_adders}
that work on extended \aop{}s rather than on \aop{}s only.
For analyzing the depth of the arising circuits,
we proceed the following way:
Fixing a depth bound $d \in \N$ and a number $n$ of symmetric inputs,
we determine up to which number $m$ of alternating inputs
an extended \aop{} can be realized by a circuit with depth $d$.
From this, we will derive the claimed depth bound in \cref{sec-grinchuk-depth}.

\begin{remark_no_num}
 The notation for (extended) \aop{}s used by \citet{Grinchuk} differs from ours:
 We always consider an \aop{} $f(t) = f\DBkt{t_0, \dotsc, t_{m-1}}$ with $m$ inputs,
 while Grinchuk analyzes an \aop{} with $m+1$ inputs,
 i.e., the \aop{} $f(t) = f\DBkt{t_0, \dotsc, t_{m}}$.
\end{remark_no_num}

\begin{definition}[\citet{Grinchuk}] \label{def m}
 Given $d, n \in \N$, the \emph{capacity} of $d$ and $n$
 is
 \[
  m(d, n) := \max \big\{m \in \N \where{}{} \text{There is a circuit for } f\Bkt{\Bkt{s_0, \dotsc, s_{n-1}}, \Bkt{t_0, \dotsc, t_{m-1}}}
                                            \text{ with depth at most } d\,.\big\}\,,
 \]
 where $m(d, n) := - \infty$ in case there is no such $m$.
\end{definition}

Note that we also allow $m(d, n) = 0$ in \cref{def m}.
\cref{sym-log-n} implies that for $d, n \in \N$, 
$m(d, n)$ is a natural number if and only if $n \leq 2^d$.
Furthermore, by dualization, we have
\[
  m(d, n) = \max \big\{m \in \N \where{}{} \text{There is a circuit for } f^*\Bkt{\Bkt{s_0, \dotsc, s_{n-1}}, \Bkt{t_0, \dotsc, t_{m-1}}}
                                           \text{ with depth at most } d\,.\big\}\,.
\]

Finding a closed formula for $m(d, n)$ for general $d, n \in \N$ is an open problem,
but we will derive a lower bound on $m(d, n)$ in \cref{Grin-main}.
\citet{Grinchuk} gives a lower bound on $m(d, n)$ in two steps:
First, he bounds $m(d, n)$ from below by a recursively defined function $M(d, n)$
which is directly connected to his recursion formulas.
Secondly, he gives a numerical lower bound on the values $M(d, n)$
from which he can derive this depth bound.
However, we skip the intermediate step of defining $M(d, n)$ and directly
give a numerical lower bound on $m(d, n)$
which is marginally stronger than Grinchuk's bound.
This leads to a slightly better depth bound
and simplifies the size analysis of the arising circuit.
Moreover, in order to prove a linear size bound in \cref{depth-opt-size},
we use slightly different recursion formulas than Grinchuk.

For the following definition, the parameter $\xi$ could be set to any
value above $1$ and strictly below $2$. However, the larger its value 
is the better bounds we get on the depth of our circuits. Thus, in order
to work with a fixed value, we set it in the following to $1.999$.

\begin{definition} \label{def-mu}
 Let $\xi := 1.999$.
 We define the function $\mu \colon \N_{\geq 1} \times \N \to \R$ by
 $
  \mu(d, n) := \xi \frac{2^d - n - 2}{d} + 2
 $.
\end{definition}


\begin{definition} \label{def-dmin}
Given $n, m \in \N$ with $m \geq 1$, we define
 $\dmin(n, m) := \min \set{d \in \N_{\geq 1} \where{}{} m \leq \mu(d, n)}$.
\end{definition}

As $\lim_{d \to \infty} \mu(d, n) = \infty$, the value $\dmin(n, m)$ is well-defined.

Given $m$ and $n$,
our depth optimization algorithm (see \cref{grinchuk-alg})
will compute a circuit for an extended \aop{}
with $n$ symmetric and $m$ alternating inputs with depth at most $d$,
where $d$ is minimum with $m \leq \mu(d, n)$.
Before being able to state our algorithm,
we need some basic properties about $\dmin(n, m)$ and $\mu(d, n)$.

\begin{lemma} \label{small-values-of-d}
\cref{table-d-m-n} shows the value $\dmin(n, m)$
for all $1 \leq m \leq 9$ and $0 \leq n \leq 11$.
\begin{proof}
See \cref{app-proofs}, \cpageref{proof-small-values-of-d}.
\end{proof}
\end{lemma}

\begin{table}
\begin{center}
\newcommand{\dcolorone}{\cellcolor{red!25} 1}
\newcommand{\dcolortwo}{\cellcolor{mid_orange!25} 2}
\newcommand{\dcolorthree}{\cellcolor{yellow!25} 3}
\newcommand{\dcolorfour}{\cellcolor{green!25} 4}
\newcommand{\dcolorfive}{\cellcolor{cyan!25} 5}
\newcommand{\dcolorsix}{\cellcolor{blue!25} 6}
\setlength{\tabcolsep}{12pt}
\begin{tabular}{c|rrrrrrrrrrrrr}
  $\tikz{\node[below left, inner sep=2pt] (def) {$m$};%
      \node[above right,inner sep=2pt] (abc) {$n$};%
      \draw (def.north west|-abc.north west) -- (def.south east-|abc.south east);}$
        & 0 & 1 & 2 & 3 & 4 & 5 & 6 & 7 & 8 & 9 & 10 & 11 & 12 \\
      \midrule
      1 & \dcolorone  & \dcolortwo  & \dcolortwo  & \dcolortwo  & \dcolorthree  & \dcolorthree& \dcolorthree& \dcolorthree& \dcolorfour & \dcolorfour & \dcolorfour & \dcolorfour & \dcolorfour \\
      2 & \dcolorone  & \dcolortwo  & \dcolortwo  & \dcolorthree& \dcolorthree& \dcolorthree& \dcolorthree& \dcolorfour & \dcolorfour & \dcolorfour & \dcolorfour & \dcolorfour & \dcolorfour \\
      3 & \dcolortwo  & \dcolorthree& \dcolorthree& \dcolorthree& \dcolorthree & \dcolorfour & \dcolorfour & \dcolorfour & \dcolorfour & \dcolorfour & \dcolorfour & \dcolorfour & \dcolorfive\\
      4 & \dcolorthree& \dcolorthree & \dcolorthree & \dcolorfour & \dcolorfour & \dcolorfour & \dcolorfour & \dcolorfour & \dcolorfour & \dcolorfour & \dcolorfive & \dcolorfive & \dcolorfive\\
      5 & \dcolorthree & \dcolorthree & \dcolorfour & \dcolorfour & \dcolorfour & \dcolorfour & \dcolorfour & \dcolorfour & \dcolorfive & \dcolorfive & \dcolorfive & \dcolorfive & \dcolorfive\\
      6 & \dcolorfour & \dcolorfour & \dcolorfour & \dcolorfour & \dcolorfour & \dcolorfour & \dcolorfive & \dcolorfive & \dcolorfive & \dcolorfive & \dcolorfive & \dcolorfive & \dcolorfive\\
      7 & \dcolorfour & \dcolorfour & \dcolorfour & \dcolorfour & \dcolorfive & \dcolorfive & \dcolorfive  & \dcolorfive  & \dcolorfive  & \dcolorfive  & \dcolorfive  & \dcolorfive & \dcolorfive\\
      8 & \dcolorfour & \dcolorfour  & \dcolorfive  & \dcolorfive  & \dcolorfive  & \dcolorfive  & \dcolorfive  & \dcolorfive  & \dcolorfive  & \dcolorfive  & \dcolorfive  & \dcolorfive & \dcolorfive\\
      9 & \dcolorfive  & \dcolorfive  & \dcolorfive  & \dcolorfive  & \dcolorfive  & \dcolorfive  & \dcolorfive  & \dcolorfive  & \dcolorfive  & \dcolorfive  & \dcolorfive  & \dcolorfive & \dcolorfive\\ \bottomrule
\end{tabular}
\caption{The value $\dmin(n, m)$ for all $1 \leq m \leq 9$ and $0 \leq n \leq 11$
according to \cref{small-values-of-d}.
Cells are colored by the containing number $\dmin(n, m)$.
}
\label{table-d-m-n}
\end{center}
\end{table}


\begin{lemma} \label{n < 2^d}
Given $d, n, m \in \N$ with $d \geq 1$ and $2 \leq m \leq \mu(d, n)$, we have $n \leq 2^d - 2$.
\begin{proof}
We have $m \leq \mu(d, n) = \xi \frac{2^d - n - 2}{d} + 2$ and thus
$n \leq 2^d - \frac{d(m-2)}{\xi} - 2 \overset{m \geq 2}{\leq} 2^d - 2$.
\end{proof}
\end{lemma}

The next two lemmas give concrete realizations for $f(s, t)$
when either the number $m$ of alternating inputs is small
or the expected depth is small.

\begin{lemma}\label{depth-m-small}
Let integers $d, n, m \in \N$ with $d \geq 1$, $0 \leq n < 2^d$
and $m \leq \mu(d, n)$ be given.
Then, for $m \leq 2$,
there is a circuit for $f(\Bkt{s_0, \dotsc, s_{n-1}}, \Bkt{t_0, \dotsc, t_{m-1}})$ with depth at most $d$.
\begin{proof}
 If $d=1$ then we  have
 \begin{equation*}
  m + n \leq \mu(1, n) + n \leq \xi \frac{2^1 - n - 2}{1} + 2 + n = 2 + n(1-\xi) 
    \overset{\xi \geq 1}{\leq} 2^1
 \end{equation*}
 And if $d \geq 2$, under the assumptions of this \lcnamecref{depth-m-small}, we have
 \begin{equation*}
   m + n \leq \xi \frac{2^d - n - 2}{d} + 2 + n
    = \xi \frac{2^d + \left(\frac{d}{\xi} - 1\right) n + 2 \frac{d}{\xi} - 2}{d}
    \overset{\substack{n \leq 2^d - 1, \\ d \geq \xi}}{\leq} 
          \xi \frac{2^d + \left(\frac{d}{\xi} - 1\right) (2^{d}-1) + 2 \frac{d}{\xi} - 2}{d}
    = \frac{d 2^d + d - \xi}{d}
    < 2^d + 1\,,
 \end{equation*}
 and as both $m + n$ and $2^d$ are natural numbers, we conclude $m + n \leq 2^d$.
 For $m \leq 2$, by \cref{sym-log-n}, this implies that $f(s, t)$
 is a symmetric tree that can be realized with depth $d$.
\end{proof}
\end{lemma}

\begin{figure}[!h]
\begin{center}
  \begin{subfigure}[t]{0.48\textwidth}
  \begin{center}
  \adjustbox{max width=.7\textwidth}{
  \centering{\begin{tikzpicture}

\node[input] (s0) at (1.5, 3.2){$s_0$};
\node[input] (s1) at (2.5, 3.2){$s_1$};
\node[input] (t0) at (3.5, 3.2){$t_0$};
\node[input] (t1) at (4.5, 3.2){$t_1$};
\node[input] (t2) at (5.5, 3.2){$t_2$};
\node[input] (t3) at (6.5, 3.2){$t_3$};

\node[and-gate] at (2,2) (and1){};
\draw (s0) -- (and1.input 2);
\draw (s1) -- (and1.input 1);

\node[and-gate] at (6,2) (and2){};
\draw (t2) -- (and2.input 2);
\draw (t3) -- (and2.input 1);

\node[and-gate] at (3,1) (and3){};
\draw (and1.output) -- (and3.input 2);
\draw (t0) -- (and3.input 1);

\node[or-gate] at (5,1) (or1){};
\draw (t1) -- (or1.input 2);
\draw (and2.output) -- (or1.input 1);

\node[and-gate] at (4,0) (and4){};
\draw (and3.output) -- (and4.input 2);
\draw (or1.output) -- (and4.input 1);





\draw[output-edge] (and4.output) -- ($(and4.output) - (0, .5)$) {};


\end{tikzpicture}
  }
  }
  \end{center}
 \caption{Realization of $s_0 \land s_1 \land t_0 \land (t_1 \lor (t_2 \land t_3 ))$
    with depth 3.}
 \label{img-proof_small_1}
\end{subfigure}
\hspace*{0.02\textwidth}
\begin{subfigure}[t]{0.48\textwidth}
  \begin{center}
  \adjustbox{max width=.7\textwidth}{
  \centering{\begin{tikzpicture}

\node[input] (s0) at (1.5, 3.2){$s_0$};
\node[input] (t0) at (2.5, 3.2){$t_0$};
\node[input] (t1) at (3.5, 3.2){$t_1$};
\node[input] (t2) at (4.5, 3.2){$t_2$};
\node[input] (t3) at (5.5, 3.2){$t_3$};
\node[input] (t4) at (6.5, 3.2){$t_4$};

\node[and-gate] at (2,2) (and1){};
\draw (s0) -- (and1.input 2);
\draw (t0) -- (and1.input 1);

\node[or-gate] at (4,2) (or1){};
\draw (t1) -- (or1.input 2);
\draw (t2) -- (or1.input 1);

\node[or-gate] at (5,2) (or2){};
\draw (t1) -- (or2.input 2);
\draw (t3) -- (or2.input 1);

\node[and-gate] at (2.5,1) (and2){};
\draw (and1.output) -- (and2.input 2);
\draw (or1.output) -- (and2.input 1);

\node[or-gate] at (6,1) (or3){};
\draw (or2.output) -- (or3.input 2);
\draw (t4) -- (or3.input 1);

\node[and-gate] at (4,0) (and3){};
\draw (and2.output) -- (and3.input 2);
\draw (or3.output) -- (and3.input 1);





 \draw[output-edge] (and3.output) -- ($(and3.output) - (0, .5)$) {};

\end{tikzpicture}
  }
  }
  \end{center}
 \caption{Realization of $s_0 \land t_0 \land (t_1 \lor (t_2 \land (t_3 \lor t_4)))$ 
     with depth 3.}
 \label{img-proof_small_2}
\end{subfigure}
\end{center}
\caption{Circuits for special cases in the proof of \cref{depth-d-small}.}
\end{figure}
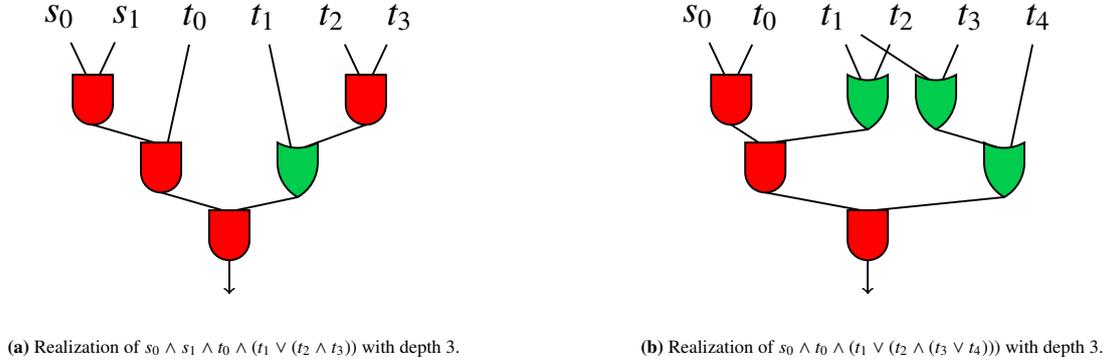

\begin{lemma} \label{depth-d-small}
Let integers $d, n, m \in \N$ with $1 \leq d \leq 3$, $0 \leq n < 2^d$, and $m \leq \mu(d, n)$ be given.
Then, there is a circuit for $f(\Bkt{s_0, \dotsc, s_{n-1}}, \Bkt{t_0, \dotsc, t_{m-1}})$ with depth at most $d$.
\begin{proof}
 \cref{depth-m-small} proves the statement for $m \leq 2$,
 so assume $m \geq 3$.
 It suffices to show the \lcnamecref{depth-d-small} for $d = \dmin(n, m)$.
 From \cref{table-d-m-n}, we can read off the values of $n$ and $m$ for which must verify
 that a circuit for $f(s, t)$ with depth $d$ exists.

 For $d = 1$, by \cref{table-d-m-n}, there is no $m \geq 3$ fulfilling the conditions of this \lcnamecref{depth-d-small}.

 For $d = 2$, \cref{table-d-m-n} and $m \geq 3$ imply $m = 3$ and $n = 0$,
 and the standard realization of $f(s, t)$ has depth $m - 1 = 2$.

 Hence, let $d = 3$. 

 If $m=3$, then \cref{table-d-m-n} implies $n \in \set{1, 2, 3, 4}$.
 For $m = 3$, note that the definition of extended \aop{}s (cf.~\cref{def-extended-aop}) implies
 $f(s, t) = \sym\DBkt{s_0, \dotsc, s_{n-1}, t_0, t_1 \lor t_2}$.
 A depth-optimum circuit for $f(s, t)$ has depth at most
 $\BktU{\log_2 (n + m)} \leq \BktU{\log_2 7} = 3$ by \cref{almost-sym-log-n}.
 
 If $m=4$, then \cref{table-d-m-n} implies $n \in \set{0, 1, 2}$.
 For $n=2$ we have to realize 
 $s_0 \land s_1 \land t_0 \land (t_1 \lor (t_2 \land t_3 ))$ with
 a circuit of depth 3. See \cref{img-proof_small_1} for such a realization.
 Obviously, this also gives a realization with depth $3$ if $n \in \set{0, 1}$.

 If $m=5$, then \cref{table-d-m-n} implies $n \in \set{0, 1}$.
 For $n=1$ we have to realize 
 $s_0 \land t_0 \land (t_1 \lor (t_2 \land (t_3 \lor t_4)))$ with
 a circuit of depth 3. See \cref{img-proof_small_2} for such a realization.
 Obviously, this also gives a realization with depth $3$ if $n = 0$.
\end{proof}
\end{lemma}

For the following main proposition of this \lcnamecref{sec-grinchuk},
we need another calculation lemma.

\begin{lemma} \label{m-k-small-enough}
  Consider $d, m, n, k \in \N$ with
  \begin{multicols}{2}
  \begin{enumerate}
   \item $d \geq 3$, \label{cond-d-large}
   \item $m \leq \mu(d+1, n)$, \label{cond-m-small}
   \item $\mu(d, n) \geq 1$, and \label{cond-mu-large}
   \item $k$ the maximum odd integer such that $k \leq \mu(d, n)$. \label{cond-k-odd}
  \end{enumerate}
  \end{multicols}
  Then, we have $0 \leq \frac{k-1}{2} < 2^d$ and
  \begin{equation} \label{to-show}
   m - k \leq \mu \Bkt{d, \frac{k-1}{2}}\,.
  \end{equation}
\begin{proof}
See \cref{app-proofs}, \cpageref{proof-m-k-small-enough}.
\end{proof}
\end{lemma}

Finally, we will now see that $\BktD{\mu(d, n)}$ is a lower bound on $m(d, n)$ when $n < 2^d$.

\linebox{
\begin{proposition} \label{Grin-main}
 Consider $d, n \in \N$ with $d \geq 1$ and $0 \leq n < 2^d$.
 For all inputs $s = \Bkt{s_0, \dotsc, s_{n-1}}$ and $t = \Bkt{t_0, \dotsc, t_{m-1}}$
 with $m \leq \mu(d, n)$, there is a circuit for
 the extended \aop{} $f(s, t)$ with depth at most $d$.
\end{proposition}
}
 \begin{proof}
  We prove the statement by induction on $d$.

  In the base case we assume that $d \leq 3$.
  Then, \cref{depth-d-small} proves the statement.

  For the induction step, assume that the \lcnamecref{Grin-main}
  is true for some $d \geq 3$ and all $0 \leq n < 2^d$.
  Given $d, n, m \in \N$ with $0 \leq n < 2^{d+1}$ and
  \begin{equation} \label{req m}
   m \leq \mu(d+1, n)
  \end{equation}
  and inputs $s = \Bkt{s_0, \dotsc, s_{n-1}}$ and $t = \Bkt{t_0, \dotsc, t_{m-1}}$,
  we need to find a circuit for $f(s, t)$ with depth at most $d+1$.

  \textbf{Case 1:} Assume that $m \leq 2$.

  In this case, a circuit for $f(s, t)$ with depth $d+1$ is provided by \cref{depth-m-small}.

  \textbf{Case 2:} Assume that $m \geq 3$ and $m \leq \mu(d, n)$.

  As $2 \leq m \leq \mu(d, n)$, \cref{n < 2^d} yields $n \leq 2^d - 2$.
  Thus, by induction hypothesis,
  we obtain a circuit for $f(s, t)$ with depth $d \leq d+1$.

  \textbf{Case 3:} Assume that
  \begin{equation}
   m > \mu(d, n)
   \quad \text{and} \quad
   m \geq 3\,. \label{m-not-small-m-at-least-3}
   \end{equation}

  \textbf{Case 3.1:} Assume that $n \geq 2^d$.

  In this case, we apply the symmetric split
  \begin{equation} \label{Grin case n large}
  f\Bkt{s, t} = \sym(s') \land f\Bkt{s \backslash s', t}
  \end{equation}
  from \cref{symmetric split subset} with $k := 2^d \leq n$
  and an arbitrary $k$-length sub-tuple $s'$ of $s$.
  In order to show that \cref{Grin case n large} yields depth $d$,
  it suffices to show that both $\sym(s')$ and $f\Bkt{s \backslash s', t}$
  can be realized with depth at most $d$.

  A circuit for $\sym(s')$ with depth $d = \bktU{\log_2 k}$
  can be constructed by \cref{sym-log-n} as $s'$ has $k$ entries.

  As $m \leq \mu(d+1, n)$ by assumption \labelcref{req m} and
  $m \geq 3$ by assumption \labelcref{m-not-small-m-at-least-3},
  \cref{n < 2^d} implies $n \leq 2^{d+1} - 2$.
  Hence, we have
  $|s \backslash s'| = n - k \leq 2^{d+1} - 2 - 2^d = 2^d - 2$
  and
  {
  \newcommand{\aw}{6}
  \begin{align*}
   m \overset{\labelcref{req m}}{\leq} \mu(d+1, n)
     = \xi\frac{2^{d+1} - n - 2}{d+1} + 2
     \overset{k=2^d}{=} \xi\frac{2^{d} - (n-k) - 2}{d+1} + 2
     \overset{n - k \leq 2^d - 2}{\leq} \xi \frac{2^{d} - (n-k) - 2}{d} + 2
     \overset{\shortcref{def-mu}}{=} \mu(d, n - k)\,.
  \end{align*}
  }
  By induction hypothesis,
  we can thus find a circuit for $f\Bkt{s \backslash s', t}$ with depth $d$.

  Together, this shows that the split \labelcref{Grin case n large}
  yields a circuit for the Boolean function $f(s, t)$ with depth $d+1$.

  \textbf{Case 3.2:} Assume that $n < 2^d$.

  \textbf{Case 3.2.1:} Assume that $m \leq \mu(d, 0)$.

  Here, the \aop{} $g(t) = f((), t)$ can be realized with depth $d$ by induction hypothesis.
  Since $n < 2^d$, this means that the symmetric split
  \[f(s, t) = \sym(s) \land g(t)\]
  from \cref{symmetric split A}
  yields depth $d+1$.

  \textbf{Case 3.2.2:} Assume that $m > \mu(d, 0)$.

   We have 
   $\mu(d, n) = \xi \frac{2^d - n - 2}{d} + 2 \overset{n \leq 2^d - 1}{\geq} -\frac{\xi}{d} + 2 \overset{d \geq 3}{\geq} 1$,
   so we may choose a maximum odd integer $k$ with
  \begin{align} \label{choice of k prop}
   k \leq \mu(d, n)\,.
  \end{align}
  Assumption \labelcref{m-not-small-m-at-least-3} implies that $k < m$.
  This allows us to apply the alternating split
  \begin{align}
   f(s, t) = f\Bkt{s, \prefix{t}{2k+1}} \land f^*\Bkt{\everysecond{\prefix{t}{2k+1}}, \suffix{t}{2k+1}} \label{Grinchuk alternating split}
  \end{align}
  from \cref{aop-split} with $k$ as length of the odd-length prefix $\prefix{t}{2k+1} = \Bkt{t_0, \dotsc, t_{k-1}}$.
  Recall that
  $\widehat{t'} = \Bkt{t_1, t_3, \dotsc, t_{k-2}}$
  and $t'' = \Bkt{t_{k}, \dotsc, t_{m-1}}$.
  Due to $n < 2^d$ and \cref{choice of k prop},
  the induction hypothesis allows us to realize $f\Bkt{s, \prefix{t}{2k+1}}$ with depth $d$.
  As all requirements of \cref{m-k-small-enough} are fulfilled,
  we obtain $0 \leq \frac{k-1}{2} < 2^d$ and
  $m - k \leq \mu \Bkt{d, \frac{k-1}{2}}$.
  As the number of inputs of $\everysecond{\prefix{t}{2k+1}}$ and $\suffix{t}{2k+1}$
  is exactly $\frac{k-1}{2}$ and $m - k$, respectively,
  this implies that by induction hypothesis,
  we can realize $f^*\Bkt{\everysecond{\prefix{t}{2k+1}}, \suffix{t}{2k+1}}$
  with depth $d$.
  We conclude that realization \labelcref{Grinchuk alternating split} yields
  a circuit for $f(s, t)$ with depth at most $d+1$.

  This finishes the proof of the induction step (case 2) and hence of the \lcnamecref{Grin-main}.
  \end{proof}

 \begin{figure}[t]
 \begin{algorithm}[H]
 \caption{Depth optimization for extended \aop{}s}
 \label{grinchuk-alg}
    \DontPrintSemicolon
    \KwIn{$n, m \in \N$, symmetric inputs $s = (s_0, \dotsc, s_{n-1})$ and alternating inputs $t = (t_0, \dotsc, t_{m-1})$.}
    \KwOut{Circuit $\depthckt\Bkt{s, t}$ computing $f(s, t)$.}
    \BlankLine

    \If {$m \leq 2$}
    { \label{algo base start}
     \Return Optimum circuit for $\sym\DBkt{s_0, \dotsc, s_{n-1}, t_0, \dotsc, t_{m-1}}$. \tcp*{See \cref{sym-log-n}}  \label{algo m 2}
    }
    $d \gets \dmin(n, m)$. \tcp*{Hence, $n < 2^d$.\hspace*{11.4mm}} \label{algo compute d}
    \If(\tcp*[f]{Hence, $m \in \{3, 4, 5\}$.\hspace*{4mm}}){$d \leq 3$}
    {
     \If{$m = 3$}
     {
       \Return Optimum circuit for $\sym\DBkt{s_0, \dotsc, s_{n-1}, t_0, t_1 \lor t_2}$. \tcp*{See \cref{almost-sym-log-n}\hspace*{4mm}} \label{algo m 3}
     }
     \If{$m \in \{4,5\}$}
     {
       \Return Circuit as in \cref{img-proof_small_1} (for $m = 4$) or 
       \cref{img-proof_small_2} (for $m = 5$).\; \label{algo base end}
     }
    }
    \ElseIf{$n \geq 2^{d-1}$}
    {
       $k \gets 2^{d-1}$.\;
       Choose $s' \subseteq s$ with $|s'| = k$.\; \label{algo det s'}
       Compute an optimum circuit $S'$ for $\sym\Bkt{s'}$.\;
       \Return $S' \land \depthckt\Bkt{s \backslash s', t}$.\; \label{algo sym split}
    }
    \Else
    {
       \If{$m \leq \mu(d-1, 0)$}
       {
       Compute an optimum circuit $S$ for $\sym\DBkt{s_0, \dotsc, s_{n-1}}$.\;
       \Return $S \land \depthckt\Bkt{\Bkt{}, t}$.\; \label{algo simple split}
       }
       Choose $1 \leq k < m$ maximum with $k$ odd and $k \leq \mu(d-1, n)$.\; \label{algo compute k}
       $\prefix{t}{k} \gets \Bkt{t_0, \dotsc, t_{k-1}}$ and $\suffix{t}{k} \gets t \backslash \prefix{t}{k}$.\;
       \Return $\depthckt\Bkt{s, \prefix{t}{k}} \land \Bkt{\depthckt\Bkt{\everysecond{\prefix{t}{k}}, \suffix{t}{k}}}^*$.\; \label{Grinchuk algo end}
    }
 \end{algorithm}
 \end{figure}


\cref{grinchuk-alg} states our algorithm for computing a
circuit for $f(s, t)$ which arises from the proof of \cref{Grin-main}.
Note that we do not explicitly state how the occurring optimum symmetric circuits are constructed.
E.g., we could apply \cref{sym-log-n}
to construct each symmetric circuit as a formula circuit on the inputs.
Then, the circuit computed by \cref{grinchuk-alg}
would be a formula circuit with a size in $\mathcal O(m \log_2(m + n) + n)$,
see \cref{Grinchuk-formula-size}.
As we need various symmetric trees during \cref{grinchuk-alg},
a better idea is to compute a non-formula circuit
that re-uses symmetric trees in different sub-functions.
We shall see in \cref{grinchuk-improved-size}
that this leads to a size of $\mathcal O(m + n)$.
As long as we always construct optimum symmetric circuits,
this does not make a difference regarding the depth analysis of the arising circuit.
Relatedly, we do not specify how the subset $s'$ of $s$ is chosen in \cref{algo det s'}
as for the depth analysis, this is irrelevant.
Thus, we postpone these topics until the size discussion in \cref{sec-grinchuk-size}.

\subsection{Depth Analysis} \label{sec-grinchuk-depth}

For bounding the depth of the circuits computed by \cref{grinchuk-alg},
we need several technical lemmas.

\begin{lemma} \label{d / ln d incr}
 For $x \in \R$ with $x > e$, the function $\zeta(x) = \frac{x}{\ln x}$ is strictly monotonely increasing.
 \begin{proof}
  We show that the first derivative of $\zeta$ is strictly positive for $x > e$.
  We have
  $\frac{d}{dx} \zeta(x) = \frac{\ln x - 1}{\ln^2 x} = \frac{1}{\ln x} \Bkt{ 1 - \frac{1}{\ln x}}$,
  and this is positive as both factors of the right-hand side function are positive for $x > e$.
 \end{proof}
\end{lemma}

\begin{lemma} \label{d / log d incr}
 For $x \geq \frac{1}{\ln (2)} \approx 1.443$, the function $\zeta(x) = \frac{2^x}{x}$ 
 is strictly monotonely increasing.
 \begin{proof}
  As $x \geq \frac{1}{\ln (2)}$, we may equivalently show that the function $y \mapsto \frac{y}{\log_2 y}$
  is monotonely increasing for $y \geq 2^{\frac{1}{\ln (2)}} = e$.
  This follows from \cref{d / ln d incr} because ${\frac{y}{\log_2 y} = \ln 2 \frac{y}{\ln y}}$
  and $\ln 2 > 0$.
 \end{proof}
\end{lemma}

\begin{lemma} \label{lem-theta-0-m}
Let $c := -0.35$.
Consider (again with $\xi:=1.999$) the function
\[\vartheta(n, m) := 2^{c} \xi (m + n) \log_2 m - (m-2) \Bkt{\log_2(m + n) + \log_2 \log_2 m + c} - \xi n - 2\xi\,.\]
We have $\vartheta(n, m) \geq 0$ for all $n, m \in \N$ with $m \geq 3$.
\begin{proof}
 See \cref{app-proofs}, \cpageref{proof-lem-theta-0-m}.
\end{proof}
\end{lemma}

Finally, we are set to analyze the depth of the circuits computed by \cref{grinchuk-alg}.

\linebox{
\begin{theorem} \label{Grinchuk-own-thm}
 Given input variables $s = \Bkt{s_0, \dotsc, s_{n-1}}$ and $t = \Bkt{t_0, \dotsc, t_{m-1}}$ with $m + n \geq 1$,
 \cref{grinchuk-alg} computes a circuit $\depthckt\Bkt{s, t}$
 for the extended \aop{} $f(s, t)$ with depth
\[
 \depth(\depthckt\Bkt{s, t})
 \begin{cases}
 = \BktU{\log_2 (m + n)}
  & \text{for $m \leq 2$ and} \\
 \leq \log_2(m + n) + \log_2 \log_2 m + 0.65
 & \text{for $m \geq 3$.}
 \end{cases}
 \]
\end{theorem}
}
 \begin{proof}
  For $m \leq 2$, \cref{grinchuk-alg} computes an optimum circuit for $f(s, t)$
  with depth $\BktU{\log_2(m + n)}$ in \cref{algo m 2}.

  For $m \geq 3$, we shall see that the algorithm computes a realization with depth at most $d := \dmin(n, m)$,
  cf.~\cref{algo compute d}.
  Since $m \geq 2$, we have $n < 2^d$ by \cref{n < 2^d}.
  Hence, \cref{Grin-main} yields a circuit for $f(s, t)$ with depth $d$.
  The rest of \cref{grinchuk-alg} performs the same steps as the proof of \cref{Grin-main},
  hence computes a realization of $f(s, t)$ with depth at most $d$.
  Now let
  \[\tilde{d} := \BktD{\log_2(m + n) + \log_2 \log_2 m + 0.65}\,.\]
  If we can show that
  \begin{equation}
   m \leq \mu\Bkt{\tilde{d}, n}\,, \label{own-grin-proof-statement}
  \end{equation}
  then, by definition of $d = \dmin(n, m )$, we have $d \leq \tilde d$
  and the \lcnamecref{Grinchuk-own-thm} is proven.

  Recall that $\mu\Bkt{\tilde{d}, n} = \xi \frac{2^{\tilde{d}} - n - 2}{\tilde{d}} + 2$.
  By \cref{d / log d incr},
  for fixed $n$, the function $x \mapsto \frac{2^{x} - n - 2}{x} + 2$
  is monotonely increasing in $x$ for $x \geq \frac{1}{\ln (2)}$.
  As $\tilde d \geq \log_2(m+n) + \log_2 \log_2 m - 0.35 \overset{m \geq 3}{\geq} \frac{1}{\ln (2)}$,
  for proving \cref{own-grin-proof-statement}, it hence suffices to show
  \begin{equation} \label{own-grin-proof-statement-2}
   m \leq \xi \frac{2^{\log_2(m + n) + \log_2 \log_2 m - 0.35} - n - 2}{\log_2(m + n) + \log_2 \log_2 m - 0.35} + 2\,.
  \end{equation}
  The right-hand side of \labelcref{own-grin-proof-statement-2} can be simplified as
  \begin{align*}
     \xi\frac{2^{\log_2(m + n) + \log_2 \log_2 m - 0.35} - n - 2}{\log_2(m + n) + \log_2 \log_2 m - 0.35} + 2
   = \xi\frac{2^{-0.35} (m + n) \log_2 m - n - 2}{\log_2(m + n) + \log_2 \log_2 m - 0.35} + 2\,.
  \end{align*}
  Hence, \labelcref{own-grin-proof-statement-2} is implied if we can prove that
  for $n, m \in \N$ with $m \geq 3$ and $c := -0.35$, we have
  \[\xi 2^{c} (m + n) \log_2 m - (m-2) \Bkt{\log_2(m + n) + \log_2 \log_2 m + c} - \xi n - 2\xi \geq 0\,.\]
  This is precisely the statement of \cref{lem-theta-0-m}.
  Hence, \cref{own-grin-proof-statement-2} is fulfilled
  and the circuit computed by \cref{grinchuk-alg}
  has depth at most $\tilde d$.
 \end{proof}

For \aop{}s, \cref{Grinchuk-own-thm} yields the following depth guarantee.

\linebox{
\begin{qedcorollary} \label{cor-aop-depth}
 Given input variables $t = \Bkt{t_0, \dotsc, t_{m-1}}$
 with $m \geq 2$,
 \cref{grinchuk-alg} computes a circuit $C$ for
 the \aop{} $g(t)$ with
 \[d(C) \leq \log_2 m + \log_2 \log_2 m + 0.65\,.\]
\end{qedcorollary}
}

This is a similar asymptotic depth bound as for the circuits by \citet{Grinchuk},
but with a better additive constant.
An improvement by more than a constant is not possible due to the matching lower bound
given by \citet{Commentz-Walter-mon}.
As \citet{Hitzschke2018} made these lower bounds more precise,
for sufficiently large instances,
our circuits are optimum up to an additive constant of $5.7$.
Even more, comparing our results with the results of the exact algoritm
by \citet{exact_aop},
we see that for up to 109 inputs,
our algorithm always finds a depth-optimum \aop{} circuit.
In fact, we conjecture that our results are always depth-optimum.

\subsection{Size-Optimization of \AOP{}s: General Framework} \label{sec-grinchuk-size}

We shall now analyze \totalref{grinchuk-alg} with respect to the
size of the resulting circuits.
The description of \cref{grinchuk-alg} suggests to construct each occurring symmetric circuit on $r$ inputs
with $r - 1$ gates
independently from other symmetric circuits which may compute the same sub-function at intermediate stages.
This yields a formula circuit for $f(s, t)$, and its properties are examined in \cref{Grinchuk-formula-size}.
Afterwards,
we will give an alternative implementation with a significantly lower size.
For \cref{Grinchuk-formula-size}, the size bound
is proven via a bound on the \define{circuit fanout},
the maximum \define{fanout}
(i.e., number of successors) of any vertex.

We use the following notation. For a circuit $C$ and a vertex $v \in \cktnodes(C)$,
the set $\cktnodes_v(C) \subseteq \cktnodes(C)$\label{inputcone-page}
of all vertices $w \in \cktnodes(C)$ such that there is a directed path
from $w$ to $v$ is called the \define{input cone} of $v$.
By $\cktins_v(C)$,\label{inputconeinputs-page} we denote the set of inputs
in the input cone of $v$.
The circuit $C_v$ with $\cktedges(C_v) = \cktedges(C) \cap \Bkt{\cktnodes(C_v) \times \cktnodes(C_v)}$,
inputs $\cktins_v(C)$, gates $\cktgates(C) \cap \cktnodes_v(C)$
and a single output $v$ is called the \define{circuit subordinate to} $v$.
Any circuit whose gates and edges are a subset of $\cktgates(C)$ and $\cktedges(C)$, respectively,
is called a \define{sub-circuit} of $C$.

\linebox{
\begin{theorem} \label{Grinchuk-formula-size}
 Assume that in \totalref{grinchuk-alg}, all symmetric circuits are constructed as formula circuits on the inputs.
 Given input variables $s = \Bkt{s_0, \dotsc, s_{n-1}}$ and $t = \Bkt{t_0, \dotsc, t_{m-1}}$ with $m \geq 2$,
 the circuit $\depthckt\Bkt{s, t}$ computed by \cref{grinchuk-alg} is a formula circuit with
 \begin{equation*}
 \size(\depthckt\Bkt{s, t}) \leq m \depth(\depthckt\Bkt{s, t}) + n - 1
 \quad \text{and} \quad
 \fanout(\depthckt\Bkt{s, t}) \leq \depth(\depthckt\Bkt{s, t})\,.
 \end{equation*}
\end{theorem}
}
 \begin{proof}
  For $m = 2$, both bounds are fulfilled
  since the symmetric tree constructed in \cref{algo m 2}
  has size at most $m + n -1$ and fanout $1$.
  Thus, assume $m \geq 3$ and write $\cktins := \cktins\Bkt{\depthckt\Bkt{s, t}}$.
  Since $\depthckt\Bkt{s, t}$ is a formula circuit,
  double counting yields $\size(\depthckt\Bkt{s, t}) = \sum_{v \in \cktins} \fanout(v) - 1$.
  Hence, both size and fanout bound are implied by the following claim:
  \begin{claim_no_num}
   Assume that $m \geq 3$ and choose $d := \dmin(n, m) \in \N$ as computed in \cref{algo compute d} of \cref{grinchuk-alg}.
   In the circuit $\depthckt\Bkt{s, t}$,
   each of the symmetric inputs has fanout exactly $1$
   and each of the alternating inputs has fanout at most $d$.
   \begin{proof_of_claim}
    We proceed by induction on $d$.
    For $d \leq 2$ (which implies $m \leq 3$), each computed realization has fanout $1$.
    For $d = 3$, each of the symmetric inputs has fanout exactly $1$ und each of the
    alternating inputs has fanout at most $2$.
    Thus, assume that $d > 3$.
    Both in \cref{algo sym split,algo simple split},
    the returned realization is the disjunction of two circuits on disjoint input sets.
    As any symmetric circuit has fanout $1$
    and by induction hypothesis, $\depthckt\Bkt{s, t}$ fulfills the claimed fanout bounds
    in these two cases.
    In \cref{Grinchuk algo end}, we return
    $\depthckt\Bkt{s, t} = \depthckt\Bkt{s, \prefix{t}{k}} \land \Bkt{\depthckt\Bkt{\everysecond{\prefix{t}{k}}, \suffix{t}{k}}}^*$.
    By induction hypothesis applied to the two sub-circuits,
    every input in $s$ has fanout $1$ in $\depthckt\Bkt{s, t}$,
    every input in $\prefix{t}{k}$ has fanout at most $(d-1) + 1 = d$ in $\depthckt\Bkt{s, t}$
    and every input in $\suffix{t}{k}$ has fanout at most $d-1 < d$ in $\depthckt\Bkt{s, t}$.
    This proves the induction step and hence the claim.
   \end{proof_of_claim}
  \end{claim_no_num}
 \end{proof}

Thus, the formula circuit $\depthckt\Bkt{s, t}$ computed by \cref{grinchuk-alg} has
size $\mathcal O(m \log_2(m + n) + n)$.

In the following sections,
we will show that with a different approach, the circuits constructed in \cref{grinchuk-alg}
can be implemented with linear size.

\subsection{Leftist Circuits and Triangular Sets} \label{sec-sym-share}

As we construct numerous symmetric trees during \cref{grinchuk-alg},
we cannot afford each of them to have a size linear in the number of inputs independently.
Hence, the key idea of our construction is to build two so-called leftist symmetric circuits,
one \AND{} circuit and one \OR{} circuit,
and to use these when constructing symmetric \AND{} and \OR{} trees in \cref{grinchuk-alg}.
This way, we can show that the amount of additional gates needed to construct a single symmetric tree
is logarithmic in the number of its inputs.



\begin{definition}
 Let $n \in \N$ and a commutative and associative operator $\circ$ be given.
 A circuit $S$ on inputs $x_0, \dotsc, x_{n-1}$ over the basis $\set{\circ}$
 is called \define{ordered} if
 \begin{itemize}
  \item the underlying undirected graph of $S$ is acyclic and
  \item for each vertex $v \in \cktnodes(S)$,
    there is an interval $I_v \subseteq \set{0, \dotsc, n-1}$
    such that $\cktins_v(S) = \{x_i \where i \in I_v\}$.
 \end{itemize}
\end{definition}

This yields a partial order on $\cktnodes(S)$:
We say that a vertex $v \in \cktnodes(S)$ is \define{left} of a vertex $w \in \cktnodes(S)$ if
$I_v \cap I_w = \emptyset$ and $\max\set{i \in I_v} < \min \set{i \in I_w}$,
and \define{right} of $w$ if $w$ is left of $v$.
The two predecessors of any gate vertex are related with respect to this partial order.
Thus, we may call the predecessors left and right, respectively.
We extend the definitions of left and right from vertices to sets of vertices
$W_1, W_2 \subseteq \cktnodes(S)$ with $W_1 \cap W_2 = \emptyset$:
We say that $W_1$ is left (right) of $W_2$
if for any pair of vertices $(w_1, w_2) \in W_1 \times W_2$,
the vertex $w_1$ is left (right) of the vertex $w_2$.

\begin{definition}\label{leftist-def}
 A circuit arising from \cref{alg-leftist-symtree} is called \define{leftist}.
\end{definition}

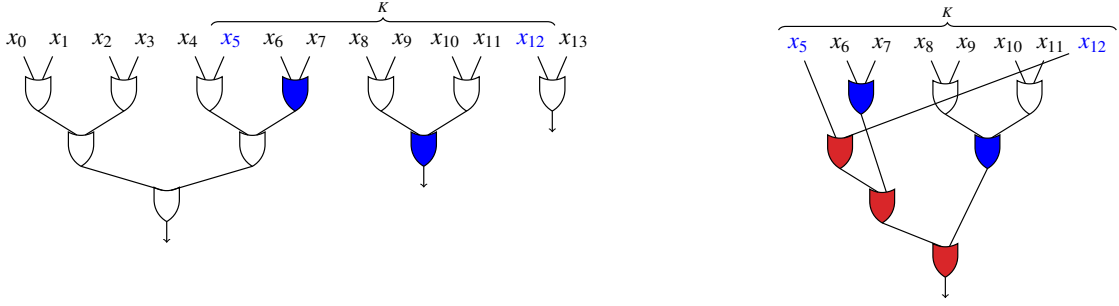
\begin{figure}[t]
 \begin{subfigure}[t]{0.61\textwidth}
 \begin{center}
 \adjustbox{max width=0.8\textwidth}{
  \centering{\begin{tikzpicture}
\node[input] (x_0) at (0.5, 0.2){$x_0$};
\node[input] (x_1) at (1.5, 0.2){$x_1$};
\node[input] (x_2) at (2.5, 0.2){$x_2$};
\node[input] (x_3) at (3.5, 0.2){$x_3$};
\node[input] (x_4) at (4.5, 0.2){$x_4$};
\node[input, blue] (x_5) at (5.5, 0.2){$x_5$};
\node[input] (x_6) at (6.5, 0.2){$x_6$};
\node[input] (x_7) at (7.5, 0.2){$x_7$};
\node[input] (x_8) at (8.5, 0.2){$x_8$};
\node[input] (x_9) at (9.5, 0.2){$x_9$};
\node[input] (x_10) at (10.5, 0.2){$x_{10}$};
\node[input] (x_11) at (11.5, 0.2){$x_{11}$};
\node[input, blue] (x_12) at (12.5, 0.2){$x_{12}$};
\node[input] (x_13) at (13.5, 0.2){$x_{13}$};

\draw[decorate,decoration={brace, amplitude=5pt, raise=10pt}] (x_5.west) -- (x_12.east) node [midway, above, sloped, yshift=15pt] {$K$};


\node[or-gate, fill=white, fill=white] at (1, -1) (or1) {};
\draw (x_0) -- (or1.input 2);
\draw (x_1) -- (or1.input 1);

\node[or-gate, fill=white] at (3, -1) (or2) {};
\draw (x_2) -- (or2.input 2);
\draw (x_3) -- (or2.input 1);

\node[or-gate, fill=white] at (5, -1) (or3) {};
\draw (x_4) -- (or3.input 2);
\draw (x_5) -- (or3.input 1);

\node[or-gate, fill=blue] at (7, -1) (or4) {};
\draw (x_6) -- (or4.input 2);
\draw (x_7) -- (or4.input 1);

\node[or-gate, fill=white] at (9, -1) (or5) {};
\draw (x_8) -- (or5.input 2);
\draw (x_9) -- (or5.input 1);

\node[or-gate, fill=white] at (11, -1) (or6) {};
\draw (x_10) -- (or6.input 2);
\draw (x_11) -- (or6.input 1);

\node[or-gate, fill=white] at (13, -1) (or7) {};
\draw (x_12) -- (or7.input 2);
\draw (x_13) -- (or7.input 1);

\draw[->] (or7.output) -- ($(or7.output) - (0, 0.5)$);


\node[or-gate, fill=white] at (2, -2.3) (or8) {};
\draw (or1.output) -- (or8.input 2);
\draw (or2.output) -- (or8.input 1);

\node[or-gate, fill=white] at (6, -2.3) (or9) {};
\draw (or3.output) -- (or9.input 2);
\draw (or4.output) -- (or9.input 1);

\node[or-gate, fill=blue] at (10, -2.3) (or10) {};
\draw (or5.output) -- (or10.input 2);
\draw (or6.output) -- (or10.input 1);

\draw[->] (or10.output) -- ($(or10.output) - (0, 0.5)$);


\node[or-gate, fill=white] at (4, -3.6) (or11) {};
\draw (or8.output) -- (or11.input 2);
\draw (or9.output) -- (or11.input 1);

\draw[->] (or11.output) -- ($(or11.output) - (0, 0.5)$);


\phantom{
\node[or-gate, fill=red] at (9, -4.9) (or13) {};
\draw[->] (or13.output) -- ($(or13.output) - (0, 0.5)$);
}
\end{tikzpicture}}
  }
 \end{center}
  \caption{A leftist \OR{} tree $S$ on $n = 14$ inputs
  with the $4$-element set $B = B(K, S)$ of boundary vertices for $K = \set{x_5, \dotsc, x_{12}}$ marked blue.}
  \label{fig-leftist-1}
 \end{subfigure}
 \hfill
\begin{subfigure}[t]{0.35\textwidth}
 \begin{center}
 \adjustbox{max width=0.8\textwidth}{
  \centering{\begin{tikzpicture}
\node[input, blue] (x_5) at (5.5, 0.2){$x_5$};
\node[input] (x_6) at (6.5, 0.2){$x_6$};
\node[input] (x_7) at (7.5, 0.2){$x_7$};
\node[input] (x_8) at (8.5, 0.2){$x_8$};
\node[input] (x_9) at (9.5, 0.2){$x_9$};
\node[input] (x_10) at (10.5, 0.2){$x_{10}$};
\node[input] (x_11) at (11.5, 0.2){$x_{11}$};
\node[input, blue] (x_12) at (12.5, 0.2){$x_{12}$};

\draw[decorate,decoration={brace, amplitude=5pt, raise=10pt}] (x_5.west) -- (x_12.east) node [midway, above, sloped, yshift=15pt] {$K$};

\node[or-gate, fill=blue] at (7, -1) (or4) {};
\draw (x_6) -- (or4.input 2);
\draw (x_7) -- (or4.input 1);

\node[or-gate, fill=white] at (9, -1) (or5) {};
\draw (x_8) -- (or5.input 2);
\draw (x_9) -- (or5.input 1);

\node[or-gate, fill=white] at (11, -1) (or6) {};
\draw (x_10) -- (or6.input 2);
\draw (x_11) -- (or6.input 1);


\node[or-gate, fill=blue] at (10, -2.3) (or10) {};
\draw (or5.output) -- (or10.input 2);
\draw (or6.output) -- (or10.input 1);

\node[or-gate, fill=red] at (6.5, -2.3) (or11) {};
\draw (x_5) -- (or11.input 2);
\draw (x_12) -- (or11.input 1);

\node[or-gate, fill=red] at (7.5, -3.6) (or12) {};
\draw (or11.output) -- (or12.input 2);
\draw (or4.output) -- (or12.input 1);

\node[or-gate, fill=red] at (9, -4.9) (or13) {};
\draw (or12.output) -- (or13.input 2);
\draw (or10.output) -- (or13.input 1);

\draw[->] (or13.output) -- ($(or13.output) - (0, 0.5)$);

\end{tikzpicture}}
  }
 \end{center}
  \caption{Circuit resulting from Huffman coding on the set $B$ from \cref{fig-leftist-1}.}
  \label{fig-leftist-2}
 \end{subfigure}
 \caption{Illustration of the proof of \cref{sym-prep} with $K = \set{x_5, \dotsc, x_{12}}$
 and $L = \emptyset$.
  Note that $|B| = 4 = 2 \log_2(|K|) - 2$.
 }
 \label{fig-leftist}
\end{figure}

\totalref{fig-leftist-1} depicts a leftist \OR{} tree on $n = 14$ inputs.

  \begin{figure}
  \begin{algorithm}[H]
    \caption{Leftist circuit construction}
    \label{alg-leftist-symtree}
    \DontPrintSemicolon
    \KwIn{A commutative and associative operation $\circ \colon \I \times \I \to \I$,
          $n \in \N_{\geq 1}$.}
    \KwOut{A (leftist) circuit $S$ on inputs $x = (x_0, \dotsc, x_{n-1})$ over the basis $\basis = \set{\circ}$.}
    \BlankLine
    Let $S$ be a circuit on inputs $x = (x_0, \dotsc, x_{n-1})$ with $\cktgates(S) = \emptyset$.\;
    $i \gets 0$.\;
    \While{$n \geq 2$}
    { \label{leftist-step1-start}
     Let $k \in \N$ be maximum with $k \leq \log_2 n$.\; \label{leftist-choose-k}
     Add an ordered full symmetric $\circ$-tree $T_k$ on inputs $x_i, \dotsc, x_{i + 2^k - 1}$ to $S$.\;
     $i \gets i + 2^k$.\;
     $n \gets n - 2^k$. \label{leftist-step1-end}\;
    }
    $\cktouts(S) \gets \set{v \in \cktnodes \where |\delta^+(v)| = 0}$.\;
    \KwRet{$S$}\;
  \end{algorithm}
  \end{figure}

Note that if $n = 2^d$ for $d \in \N$,
then there is a unique connected ordered tree with optimum depth $d$ on inputs $x_0, \dotsc, x_{n-1}$.
Otherwise, if $2^{d-1} < n < 2^d$ for some $d \in \N$,
we consider the binary decomposition $n = \sum_{k = 0}^{\bktD{\log_2 n}} a_k 2^k$ with $a_k \in \set{0, 1}$ of $n$.
We partition the $n$ inputs into groups of $2^k$ inputs for each $k$ with $a_k = 1$.
On each input group, we construct a connected ordered tree $T_k$ with depth $k$.
This yields an ordered (but unconnected) circuit.
\cref{alg-leftist-symtree} does exactly this, where the trees $T_k$ are sorted
from left to right by decreasing $k$.

\begin{definition} \label{def-consec}
 Let $n \in \N$ and a leftist circuit $S$ on inputs $x_0, \dotsc, x_{n-1}$ be given.
 A subset $K \subseteq \cktins(S)$ of the inputs is called \define{consecutive}
 if there are $0 \leq a < b \leq n-1$ with $K = \set{x_i \where a \leq i \leq b}$.
\end{definition}

\begin{definition} \label{def-boundary}
 Consider $n \in \N$ and a leftist circuit $S$ on inputs $x_0, \dotsc, x_{n-1}$.
 Given a subset $K \subseteq \cktins(S)$,
 let $B(K, S) \subseteq \cktnodes(S)$ be defined as
 \[B(K, S) := \set{v \in \cktnodes(S) \where \cktins_v(S) \subseteq K \text{ and }
  \cktins_w(S) \nsubseteq K \text{ for all } w \in \delta^+(v)}\,.\]
We call the elements of $B(K, S)$ \define{boundary vertices} of $K$
with respect to $S$.
\end{definition}

In \totalref{fig-leftist-1}, the boundary vertices of $K$ with respect to $S$ are marked blue.

Note that in a leftist circuit $S$,
for every vertex $v \in \cktnodes(S)$,
the sub-circuit $S\!_v$ subordinate to $v$ is a full binary tree.
Furthermore, given a subset $K$ of the inputs with boundary set $B := B(K, S)$,
we have
\begin{equation}
\Cupdot_{v \in B} \cktins_v(S) = K\,. \label{B-covers-K}
\end{equation}
In particular, there is an ordering $b_0, \dotsc, b_{|B|- 1}$
of $B$
such that $b_j$ is left of $b_{j+1}$
for all $j \in \set{0, \dotsc, {|B| - 2}}$.

In order to get an implementation of \cref{grinchuk-alg}
which guarantees a linear size of the output, we mainly have
to bound the number of gates used for the symmetric parts.
The reason is that inputs can be part of many different 
symmetric trees while by construction they are only once input
of alternating parts.
A crcucial step is the choice of the set $s'$ in \cref{algo det s'}
of \cref{grinchuk-alg}. For the depth bound, $s'$ can be an arbitrary
subset of $s$ with $k$ elements. However, for the size analysis, it
is important that both $s'$ and $s\setminus s'$ have a special 
structure that allows the re-use of subtrees realizing the 
corresponding symmetric functions. 
It turns out that being triangular as defined in the following 
definition is a desirable property for this purpose.

\begin{definition} \label{def-triangular}
 Let $n \in \N$ and a leftist circuit $S$
 on inputs $x_0, \dotsc, x_{n-1}$ be given.
 Consider a subset $K \subseteq \cktins(S)$ of the inputs
 with boundary vertices $B := B(K, S)$.
 The \define{boundary tree sequence} of $K$ with respect to $S$
 is the sequence $T_0, \dotsc, T_{|B| - 1}$ of binary trees subordinate to the boundary vertices,
 sorted from left to right.
 We call $K$ \define{triangular}
 if there is some $J \in \set{0, \dotsc, {|B|}-1}$ such that
 the inputs of $T_0, \dotsc, T_J$ are consecutive,
 the inputs of $T_{J+1}, \dotsc, T_{|B| - 1}$ are consecutive, and
  \begin{equation}
   \depth(T_j) < \depth(T_{j+1}) \, \text{ for } \, 0 \leq j < J - 1
   \qquad \text{and} \qquad
   \depth(T_j) > \depth(T_{j+1}) \, \text{ for } \, J + 1 \leq j < |B| - 1\,. \label{triangular-card}
  \end{equation}
 We call $T_0, \dotsc, T_J$ the \define{increasing} part of the boundary tree sequence,
 and $T_{J+1}, \dotsc, T_{|B| - 1}$ its \define{decreasing} part.
\end{definition}

For a given set, the boundary tree sequence is unique;
but the value $J$ in the definition of a triangular set is not necessarily unique.
The consecutive set $K$ in \totalref{fig-leftist-1} is triangular,
where the boundary vertices are marked blue
and we can choose $J = 1$ or $J = 2$.
The set $N \backslash K$ in \cref{fig-leftist-1} is not triangular
as condition \labelcref{triangular-card} is not fulfilled for the boundary tree sequence,
which contains trees with depths $2$, $0$, and $0$.
In \totalref{img-triangular}, the sets $N$, $K$ and $N \backslash K$
are all triangular.
The following \lcnamecref{lem-tri-easy}, which is easy to prove,
gives some criteria for when a set $K$ is triangular.
Later, in \cref{lem-triangular-remains},
we shall see a sufficient condition for $N \backslash K$ to be triangular
when $K$ is triangular.

\begin{lemma} \label{lem-tri-easy}
 Let $n \in \N$ and a leftist circuit $S$
 on inputs $x_0, \dotsc, x_{n-1}$ be given.
 Then, the following statements hold:
 \begin{enumerate}
  \item The empty set is triangular.
  \item The set $K = \set{x_0, \dotsc, x_{n-1}}$ is triangular.
  \item If $\emptyset \neq K \subseteq \cktins(S)$ is triangular and $x_i$ is the right-most or left-most vertex
  in $K$, then $K \backslash \set{x_i}$ is triangular.
  \item Let $K \subseteq \cktins(S)$ be a consecutive subset of the inputs.
  Then, $K$ is triangular.
 \end{enumerate}
\end{lemma}
\begin{proof}
 See \cref{app-proofs}, \cpageref{lem-tri-easy-proof}.
\end{proof}

In order to show in \cref{sym-prep} how we can use leftist circuits in order to save gates,
we need some basic properties about leftist circuits.

\begin{lemma} \label{reduction-leftist}
 Let $S$ be a leftist circuit on inputs $x_0, \dotsc, x_{n-1}$.
 Then, the circuit $S'$ arising from deleting $x_0, \dotsc, x_{n-1}$
 (with their successors being the inputs of $S'$) is again a leftist circuit.
 \begin{proof}
  By definition, $S$ is a collection of connected ordered trees $T_k$ of depth $k$ for each $k$
  with non-trivial coefficient in the binary decomposition of $n$.
  When removing $x_0, \dotsc, x_{n-1}$,
  every tree $T_k$ with $k \geq 1$ is transformed into a connected ordered tree $T'_{k-1}$ with depth $k-1$;
  and trees with depth $0$ are removed.
  This is exactly the leftist circuit that \cref{alg-leftist-symtree} produces
  with the successors of $x_0, \dotsc, x_{n-1}$ as inputs.
 \end{proof}
\end{lemma}

\begin{lemma} \label{leftist-properties}
 Let $S$ be a leftist circuit on inputs $x_0, \dotsc, x_{n-1}$
 and $K \subseteq \cktins(S)$ be triangular with $|K| = k$ and $B := B(K, S)$.
 Furthermore, let $S'$ be the leftist circuit arising from $S$ by deleting the input vertices
 (see \cref{reduction-leftist}),
 and let 
 $K' := \set{v \in \cktnodes(S) \where \depth(v) = 1 \text{ and }  w \in K \text{ for all } w \in \delta^-(v)}$.
 Then, the following statements are fulfilled:
 \begin{enumerate}
  \item We have $|B| \leq k$. \label{simple-bound-B}
  \item We have $B = B(K', S') \cupdot (K \cap B)$. \label{B-part}
  \item The set $K'$ is triangular with respect to $S'$. \label{K'-tri}
  \item \label{B-modulo}
  For $k \geq 2$, we have $|K \cap B| \leq 2$, and  $|K \cap B| \equiv k \mod 2$, and
 $|K'| =
     \begin{cases}
      \frac{k - 1}{2} & \text{ if } k \text{ odd,} \\
      \frac{k}{2} & \text{ if } k \text{ even, } |K \cap B| = 0\,,\\
      \frac{k - 2}{2} & \text{ if } k \text{ even, } |K \cap B| = 2\,.
     \end{cases}
     $
 \end{enumerate}
 \begin{proof}
  The first two statements follow directly from the definitions.

  The third statement follows from the second:
  The boundary tree sequence for $K$ with respect to $S$
  can be transformed into a boundary sequence for $K'$ with respect to $S'$
  as in the proof of \cref{reduction-leftist}.

  For proving the fourth statement,
  note that as $K$ is triangular,
  there are at most $2$ boundary trees with depth $0$,
  hence $|K \cap B| \leq 2$.
  Furthermore, by \cref{B-covers-K}, the inputs of the boundary trees
  form a partition of $K$,
  and each boundary tree with depth at least $1$ contains an even number of vertices.
  Hence, we have $|K \cap B| \equiv k \mod 2$.
  Together with the second statement, this implies the estimation of $|K'|$.
 \end{proof}
\end{lemma}

\begin{lemma} \label{B-lemma}
 Consider $k, n \in \N$ with $n \geq k \geq 3$.
 Let a leftist circuit $S$ on inputs $x_0, \dotsc, x_{n-1}$
 and a triangular subset $K \subseteq \cktins(S)$ with $|K| = k$ be given.
 Then, for $B := B(K, S)$, we have
 \[|B| \leq 2 \log_2 k - 1\,.\]
 \begin{proof}
   As in \cref{reduction-leftist,leftist-properties},
   let $S'$ be the leftist circuit arising from $S$ by deleting the input vertices,
   and let
   $K' := \set{v \in \cktnodes(S) \where \depth(v) = 1 \text{ and } w \in K \text{ for all } w \in \delta^-(v)}$.
  We prove the statement by induction on $k$.

    For $3 \leq k \leq 6$, we show the statement explicitly.
    We have
    {
    \newcommand{\aw}{8}
    \begin{align*}
    |B| \overset{\shortcref{leftist-properties}, \labelcref{B-part}}{=} |B(K', S')| + |K \cap B|
     \ca{\aw}{\overset{\shortcref{leftist-properties}, \labelcref{simple-bound-B}}{\leq}} |K'| + |K \cap B| \\
     \ca{\aw}{\overset{\shortcref{leftist-properties}, \labelcref{B-modulo}}{=}}
     \begin{cases}
      \frac{k - 1}{2} + 1 & \text{ if } k \text{ odd} \\
      \frac{k}{2} + 0 & \text{ if } k \text{ even, } |K \cap B| = 0\\
      \frac{k - 2}{2} + 2 & \text{ if } k \text{ even, } |K \cap B| = 2
     \end{cases} \\
     \ca{\aw}{\overset{k \in \set{3, 4, 5, 6}}{\leq}} 2 \log_2 k - 1\,.
    \end{align*}
    }
    Now, we may assume $k \geq 7$.
    For $k \geq 7$, property \labelcref{B-modulo} of \cref{leftist-properties} implies $|K'| \geq 3$.
    As $K'$ is triangular by \cref{K'-tri} of \cref{leftist-properties},
    we may apply the induction hypothesis to the set $K'$ and the circuit $S'$,
    which yields $|B(K', S')| \leq 2 \log_2 (|K'|) - 1$.
    This implies
    {
    \newcommand{\aw}{8}
    \begin{align*} \label{B-from-ind}
     |B| \overset{\shortcref{leftist-properties}, \labelcref{B-part}}{=} |B(K', S')| + |K \cap B|
         \ca{\aw}{\overset{\IH}{\leq}} 2 \log_2\Bkt{|K'|} - 1 + |K \cap B| \\
         \ca{\aw}{\overset{\shortcref{leftist-properties}, \labelcref{B-modulo}}{=}}
         \begin{cases}
      2 \log_2\Bkt{\frac{k - 1}{2}} - 1 + 1 & \text{ if } k \text{ odd} \\
      2 \log_2\Bkt{\frac{k}{2}} - 1 + 0 & \text{ if } k \text{ even, } |K \cap B| = 0\\
      2 \log_2\Bkt{\frac{k - 2}{2}} - 1 + 2 & \text{ if } k \text{ even, } |K \cap B| = 2
     \end{cases} \\
     \ca{\aw}{\leq} 2 \log_2 k -1\,.
    \end{align*}
    }
    This proves the induction step and hence the \lcnamecref{B-lemma}.
 \end{proof}
\end{lemma}

For the proof of the next main result, \cref{sym-prep},
we need an excursion about delay, a generalization of depth:
When each input $x_i$ of a circuit $C$ is associated with a time $a(x_i) \in \N$
specifying when the input is available, we define the \define{delay}
of a circuit $C$ on inputs $x_0, \dotsc, x_{n-1}$ by\label{delay-page}
\[\delay(C) := \max_{i} \bktCfixed{big}{ a(x_i) +
     \max \set{|P| \where P \text{ is a directed path in } C \text{ starting in } x_i}}\,.\]
When the input arrival times are all $0$,
the notions of delay and depth coincide.
Delay-optimum symmetric \AND{} trees or \OR{} trees
can be constructed efficiently:

 \begin{theorem}[\citet{Huffman}, \citet{Golumbic} and Van Leeuwen \citep{vanLeeuwen}] \label{symtree-opt}
 Let $\circ \in \set{\AND, \OR}$,
 inputs $x_0, \dotsc, x_{n-1}$ with $n \geq 1$ and arrival times $a(x_0), \dotsc, a(x_{n-1}) \in \N$ be given.
 There is an algorithm called \define{Huffman coding} that computes an optimum solution $C$
 to the \prsymdelayopt{} with
 $ \delay\Bkt{C} = \bktUfixed{big}{\log_2\Bkt{\sum_{i = 0}^{n-1} 2^{a(x_i)}}}$.
 If we assume that the inputs are sorted by increasing arrival time,
 then the algorithm can be implemented to run in time $\mathcal O(n)$;
 otherwise, in time $\mathcal O(n \log_2 n)$.
\end{theorem}

Huffman coding \citep{Huffman} is known in a more general context;
the proof of the delay bound is due to \citet{Golumbic}
and Van Leeuwen~\citep{vanLeeuwen} showed that the algorithm can be implemented
to run in linear time after sorting.
Note that the formula for the optimum delay is closely related to
Kraft's inequality \citep{Kraft}.

Using Huffman coding, we now show that, given a leftist circuit $S$ on all inputs,
we can construct an optimum symmetric tree for any triangular subset $K$ of the inputs
(plus, possibly, some more inputs)
while adding only few gates to $S$.
\cref{fig-leftist-2} shows the tree constructed for $S$ and $K$ from \cref{fig-leftist-1}.

\begin{proposition} \label{huff-on-B}
 Let $n \in \N$ and a leftist circuit $S$ on $n$ inputs $x_0, \dotsc, x_{n-1}$ be given.
 Let $K \subseteq \cktins(S)$ be a triangular set of inputs with
 boundary vertices $B := B(K, S)$.
 Furthermore, let inputs $L$ (not necessarily inputs of $S$)
 with $K \cap L = \emptyset$ be given.
 Consider arrival times $a(y) = 0$ for each $y \in K$ and
 $a(y) \in \N$ for each $y \in L$.
 A delay-optimum symmetric tree on $K \cupdot L$
 with respect to arrival times $a$ can be constructed
 (while possibly reusing the gates in $S$)
 using at most $|B| + |L| - 1$ additional gates.
 \begin{proof}
  Let $H$ denote the circuit arising from Huffman coding (see \cref{symtree-opt}) on input set $B \cupdot L$,
  where an input $v \in B$ has arrival time $a(v) = \log_2 |\cktins_v(S)|$ and
  an input $v \in L$ has arrival time $a(v)$.
  Since $S$ is leftist,
  we have $a(v) \in \N$ for all $v \in B$.
  Thus, the delay of $H$ with respect to arrival times $a$ is exactly
  \begin{equation*}
  \BktU{\log_2 \bktRfixed{Bigg}{\sum_{v \in B \cupdot L} 2^{a(v)}}}
    = \BktU{\log_2 \bktRfixed{Bigg}{\sum_{v \in B} |\cktins_v(S)| + \sum_{v \in L} 2^{a(v)}}} \\
    \overset{\labelcref{B-covers-K}}{=} \BktU{\log_2 \bktRfixed{Bigg}{|K|+\sum_{v \in L} 2^{a(v)}}} \\
    = \BktU{\log_2 \bktRfixed{Bigg}{\sum_{v \in K \cup L} 2^{a(v)}}}\,,
  \end{equation*}
  which is the optimum possible delay of a symmetric tree on inputs $K$ and $L$
  by \cref{symtree-opt}.
  Hence, $S \cup H$ contains a delay-optimum symmetric tree on $K \cupdot L$.
  Since $H$ contains exactly $|B| + |L| - 1$ gates,
  this proves the \lcnamecref{huff-on-B}.
 \end{proof}
\end{proposition}

 \begin{figure}
  \begin{algorithm}[H]
    \caption{Computation of boundary vertices}
    \label{det-B}
    \DontPrintSemicolon
    \KwIn{A leftist circuit $S$ on inputs $x_0, \dotsc, x_{n-1}$,
          a consecutive subset $K = \Bkt{x_{a}, \dotsc, x_{b}} \subseteq \cktins(C)$,
          precomputed data from \cref{runtime-det-B}.}
    \KwOut{The set $B:= B(K, S)$.}
      $B \gets \emptyset$\;
      $i \gets a$\; \label{det-B-choose-i-init}
      \While{$i \leq b$}
      {
          \label{begin-while}
          Choose $j$ maximum such that a right \nachfahre{} $s_j(x_i)$ of depth $j$ exists
          and $\cktins_{s_j(x_i)}(S) \subseteq K$. \label{choose-j}\;
          $s \gets s_j(x_i)$\; \label{consider-s}
          $B \gets B \cup \set{s}$ \label{add-gate-to-B}\;
          $i \gets r(s) + 1$ \label{end-while} \label{det-B-choose-i}
      }
      \KwRet{$B$}\;
  \end{algorithm}
 \end{figure}

Hence, given an input set $K$ which is triangular with respect to a
leftist circuit $S$,
once the set $B(K, S)$ of boundary vertices is known,
sorted by increasing depth,
we can construct an optimum symmetric tree on $S$
(plus a constant number $|L|$ of additional inputs)
in time $\mathcal O(|B(K, S)|)$.
It remains to determine the boundary vertices.

Given a consecutive subset $K \subseteq \cktins(S)$ of a leftist circuit $S$,
we use \cref{det-B} to compute the boundary vertices
(the general case follows in \cref{sym-prep}).
Here, given a vertex $v \in \cktnodes(S)$,
we denote the indices of the left-most and right-most input in $\cktins_v(S)$
by $l(v)$ and $r(v)$, respectively.
Furthermore, given an input $x_i$ and a vertex $v \in \cktnodes(S)$,
we call $v$ a \define{right \nachfahre{}} of $x_i$
if there is a set of vertices
$\bktCfixed{big}{s_j(x_i) \in \cktnodes(S) \where j \in \bktC{0, \dotsc, r}}$ for some $r \geq 0$
with $s_{j}(x_i)$ being a left predecessor of $s_{j+1}(x_i)$ for all $j \in \set{0, \dotsc, r-1}$,
$x_i = s_0(x_i)$ and $v = s_r(x_i)$.
Note that $s_j(x_i)$ has depth $j$ in $S$ as $S$ is leftist.
Let $r_i$ denote the highest depth (and index) of any right \nachfahre{} of input $x_i$.
The following \lcnamecref{runtime-det-B} shows that \cref{det-B}
works correctly and estimates its running time.

\begin{lemma} \label{runtime-det-B}
 Let a leftist circuit $S$ on inputs $x_0, \dotsc, x_{n-1}$
 and a consecutive subset $K \subseteq \cktins(S)$ of size $k := |K|$ be given.
 Assume that the following data of size $\mathcal O(n \log_2 n)$ is available:
 \begin{itemize}
  \item For every input $x_i$ and the set $\bktCfixed{normal}{s_j(x_i)}_{0 \leq j \leq r_i}$ of right \nachfahre{}s of $x_i$ with depth $j$,
  we store a pointer to $s_j(x_i)$ in an array with constant-time access.
  \item For every vertex $v \in \cktnodes$, we store the index $r(v)$ of the right-most input in $\cktins(v)$.
  \item For every $i \in \set{0, \dotsc, n-1}$, we store the highest depth $r_i$ of any right \nachfahre{} of input $x_i$.
 \end{itemize}
 Then, \cref{det-B} computes the set $B := B(S, K)$ of boundary vertices in time $\mathcal O(\log_2 k)$.
 The algorithm can be implemented with the same running time
 if the output set $B$ has to be sorted by increasing depth.
 \begin{proof}
  As a first step, we prove that \cref{det-B} works correctly.

  We first see by induction on $i$ that whenever a vertex is added to $B$, it is a boundary vertex.
  So consider an iteration of the while-loop (\crefrange{begin-while}{end-while})
  where a vertex $s := s_j(x_i)$ is added to $B$ in \cref{add-gate-to-B}.
  In \cref{choose-j}, $j$ is chosen maximum such that a right \nachfahre{} $s_j(x_i)$ of depth $j$ exists
  and $\cktins_{s_j(x_i)}(S) \subseteq K$.
  Thus, in order to show that $s := s_j(x_i)$ is a boundary vertex,
  it remains to show that if $s$ has a successor $t$, then $\cktins_t(S) \nsubseteq K$.
  Assume that $s$ has a successor $t$ and that $\cktins_t(S) \subseteq K$.
  By maximality of $j$, the vertex $s$ must be a right predecessor of $t$.
  If $a = i$, then we immediately have $\cktins_t(S) \nsubseteq K$.
  If $i > a$, we must have $\cktins_t(S) \nsubseteq K$ since otherwise,
  we would have skipped $i$ in the while-loop (\crefrange{begin-while}{end-while}).
  Hence, $s$ is a boundary vertex.

  Assume now that there is a boundary vertex $v$ that is not added to $B$ by \cref{det-B}.
  If in some iteration of the while-loop (\crefrange{begin-while}{end-while}),
  we have $i = l(v)$, then $v$ is added to $B$.
  So assume that $l(v)$ is skipped, meaning that $x_{l(v)} \in \cktins_s(S)$ for some $s$ considered in \cref{consider-s}.
  But two \nachfahre{}s $s$ and $v$ of an input $x_{l(v)}$ cannot be in $B$ simultaneously
  by \labelcref{B-covers-K},
  so this is a contradiction.

  This proves the correctness of \cref{det-B}.
  Now, we analyze its running time.
  As in each iteration, one vertex is added to $B$, there are at most $B$ iterations,
  and by \cref{B-lemma}, we have $B \in \mathcal O(\log_2 k)$.
  The index $j$ computed in \cref{choose-j} is exactly
  $j := \min\bktCfixed{normal}{\bktDfixed{normal}{\log_2 \Bkt{b - i + 1}}, r_i}$.
  Thus, using the precomputed data, each iteration can be implemented to run in constant time,
  so the total running time of \cref{det-B} is $\mathcal O(\log_2 k)$.

  If, during the algorithm, we use buckets to store the boundary vertices
  ordered by their depth,
  the total running time does not increase as there are at most $\bktU{\log_2 k}$ possible depths.
 \end{proof}
\end{lemma}

Indeed, for the leftist circuit $S$ and the consecutive subset $K$ shown in \cref{fig-leftist-1},
the algorithm works correctly:
The vertices $x_i$ chosen in \cref{det-B-choose-i-init,det-B-choose-i} are
$x_5$, $x_6$, $x_8$, and $x_{12}$,
and exactly the blue vertices are added to $B$.

From the previous statements, we obtain the following \lcnamecref{sym-prep}.

\linebox{
\begin{theorem} \label{sym-prep}
 Let $n \in \N$ and a leftist circuit $S$ on inputs $x_0, \dotsc, x_{n-1}$ be given.
 Let $K \subseteq \cktins(S)$ be triangular with $k := |K|$.
 Furthermore, let inputs $L$ (not necessarily inputs of $S$)
 with $l := |L|$ and $K \cap L = \emptyset$ be given.
 Consider arrival times $a(y) = 0$ for each $y \in K$ and
 $a(y) \in \N$ for each $y \in L$, where $L$ is sorted by increasing arrival time.
 A delay-optimum symmetric tree on $K \cupdot L$
 with respect to arrival times $a$ can be constructed
 in time $\mathcal O(\log_2 k + l)$,
 assuming the precomputed data from \cref{runtime-det-B} is given.
 The number of additional gates (while possibly reusing the gates in $S$)
 is at most $k + l - 1$ for $k \leq 2$ and at most $2 \log_2 k + l - 2$ for $k \geq 3$.
\end{theorem}
}
 \begin{proof}
  If $k \leq 2$, we obtain a delay-optimum tree with exactly $k + l - 1$ gates
  by Huffman coding on $K \cup L$ in time $\mathcal O(l + k) \overset{k \leq 2}{=} \mathcal O(l)$, see \cref{symtree-opt}.

  Now assume that $k \geq 3$.
  By \cref{huff-on-B}, we can compute a delay-optimum symmetric tree on $K \cupdot L$
  by Huffman coding on the vertices of $B(K, S)$ and $L$.
  By \cref{B-lemma} and \cref{huff-on-B}, we have $2 \log_2 k + l - 2$ additional gates.
  By \cref{def-triangular},
  the inputs belonging to the increasing and decreasing
  part of the boundary sequence of $K$ are consecutive sets,
  hence we can apply \cref{runtime-det-B} to both to obtain $B(K, S)$
  in time $\mathcal O(\log_2 k)$, using the precomputed data.
  Huffman coding on $B(K, S)$ and $L$ can be performed
  in time $\mathcal O(\log_2 k + l)$.
 \end{proof}

\begin{figure}[!h]
  \begin{center}
  \adjustbox{max width=.5\textwidth}{
  \centering{\begin{tikzpicture}

\node[input] (x_0) at (0.5, 0.2){$x_0$};
\node[input, blue] (x_1) at (1.5, 0.2){$x_1$};
\node[input] (x_2) at (2.5, 0.2){$x_2$};
\node[input] (x_3) at (3.5, 0.2){$x_3$};
\node[input] (x_4) at (4.5, 0.2){$x_4$};
\node[input] (x_5) at (5.5, 0.2){$x_5$};
\node[input] (x_6) at (6.5, 0.2){$x_6$};
\node[input] (x_7) at (7.5, 0.2){$x_7$};
\node[input] (x_8) at (8.5, 0.2){$x_8$};
\node[input] (x_9) at (9.5, 0.2){$x_9$};
\node[input] (x_10) at (10.5, 0.2){$x_{10}$};
\node[input] (x_11) at (11.5, 0.2){$x_{11}$};
\node[input, blue] (x_12) at (12.5, 0.2){$x_{12}$};
\node[input] (x_13) at (13.5, 0.2){$x_{13}$};

\draw[decorate,decoration={brace, amplitude=5pt, raise=30pt}] (x_1.west) -- (x_12.east) node [midway, above, sloped, yshift=35pt] {$N$};

\draw[decorate,decoration={brace, amplitude=5pt, raise=10pt}] (x_4.west) -- (x_11.east) node [midway, above, sloped, yshift=15pt] {$K$};


\node[or-gate, fill=white, fill=white] at (1, -1) (or1) {};
\draw (x_0) -- (or1.input 2);
\draw (x_1) -- (or1.input 1);

\node[or-gate, fill=blue] at (3, -1) (or2) {};
\draw (x_2) -- (or2.input 2);
\draw (x_3) -- (or2.input 1);

\node[or-gate, fill=white] at (5, -1) (or3) {};
\draw (x_4) -- (or3.input 2);
\draw (x_5) -- (or3.input 1);

\node[or-gate, fill=white] at (7, -1) (or4) {};
\draw (x_6) -- (or4.input 2);
\draw (x_7) -- (or4.input 1);

\node[or-gate, fill=white] at (9, -1) (or5) {};
\draw (x_8) -- (or5.input 2);
\draw (x_9) -- (or5.input 1);

\node[or-gate, fill=white] at (11, -1) (or6) {};
\draw (x_10) -- (or6.input 2);
\draw (x_11) -- (or6.input 1);

\node[or-gate, fill=white] at (13, -1) (or7) {};
\draw (x_12) -- (or7.input 2);
\draw (x_13) -- (or7.input 1);

\draw[->] (or7.output) -- ($(or7.output) - (0, 0.5)$);

\node[or-gate, fill=white] at (2, -2.3) (or8) {};
\draw (or1.output) -- (or8.input 2);
\draw (or2.output) -- (or8.input 1);

\node[or-gate, fill=blue] at (6, -2.3) (or9) {};
\draw (or3.output) -- (or9.input 2);
\draw (or4.output) -- (or9.input 1);

\node[or-gate, fill=blue] at (10, -2.3) (or10) {};
\draw (or5.output) -- (or10.input 2);
\draw (or6.output) -- (or10.input 1);

\draw[->] (or10.output) -- ($(or10.output) - (0, 0.5)$);


\node[or-gate, fill=white] at (4, -3.6) (or11) {};
\draw (or8.output) -- (or11.input 2);
\draw (or9.output) -- (or11.input 1);

\draw[->] (or11.output) -- ($(or11.output) - (0, 0.5)$);
\end{tikzpicture}
  }
  }
  \end{center}
 \caption{Illustration of \cref{lem-triangular-remains} and \cref{det-triangular-subset}.
    The gates shown form a leftist circuit on inputs $x_0, \dotsc, x_{13}$.
    The sets $N$, $K$, $N \backslash K$ of inputs are all triangular.
    The $5$ blue vertices are the boundary vertices of $N$ with respect to $S$.}
 \label{img-triangular}
\end{figure}

 \begin{figure}
  \begin{algorithm}[H]
  \caption{Determining a triangular subset $K$ of a triangular set $N$}
  \label{det-triangular-subset}
    \DontPrintSemicolon
    \KwIn{A leftist circuit $S$ on inputs $x_0, \dotsc, x_{r-1}$,
          a triangular set $N \subseteq \cktins(S)$
          with $2^{d-1} \leq |N| < 2^d$.}
    \KwOut{A set $K \subseteq N$.}
    Let $T_0, \dotsc, T_{|B(N, S)| - 1}$ denote the boundary tree sequence for $N$ w.r.t.~$S$.\;
    \If{$\exists~j$ with $\depth(T_j) = d-1$}
    {
       \KwRet{$\cktins(T_j)$}\; \label{tri-ret-K-d-1}
    }
    \Else
    {
       $D \gets \max\set{{d' \in \N \where \exists~ j_0 < j_1 \text{ with } \depth(T_{j_0}) = \depth(T_{j_1}) = d'}}$.\; \label{tri-choose-D}
       Choose $j_0 < j_1$ such that $\depth(T_{j_0}) = \depth(T_{j_1}) = D$.\;
       \KwRet{$\bigcup_{j = j_0}^{j_1} \cktins(T_j)$}\; \label{tri-ret-K-d-2}
    }
  \end{algorithm}
 \end{figure}

We will use this \lcnamecref{sym-prep}
for the improved construction of symmetric circuits during our
\aop{} optimization algorithm \totalref{grinchuk-alg} in \cref{sec::linear_aops}.
Here, we will also need \cref{det-triangular-subset}
which, given a triangular set $N$,
extracts a triangular subset $K$ of a certain size such that
also $N \backslash K$ is triangular
and such that we can prove in \cref{lem-triangular}
that $K$ and $N \backslash K$ cannot be both large simultaneously.
\cref{img-triangular} gives an example for \cref{det-triangular-subset}.

\begin{proposition}\label{lem-triangular-remains}
 Let $r \in \N$ and a leftist circuit $S$
 on inputs $x_0, \dotsc, x_{r-1}$ be given.
 Consider a triangular subset $N \subseteq \cktins(S)$
 with $n := |N| \geq 1$.
 Let $d \in \N_{\geq 1}$ be the unique integer with
 $2^{d-1} \leq n < 2^d$.
 Then, \cref{det-triangular-subset} computes a subset $K \subseteq N$
 with $|K| = 2^{d-1}$ such that
 both $K$ and $N \backslash K$ are triangular.
 \begin{proof}
   We use the notation from \cref{det-triangular-subset}.

   First, we need to show that the value $D$ in \cref{tri-choose-D} exists.
   Thus, assume that there is no tree $T_j$ with depth $d-1$.
   As $n < 2^d$, the maximum depth of any tree $T_j$ is hence $d-2$.
   By \cref{def-triangular},
   for each $d' \in \set{0, \dotsc, d-2}$,
   the boundary tree sequence contains at most $2$ full binary trees
   with depth $d'$.
   Based on this, the value $D$ in \cref{tri-choose-D} exists as otherwise,
   there is at most one tree per depth, and
   we obtain the contradiction
   $2^{d-1} \leq n \leq \sum_{d' = 0}^{d-2} 2^{d'} = 2^{d-1} - 1 < 2^{d-1}$.
   Thus, \cref{det-triangular-subset} correctly computes a set $K \subseteq N$.

   It is easy to see that $K$ and $N \backslash K$ are triangular with respect to $S$
   as $T_{j_0}, \dotsc, T_{j_1}$ is the boundary tree sequence for $K$
   and $T_0, \dotsc, T_{j_0 - 1}, T_{j_1 + 1}, \dotsc, T_{|B(N, S) - 1|}$
   is the boundary tree sequence for $N \backslash K$.

   It remains to show that $|K| = 2^{d-1}$.
   If $K$ is constructed in \cref{tri-ret-K-d-1},
   this is certainly the case, so assume
   that $K$ is constructed in \cref{tri-ret-K-d-2} as $K = \bigcup_{j = j_0}^{j_1} \cktins(T_j)$.
   For every $i \in \set{D, \dotsc, d-2}$,
   the sequence $T_0, \dotsc, T_{|B(N, S)|-1}$ must contain a tree with depth $i$ as otherwise,
   there is $I > D$ such that no tree with depth $I$ exist, which implies
   {
   \newcommand{\aw}{4}
   \begin{equation*}
     2^{d-1} \leq n
             \leq 2 \sum_{i = 0}^{D} 2^{i} + \sum_{i = D + 1}^{d-2} 2^{i} - 2^I
             = \sum_{i = 0}^{d-2} 2^i + \sum_{i=0}^{D} 2^i - 2^I
             = 2^{d-1} + 2^{D + 1} - 2^I - 2
             \overset{I > D}{<} 2^{d-1}\,,
   \end{equation*}
   }
   a contradiction.
   From this, we conclude that
   \[|K| = \sum_{j = j_0}^{j_1} |\cktins(T_j)| =  2 \cdot 2^{D} + \sum_{i = D + 1}^{d-2} 2^{i} = 2^{D+1} + 2^{d-1} - 1 - (2^{D+1} - 1) = 2^{d-1}\,. \qedhere\]
 \end{proof}
\end{proposition}

\begin{corollary}\label{lem-triangular-remains-plus-t0}
 Let $r \in \N$ and a leftist circuit $S$
 on $x_0, \dotsc, x_{r-1}$ be given.
 Consider a triangular subset $N \subseteq \cktins(S)$ with $|N| = n \geq 1$,
 and another input vertex $x_i$ which is directly to the right of $N$
 such that $N \cup \set{x_i}$ is triangular.
 Let $K \subseteq N$ be the triangular set computed by \cref{det-triangular-subset}
 for $S$ and $N$.
 Then, $(N \backslash K) \cup \set{x_i}$ is triangular.
 \begin{proof}
  By the proof of \cref{lem-triangular-remains},
  the set $N \backslash K$ is triangular
  with boundary tree sequence consisting of the
  increasing part $T_0, \dotsc, T_{j_0 - 1}$
  and the decreasing part $T_{j_1 + 1}, \dotsc, T_{|B(N, S) - 1|}$.
  As the inputs of $T_{j_1 + 1}, \dotsc, T_{|B(N, S) - 1|}$ and $x_i$ are consecutive,
  adding $x_i$ to $(N \backslash K)$ results in a new
  tree with depth $d'$, where $d'$ is minimum such that there is no tree among
  $T_{j_1 + 1}, \dotsc, T_{|B(N, S) - 1|}$ with depth $d'$,
  and all trees with smaller depth are removed from the boundary tree sequence
  (note that $d' = 0$ may also occur).
  This new boundary tree sequence fulfills \cref{triangular-card} of \cref{def-triangular},
  hence $(N \backslash K) \cup \set{x_i}$ is triangular.
  \end{proof}
\end{corollary}

We will see in \cref{lem-triangular}
that for a triangular set $K$ computed by \cref{det-triangular-subset}
for a leftist circuit $S$ and a triangular set $N \subseteq \cktins(S)$,
either the number of trees in the boundary tree sequence of $K$
is small or $N \backslash K$ is small.
For estimating what ``small`` means here, we need the following function
and some numerical estimations.

\begin{definition} \label{def-rho}
 Define the function $\rho \colon \N \to \R_{\geq 0}$ by
 $
  \rho(n) = \begin{cases}
            n & \text{if $n \in \set{0, 1, 2}$}\,, \\
            \BktD{2 \log_2(n-1)}     & \text{if $n \geq 3$}\,.
            \end{cases}
 $
\end{definition}

\begin{observation} \label{obs-rho}
 The function $n \mapsto \rho(n)$ from \cref{def-rho} is monotonely increasing in $n$.
\end{observation}

\begin{lemma} \label{lem-log-plus-2}
 For $n \geq 3$, the following inequalities are fulfilled:
 \[
 \log_2 n - 1.9 \leq \log_2 (n-1)\,, \qquad \qquad
 \bktD{2 \log_2 n} - 1 \leq \bktD{2 \log_2(n-1)}\,, \qquad \qquad
 \bktD{2 \log_2(n+1)} - 2 \leq \bktD{2 \log_2(n-1)}\,.
 \]
  \begin{proof}
 For $n = 3$, all statements can be verified directly.

 For $n \geq 4$, we have
 $(n-1) \sqrt{2} - n = n\Bkt{\sqrt{2} - 1} - \sqrt{2} > 0$
 and thus $\log_2 n < \log_2(n-1) + 0.5$, which implies the first statement.
 Furthermore, we obtain $2 \log_2 n < 2 \log_2(n-1) + 1$.
 After rounding, this implies the second statement,
 and applied twice, it yields $2 \log_2 (n+1) < 2 \log_2(n-1) + 2$,
 from which the third statement follows by rounding.
  \end{proof}
\end{lemma}

\linebox{
\begin{theorem} \label{lem-triangular}
 Let $r \in \N$ and a leftist circuit $S$
 on $r$ inputs $x_0, \dotsc, x_{r-1}$ be given.
 Consider a triangular subset $N \subseteq \cktins(S)$ with $|N| = n \geq 1$.
 Let $d \in \N_{>0}$ be the unique integer with
 $2^{d-1} \leq n < 2^d$.
 Let $k := 2^{d-1}$.
 Let $K \subseteq N$ be the output of \cref{det-triangular-subset}
 given $S$ and $N$,
 and let $B := B(K, S)$ be the boundary vertices of $K$ with respect to $S$.
 Then, the following statements are fulfilled:
  \begin{enumerate}
  \item If $n - k \geq 2$, we have $|B| + \bktD{2 \log_2 (n-k-1)} \leq \bktD{2 \log_2 (n-1)}$. \label{tri-log-bound}
  \item For $n \geq 16$, we have $|B| + \rho(n-k) \leq \rho(n)$. \label{tri-rho-bound}
  \end{enumerate}
\end{theorem}
}
\begin{proof} See \cref{app-proofs}, \cpageref{proof-lem-triangular}. \end{proof}

This result is crucial in particular for bounding the number of gates that we 
need to construct the multi-input \AND{} in \cref{grinchuk-alg}. This is 
because the number of gates that we need in addition to the gates of the 
leftist circuit for building a multi-input \AND{} circuit is linear in the 
number of boundary vertices (see \cref{lem-Phi}).

\subsection{Linear-Size \AOP{}s with Small Depth}\label{sec::linear_aops}

In this section, we present an algorithm that computes \aop{}s meeting the depth
bound of \cref{cor-aop-depth} and having linear size.
In order to do so, we first introduce a notation for the inputs
in the outermost call of \cref{grinchuk-alg}.

\begin{definition}
Consider the application of \cref{grinchuk-alg} to
symmetric inputs $s = \Bkt{s_0, \dotsc, s_{q-1}}$
and alternating inputs $t = \Bkt{t_0, \dotsc, t_{r-1}}$.
We define the \define{global inputs} $x = \bkt{x_0, x_2, \dotsc, x_{2q - 2}, x_{2q}, x_{2q+1}, \dotsc, x_{2q + r - 1}}$ by
\[x_i = \begin{cases}
         s_{\frac{i}{2}}  & \text{ for } i < 2q \text{ even}, \\
         \text{undefined} & \text{ for } i < 2q \text{ odd},  \\
         t_{i - 2q}            & \text{ for } i \geq 2q\,. \\
        \end{cases}
\]
\end{definition}

We can now use two interchangeable ways to denote the input of \cref{grinchuk-alg}:
We may either apply the algorithm to symmetric inputs $s = (s_0, s_1, \dotsc, s_{q-1})$
and alternating inputs $t = (t_0, \dotsc, t_{r-1})$,
or to global inputs $x = \bkt{x_0, x_2, \dotsc, x_{2q - 2}, x_{2q}, x_{2q+1}, \dotsc, x_{2q + r - 1}}$,
and can easily convert one notation into the other.
When applying \cref{grinchuk-alg} recursively,
we use the global notation for the outermost call of the algorithm,
and write $s$ and $t$ for the inputs considered in the current recursion step.
This notation allows us to identify the position of the currently considered inputs of $s$ and $t$
among the global inputs $x$:
We call an input $x_i$ with $i \in \{0, \dotsc, 2q + r - 1\}$
\define{even} if $i$ is even and \define{odd} otherwise.

For the analysis, we will often have to concatenate vectors; this can be done 
in a compact way with the following notation.

\begin{notation}
For two vectors $(x_0,\dots,x_{k-1})$ and $(y_0,\dots,y_{l-1})$ denote by
\[
   (x_0,\dots,x_{k-1}) \tupleconcat (y_0,\dots,y_{l-1}) := (x_0,\dots,x_{k-1}, y_0,\dots,y_{l-1})
\]
their concatenation.\label{tupleconcat-notation}
\end{notation}

Using these notations, \cref{alg-grin-leftist} states the precise algorithm:
We construct large leftist \AND{} and \OR{} circuits on the even and odd inputs, respectively (see \cref{leftist-def}),
and use these to construct symmetric trees during \cref{grinchuk-alg} via \cref{sym-prep}.
In order to apply \cref{sym-prep}, we need to show that  $s$ and $s \tupleconcat (t_0)$ are both triangular,
which we will do in \cref{lem-s-tri}.
Furthermore, for the construction of the subset $s'$ of $s$
in \cref{algo det s'} of \cref{grinchuk-alg},
we use \cref{det-triangular-subset}.
In \cref{grinchuk-improved-size}, we shall see that this leads to a linear number of gates.
The size analysis requires a deeper understanding of \cref{grinchuk-alg}.
We start with an easy \lcnamecref{obs-t-consecutive} that can be
verified by induction on the recursion depth of the algorithm.

\begin{figure}
 \begin{algorithm}[H]
 \caption{Depth optimization for extended \aop{}s via leftist circuits}
 \label{alg-grin-leftist}
   \DontPrintSemicolon
    \KwIn{Global inputs $x = \bkt{x_0, x_2, \dotsc, x_{2q - 2}, x_{2q}, x_{2q+1}, \dotsc, x_{2q + r - 1}}$.}
    \KwOut{A circuit computing $f\bkt{(x_0, x_2, \dotsc, x_{2q-2}), (x_{2q}, \dotsc, x_{2q + r - 1})}$.}
    \BlankLine
     Construct a leftist $\AND$-circuit $S\!_0$ on all even inputs $x_0, x_2, \dotsc$ of $x$.\; \label{leftist-and}
     Construct a leftist $\OR$-circuit $S\!_1$ on all odd inputs $x_{2q + 1}, x_{2q + 3}, \dotsc$ of $x$.\; \label{leftist-or}
     Precompute the data from \cref{runtime-det-B} for both $S\!_0$ and $S\!_1$.\; \label{leftist-data}
     Apply \totalref{grinchuk-alg} to compute a circuit for $f\Bkt{(x_0, x_2, \dotsc, x_{2q-2}), (x_{2q}, \dotsc, x_{2q + r - 1})}$
     while constructing all symmetric circuits using \cref{sym-prep}
     and computing $s'$ in \cref{algo det s'} via \cref{det-triangular-subset}.\;
 \end{algorithm}
 \end{figure}

\begin{observation}\label{obs-t-consecutive}
 Assume that \cref{grinchuk-alg} is applied to global inputs $x = \bkt{x_0, x_2, \dotsc, x_{2q}, x_{2q+1}, \dotsc, x_{2q + r - 1}}$.
 In each recursive call of \cref{grinchuk-alg} for
 symmetric inputs $s = \Bkt{s_0, \dotsc, s_{n-1}}$ and alternating inputs $t = \Bkt{t_0, \dotsc, t_{m-1}}$,
 there is some $j \in \{ 0, \dotsc, r - 1\}$ with $t_i = x_{2q + j + i}$ for all $0 \leq i \leq m -1$.
\end{observation}

In other words, $t$ is a consecutive subset of $x$.

Note that \cref{grinchuk-alg} computes a circuit for the \aop{} $f(s, t)$.
In order to compute a circuit for its dual $f^*(s, t)$,
we can simply call the algorithm to compute a circuit $C$ for $f(s, t)$ and return $C^*$.
In particular, this is what happens in \cref{Grinchuk algo end}.
As we want to use gates of the leftist \AND{}-tree $S\!_0$ and
the leftist \OR{}-tree $S\!_1$ in symmetric trees built during \cref{grinchuk-alg},
we need to determine the parity of the inputs in $s$ and $t$
depending on whether $f(s, t)$ or $f^*(s, t)$ is computed.
The following statement can also be proven via induction.

\begin{qedlemma} \label{grinchuk-even-odd}
 Assume that \cref{grinchuk-alg} is applied to global inputs $x = \bkt{x_0, x_2, \dotsc, x_{2q}, x_{2q+1}, \dotsc, x_{2q + r - 1}}$.
 Consider a recursive call of \cref{grinchuk-alg} with symmetric inputs $s$
 and alternating inputs $t = (t_0, \dotsc, t_{m-1})$.
 Then, for the computation of $f(s, t)$ (or $f^*(s, t)$, respectively),
 every input in $s$ as well as $t_0$ is even (or odd, respectively).
 \end{qedlemma}

As in \cref{algo det s'} of \cref{grinchuk-alg}, we apply \cref{det-triangular-subset} to $s$,
we need to show that $s$ is triangular.
Furthermore, we will use in \cref{size-small-values} that $s \tupleconcat (t_0)$ is triangular.

\begin{lemma} \label{lem-s-tri}
  Assume that \cref{grinchuk-alg} is applied to global inputs $x = \bkt{x_0, x_2, \dotsc, x_{2q}, x_{2q+1}, \dotsc, x_{2q + r - 1}}$.
  Let a leftist \AND{} circuit $S\!_0$ on $x_0, x_2, \dotsc$
  and a leftist \OR{} circuit $S\!_1$ on $x_{2q+1}, _{2q+3}, \dotsc$ be given.
  Consider a recursive call of \cref{grinchuk-alg} with symmetric inputs $s = (s_0, \dotsc, s_{n-1})$
  and alternating inputs $t = (t_0, \dotsc, t_{m-1})$
  for the computation of $f(s, t)$ (or $f^*(s, t)$, respectively).
  Then, the sets $s$ and $s \tupleconcat (t_0)$ are triangular with respect to $S\!_0$ (or $S\!_1$, respectively).
 \begin{proof}
  By \cref{grinchuk-even-odd},
  for the computation of $f(s, t)$ (or $f^*(s, t)$),
  all inputs in $s$ and $t_0$ are even (or odd).
  Hence, they are inputs of the leftist circuit $S\!_0$ (or $S\!_1$).

  By induction, we will prove
  that $s$ and $s \tupleconcat (t_0)$ are triangular.
  In the outermost call, $s$ and $s \tupleconcat (t_0)$ are both consecutive input sets of the leftist tree $S\!_0$,
  hence, by \cref{lem-tri-easy}, triangular with respect to $S\!_0$.

  For the inductive step, we assume without loss of generality that $f(s, t)$ is computed.
  Now, we may assume that $s$ and $s \tupleconcat (t_0)$ are triangular with respect to $S\!_0$.
  We show that the statement remains true for each recursive call
  (i.e., \cref{algo sym split,algo simple split,Grinchuk algo end}).

  In \cref{algo sym split}, we recursively realize $f(s \backslash s', t)$,
  where $s'$ is computed using \cref{det-triangular-subset}.
  Hence, by \cref{lem-triangular-remains}, $s \backslash s'$ is triangular,
  and by \cref{lem-triangular-remains-plus-t0}, $(s \backslash s') \tupleconcat t_0$
  is triangular, both with respect to $S\!_0$.

  In \cref{algo simple split}, we recursively compute $f((), t)$,
  and $()$ and $\bkt{t_0}$ are both triangular with respect to $S\!_0$
  by \cref{lem-tri-easy}.

  In \cref{Grinchuk algo end}, we recursively compute $f(s, t')$
  and $f^*(\widehat{t'}, t'')$.
  For $f(s, t')$, the statement is true by induction hypothesis.
  By \cref{obs-t-consecutive}, $t$ is a consecutive subset of the global inputs,
  and by \cref{grinchuk-even-odd}, $t_0$ is even.
  Thus, the inputs of $\widehat{t'}$ and $t''_0$ are all odd and a consecutive subset of inputs of $S\!_1$.
  Hence, by \cref{lem-tri-easy}, both $\widehat{t'}$ and $\widehat{t'} \tupleconcat (t''_0)$
  are triangular with respect to $S\!_1$.

  This proves the induction step and hence the \lcnamecref{lem-s-tri}.
\end{proof}
\end{lemma}

In order to prove that the size of our \aop{} circuits is linear in the number of inputs,
we partition the gates into groups
and estimate how many gates are used per group.

\begin{definition} \label{grin-gate-partition}
 Consider the circuit $\depthckt\Bkt{s, t}$ computed by \cref{alg-grin-leftist} for
 symmetric inputs $s$ and alternating inputs~$t$.
 We distinguish five types of gates used in $\depthckt\Bkt{s, t}$:
 \begin{enumerate}
  \item gates of the leftist circuits $S\!_0$ and $S\!_1$ in \cref{leftist-and,leftist-or} 
      of \cref{alg-grin-leftist} \label{leftist-gates}
  \item one concatenation gate per any alternating split in \cref{Grinchuk algo end} 
      of \cref{grinchuk-alg} \label{alt-split-gates}
  \item gates used in base-case solutions in \crefrange{algo base start}{algo base end} 
      of \cref{grinchuk-alg}\label{base-case-gates}, except gates that can be taken from the 
      leftist circuits
  \item gates used in symmetric circuits in \cref{algo sym split} or \cref{algo simple split} 
      of \cref{grinchuk-alg} \label{symtree-gates}
  \item one concatenation gate per any split in \cref{algo sym split} or \cref{algo simple split}
      of \cref{grinchuk-alg} \label{split-gates}
 \end{enumerate}
 We also call the gates of types \labelcref{alt-split-gates,base-case-gates,symtree-gates,split-gates} \define{additional gates}.
\end{definition}

Note that this indeed defines a partition of $\cktgates\Bkt{\depthckt\Bkt{s, t}}$.
Counting the gates of types \labelcref{leftist-gates} and \labelcref{alt-split-gates}
is easy,
the important step will be counting the gates of types \labelcref{base-case-gates,symtree-gates,split-gates}.

\begin{observation} \label{lem-leftist-gates}
 Consider the circuit $\depthckt\Bkt{s, t}$ computed by \cref{alg-grin-leftist} for
 symmetric input $s$ and alternating inputs $t$.
 For $m \geq 2$, the circuit $\depthckt\Bkt{s, t}$ contains at most $m + n - 2$ gates of type \labelcref{leftist-gates}.
\end{observation}

\begin{observation} \label{m > 0}
 Assume that \cref{grinchuk-alg} is applied to global inputs 
 $x = \bkt{x_0, x_2, \dotsc, x_{2q}, x_{2q+1}, \dotsc, x_{2q + r - 1}}$ with $r \geq 1$.
 Then, any $m$ considered during recursive calls fulfills $m \geq 1$.
\end{observation}

\begin{observation} \label{grin-num-alt-splits}
 Assume that \cref{grinchuk-alg} is applied to global inputs 
 $x = \bkt{x_0, x_2, \dotsc, x_{2q}, x_{2q+1}, \dotsc, x_{2q + r - 1}}$ with $r \geq 1$.
 Then, the number of alternating splits used in \cref{grinchuk-alg} is at most $r-1$. 
\end{observation}

The last two observations can be proven easily by induction on the 
recursion depth of the algorithm. For \cref{grin-num-alt-splits}, note that any 
alternating split can only be performed between two inputs $x_i$ and $x_{i+1}$ with 
$i \in \{2q,\dots, 2q + r - 2\}$ and for any such $i$ there is at most one alternating
split between $x_i$ and $x_{i+1}$.

From \cref{grin-num-alt-splits}, the number of gates of type \labelcref{alt-split-gates}
directly follows.

\begin{qedcorollary} \label{cor-alt-split-gates}
 Consider the circuit $\depthckt\Bkt{s, t}$ computed by \cref{alg-grin-leftist} for
 symmetric input $s$ and alternating inputs~$t$.
 For $m \geq 1$, the circuit $\depthckt\Bkt{s, t}$ contains at most $m - 1$ gates of type \labelcref{alt-split-gates}.
\end{qedcorollary}

In the following \lcnamecref{size-small-values},
we estimate the number of additional gates
for small values of $n$ and $m$.
Here, we need to examine the concrete realizations constructed for $f(s, t)$.
\begin{table}
\begin{adjustbox}{width=\columnwidth,center}
\newcommand{\colwidth}{1.cm}
\begin{tabular}{c|p{\colwidth}p{\colwidth}p{\colwidth}p{\colwidth}p{\colwidth}p{\colwidth}p{\colwidth}p{\colwidth}p{\colwidth}p{\colwidth}p{\colwidth}p{\colwidth}p{11mm}}
  $\tikz{\node[below left, inner sep=2pt] (def) {$m$};%
      \node[above right,inner sep=2pt] (abc) {$n$};%
      \draw (def.north west|-abc.north west) -- (def.south east-|abc.south east);}$
       &           0                   &   1         &       2              &        3                        &          4            &          5                               &  6 &  7   &           8                &          9             &        10  &                11 &  $n> 11$\\
      \midrule
   1   & \multicolumn{2}{c}{\cellcolor{blue!25}}             & \multicolumn{11}{c}{\cellcolor{red!25}}   \\
   2   & \multicolumn{2}{c}{\cellcolor{blue!25}}             & \multicolumn{11}{c}{\cellcolor{red!25} $m + 2\log_2 (n + 1) - 3$} \\
   3   & \multicolumn{2}{c}{\cellcolor{blue!25}}             & \multicolumn{3}{c}{\cellcolor{red!25}} & \multicolumn{3}{c}{\cellcolor{yellow!25}}               & \multicolumn{1}{c}{\cellcolor{yellow!25} } & \multicolumn{2}{c}{\cellcolor{green!25}$m + 5$} &  \multicolumn{1}{c}{\cellcolor{green!50}$m + 6$} \\
   4   & \multicolumn{2}{c}{\cellcolor{blue!25} $m + n - 1$ \phantom{abc}} & \multicolumn{7}{c}{\cellcolor{yellow!25} $m + 2 \log_2 n - 2$} & \multicolumn{1}{c}{\cellcolor{green!25}}\\
   5   & \multicolumn{2}{c}{\cellcolor{blue!25}}    & \multicolumn{6}{c}{\cellcolor{yellow!25}}  \\
   6   & \multicolumn{2}{c}{\cellcolor{blue!50}}             & \multicolumn{4}{c}{\cellcolor{red!50} $m + 2\log_2 (n + 1) - 2$}       \\
   7   & \multicolumn{2}{c}{\cellcolor{blue!50}$m + n$ \phantom{abc}} & \multicolumn{2}{c}{\cellcolor{red!50}} \\
   8   & \multicolumn{2}{c}{\cellcolor{blue!50}} \\ \bottomrule
\end{tabular}
\end{adjustbox}
\caption[Number of additional gates needed for small $n, m \in \N$.]{
Number of additional gates needed for the construction of $\depthckt(s, t)$
in certain cases.
\cref{size-small-values} proves the correctness of these values.
Cells with the same color contain the same formula for the number of additional gates.}
\label{table-add-gates-m-n}
\end{table}

\begin{lemma} \label{size-small-values}
Consider the circuit $\depthckt\Bkt{s, t}$ computed by \cref{alg-grin-leftist}
for symmetric inputs $s = \Bkt{s_0, \dotsc, s_{n-1}}$ and alternating inputs $t = \Bkt{t_0, \dotsc, t_{m-1}}$
with $m \geq 1$.
The number of additional gates (types \labelcref{alt-split-gates,base-case-gates,symtree-gates,split-gates}) needed
for the construction of $\depthckt\Bkt{s, t}$ is shown in \cref{table-add-gates-m-n}
for the following values of $m$ and $n$:
\begin{itemize}
 \item $1 \leq m \leq 2$ and $n \in \N$ arbitrary
 \item $d \leq 4$ and all $m, n \in \N$ with $\dmin(n, m) = d$
\end{itemize}
\begin{proof}
 See \cref{app-proofs}, \cpageref{proof-size-small-values}.
\end{proof}
\end{lemma}

We now give a common upper bound for the number of additional gates
in the cases considered in \cref{table-add-gates-m-n}.

\begin{corollary} \label{add-gates-upper-bound}
Consider the circuit $\depthckt\Bkt{s, t}$ computed by \cref{alg-grin-leftist}
for symmetric inputs $s = \Bkt{s_0, \dotsc, s_{n-1}}$ and alternating inputs $t = \Bkt{t_0, \dotsc, t_{m-1}}$
with $m \geq 1$.
For all $m, n$ appearing in \cref{table-add-gates-m-n},
the number of additional gates in $\depthckt\Bkt{s, t}$ is at most $m + \rho(n)$
with $\rho$ defined in \cref{def-rho}
\begin{proof}
See \cref{app-proofs}, \cpageref{proof-add-gates-upper-bound}.
\end{proof}
\end{corollary}

\begin{figure}[!h]
  \begin{center}
  \adjustbox{max width=.7\textwidth}{
  \centering{\includegraphics{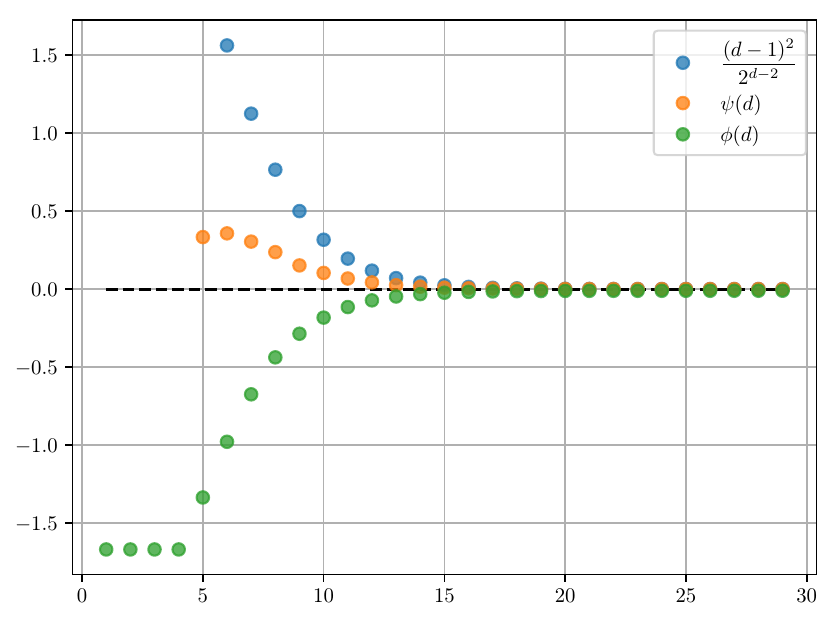}}
  }
  \end{center}
 \caption{The functions
       $\frac{(d-1)^2}{2^{d-2}}$ for $d \geq 6$ from \cref{lem-S},
       $\psi \colon \N_{\geq 5} \to \R$ from \cref{lem-psi},
       and $\phi \colon \N_{\geq 1} \to \R$ from \cref{lem-phi}.}
 \label{img-phi}
\end{figure}

For general $n$ and $m$, the number of additional gates
in our circuit for $f\Bkt{\Bkt{s_0, \dotsc, s_{n-1}}, \Bkt{t_0, \dotsc, t_{m-1}}}$
will be estimated in \cref{lem-Phi},
but we still need some technical preparations.
The following \lcnamecrefs{lem-phi} introduce important functions that are used
when proving \cref{lem-Phi}.
Note that we show three of these functions in \cref{img-phi}.

\begin{lemma} \label{lem-S}
For $n \in \N_{\geq 1}$,
consider the finite series $S\!_n := \sum_{k = 2}^n \frac{(k-1)^2}{2^{k-2}}$.
Then, the following statements are fulfilled:
\begin{multicols}{2}
\begin{enumerate}
 \item \label{explicit-S_n}
$S\!_n = 12 - \frac{4}{2^n} \Bkt{n^2 + 2 n + 3}$
\item \label{S_n-20}
$\sum_{k \geq 19} \frac{(k-1)^2}{2^{k-2}} \leq 0.006$
\end{enumerate}
\end{multicols}
\begin{proof}
See \cref{app-proofs}, \cpageref{proof-lem-S}.
\end{proof}
\end{lemma}

\begin{notation} \label{def-flodd}
 For $x \in \R$,
 let $\flodd(x) := \max \set{y \in \Z \where y \text{ odd}, y \leq x}$.
\end{notation}

\begin{lemma} \label{lem-psi}
 For $d \in \N_{\geq 5}$,
 define $\psi(d) \in \R_{\geq 0}$ by
\[\psi(d) := \frac{1 + \rho\Bkt{\frac{1}{2}\Bkt{\flodd\bktRfixed{big}{\xi\frac{2^{d-1} - 2}{d-1}} + 1}}}{\BktU{\xi\frac{2^{d-1} - 2}{d-1}} + 2}\,.\]
Then, we have
$
\psi(d) \leq \frac{(d-1)^2}{2^{d-2}}
$.
\begin{proof}
See \cref{app-proofs}, \cpageref{proof-lem-psi}.
\end{proof}
\end{lemma}

In the following \lcnamecref{lem-phi}, we will require an upper bound on $\psi(d)$ for all $d \geq 6$.
The previous \lcnamecref{lem-psi} suggests to use $\frac{(d-1)^2}{2^{d-1}}$ as this upper bound.
But in \cref{img-phi}, we see that the difference $\frac{(d-1)^2}{2^{d-1}} - \psi(d)$ is very large for small $d$,
though it quickly decreases to a value close to $0$.
Hence, in the following proof,
we evaluate $\psi(d)$ explicitly for $d \leq 18$ in \cref{table-psi}
and use the upper bound from \cref{lem-psi} only for $d \geq 19$.

{
\setlength{\tabcolsep}{15pt}
\renewcommand{\arraystretch}{1.1}

\begin{table}
 \begin{center}
 \begin{tabular}{rcc}
  \multicolumn{1}{c}{$d$} & \multicolumn{1}{c}{Upper bound on $\psi(d)$} & \multicolumn{1}{c}{Upper bound on $\sum_{d' = 5}^{d} \psi(d')$} \\ \midrule
 5 & 0.3334 & 0.3334 \\
 6 & 0.3572 & 0.6906 \\
 7 & 0.3044 & 0.9950 \\
 8 & 0.2369 & 1.2319 \\
 9 & 0.1516 & 1.3835 \\
10 & 0.1035 & 1.4870 \\
11 & 0.0677 & 1.5547 \\
12 & 0.0428 & 1.5975 \\
13 & 0.0249 & 1.6224 \\
14 & 0.0151 & 1.6375 \\
15 & 0.0090 & 1.6465 \\
16 & 0.0053 & 1.6518 \\
17 & 0.0030 & 1.6548 \\
18 & 0.0017 & 1.6565 \\ \bottomrule
\end{tabular}
 \caption[Upper bounds on $\psi(d)$ and $\sum_{d' = 5}^{d} \psi(d')$ for $d \in \set{5, \dotsc, 18}$.]{Upper bounds on $\psi(d)$ and $\sum_{d' = 5}^{d} \psi(d')$ for $d \in \set{5, \dotsc, 18}$.
          All bounds on $\psi$ have been calculated by a C++ program using floating-point interval arithmetic
          and rounding to fixed precision afterwards.
          The upper bound $1.37$ on $\sum_{d' = 5}^{18} \psi(d')$ is used in the proof of \cref{lem-phi}.}
\label{table-psi}
\end{center}
\end{table}
}

\begin{lemma} \label{lem-phi}
 Define the function $\phi \colon \N_{\geq 1} \to \R$ by
 $
  \phi(d) =
  \begin{cases}
    -1.67 & \text{for $d \leq 4$}\\
  \phi(d-1) + \psi(d)  & \text{for $d \geq 5$}
 \end{cases}\,.
 $
 Then, $\phi(d)$ is negative and monotonely increasing
 for all $d \geq 1$.
 \begin{proof}
  We have $\frac{1}{2}\Bkt{\flodd\bktRfixed{big}{\xi\frac{2^{d-1} - 2}{d-1}} + 1} \geq 3$
  and thus $\rho\Bkt{\frac{1}{2}\Bkt{\flodd\bktRfixed{big}{\xi\frac{2^{d-1} - 2}{d-1}} + 1}} \geq 2$
  for all $d \geq 5$. This implies 
  $\psi(d) > 0$, and hence the second statement holds.
  The first statement is implied by
  \newcommand{\aw}{6}
  \begin{equation*}
   \phi(4) + \sum_{d \geq 5} \psi(d)
   \overset{\shortcref{lem-psi}}{\leq} \phi(4) + \sum_{d = 5}^{18} \psi(d) + \sum_{d \geq 19} \frac{(d-1)^2}{2^{d-2}}
   \overset{\substack{\shortcref{table-psi},\\ \shortcref{lem-S}}}{\leq} \phi(4) + 1.6565 + 0.006
   < \phi(4) + 1.67
   =0\,. \tag*{\qedhere}
  \end{equation*}
 \end{proof}
\end{lemma}

The correlation of $\phi(d)$ and $\psi(d)$ for $d \geq 5$
is visualized in \cref{img-phi}.

In \cref{lem-Phi}, we finally bound the number of additional gates
in the circuit $\depthckt\Bkt{(s_0, \dotsc, s_{n-1}), (t_0,
\dotsc, t_{m-1})}$ by $\alpha m + \rho(n) - 1$
for some $\alpha \in \R_{\geq 1}$ to be defined.
We prove the statement by induction on $d := \dmin(n, m)$,
and in order to make the induction step work,
we need to show a stronger upper bound on the number of additional gates,
namely $\Bkt{\alpha + \phi(d)} m + \rho(n) - 1$.
This is a stronger bound since $\phi(d) \leq 0$ by \cref{lem-phi},
and we will see in the following proof that $\phi(d)$ and $\alpha$ are defined
in a way such that
\begin{itemize}
 \item $\alpha + \phi(d) = 1$ for $d \leq 4$,
 which ensures that the upper bound is valid in the case of small $d$
 (see also \cref{add-gates-upper-bound}),
 i.e., start of the induction of the proof, and
 \item $\phi(d) = \phi(d-1) + \psi(d)$ for $d \geq 5$,
 which will be the definition needed for $d \geq 5$
 in case of the alternating split in \cref{Grinchuk algo end} of \cref{grinchuk-alg},
 i.e., in case~2.2 of the proof.
\end{itemize}

\begin{lemma} \label{lem-Phi}
   Given input variables $s = \Bkt{s_0, \dotsc, s_{n-1}}$ and $t = \Bkt{t_0, \dotsc, t_{m-1}}$
   with $m \geq 1$,
   the number of additional gates (types \labelcref{alt-split-gates,base-case-gates,symtree-gates,split-gates} of \cref{grin-gate-partition})
   used in the circuit $\depthckt\Bkt{s, t}$
   computed by \cref{alg-grin-leftist} is at most
   \[\Phi(d, m, n) := \Bkt{\alpha + \phi(d)} m + \rho(n) \,,\]
   where $d := \dmin(n, m)$
   and $\phi(d)$ as in \cref{lem-phi},
   $\rho(n)$ as in \cref{def-rho}, and $\alpha = 2.67$.
  \begin{proof}
   We prove the statement by induction on $d$ and
   make a case distinction similarly as in \cref{grinchuk-alg}.

   For the base case, assume that $d \leq 4$ or $m \leq 2$.
   By \cref{size-small-values,add-gates-upper-bound}, we need at most $m + \rho(n)$ additional gates.
   By \cref{lem-phi}, we have $\phi(d) \geq -1.67 = - \alpha + 1$ for all $d \in \N_{> 1}$.
   Hence, we have
    $\Phi(d, m, n) = \Bkt{\alpha + \phi(d)} m + \rho(n) 
      \geq m + \rho(n)$,
   which finishes the proof in this case.

   For the induction step, assume that $m \geq 3$ and $d \geq 5$.

   We follow the course of \totalref{grinchuk-alg}.
   Note that the value $d$ in our proof coincides with the value $d$ chosen
   in \cref{grinchuk-alg},
   and recall that by \cref{n < 2^d}, we have $n < 2^d$ since $m \geq 2$.

   \textbf{Case 1:} Assume that $n \geq 2^{d-1}$.

   The assumption $d \geq 5$ implies $n \geq 2^{d-1} \geq 16$.
   In this case, we perform a symmetric split in \cref{algo sym split}.
   Recall that $k = 2^{d-1}$.
   By \cref{grinchuk-even-odd}, all symmetric inputs are even,
   and by \cref{lem-s-tri}, $s$ is triangular.
   Let $s'$ be the output of \cref{det-triangular-subset}
   when applied to input set $s$ and leftist tree $S\!_0$
   as in \cref{algo det s'} of \cref{grinchuk-alg}.
   We construct the symmetric tree on $s'$ using \cref{huff-on-B}
   with $|B| - 1$ additional gates,
   where $B := B(s', S\!_0)$ is the set of boundary vertices.
   By induction hypothesis, we need at most $\Phi(d-1, m, n-k)$
   additional gates for computing $C\Bkt{s \backslash s', t}$.
   Adding the concatenation gate, we in total use
   $|B| + \Phi(d-1, m, n-k)$
   additional gates.
   Note that
   {
   \newcommand{\aw}{8}
   \begin{align*}
    \Phi(d, m, n) - \Bkt{|B| + \Phi(d-1, m, n-k)}
    \ca{\aw}{=} \Bkt{\alpha + \phi(d)} m + \rho(n) 
                 - |B| - \Bkt{\alpha + \phi(d-1)} m - \rho(n-k)  \\
    \ca{\aw}{\overset{\shortcref{lem-phi}}{\geq}} \rho(n) - \rho(n-k) - |B| \\
    \ca{\aw}{\overset{\substack{n \geq 16, \\ \shortcref{lem-triangular},\labelcref{tri-rho-bound}}}{\geq}} 0 \,.
   \end{align*}
   }
   Hence, the number of additional gates used in this case is at most $\Phi(d, m, n)$.

   \textbf{Case 2:} Assume that $n < 2^{d-1}$.

   \textbf{Case 2.1:} Assume that $m \leq \mu(d-1, 0)$.

   Note that $n \geq 1$ as otherwise,
   we have $m > \mu(d-1, 0)$ by the choice of $d = \dmin(m, n)$.
   We construct a realization in \cref{algo simple split}
   using the simple split
   \[f(s, t) = \sym\DBkt{s_0, \dotsc, s_{n-1}} \land f\Bkt{\Bkt{}, t} \,. \]

   By induction hypothesis, the circuit for $f\Bkt{\Bkt{}, t}$
   requires $\Phi(d-1, m, 0)$ additional gates.
   Denoting the number of additional gates needed in the circuit for
   $\sym\DBkt{s_0, \dotsc, s_{n-1}}$ by $\sigma$,
   we need at most $\sigma + \Phi(d-1, m, 0) + 1$ additional gates in this case.
   We need to show that this is at most $\Phi(d, m, n)$.
   As we have
   {
   \newcommand{\aw}{6}
   \begin{align*}
    \Phi(d, m, n) - \Bkt{\sigma + \Phi(d-1, m, 0) + 1}
    \ca{\aw}{=} (\alpha + \phi(d)) m + \rho(n) - \Bkt{\sigma + (\alpha + \phi(d-1))m + \rho(0) + 1} \\
    \ca{\aw}{\overset{\rho(0) = 0}{=}} (\phi(d) - \phi(d-1))m + \rho(n) - \sigma - 1 \\
    \ca{\aw}{\overset{\shortcref{lem-phi}}{\geq}} \rho(n) - \sigma - 1\,,
   \end{align*}
   }
   it remains to show that $\rho(n) - \sigma - 1 \geq 0$.

   As $s$ is triangular by \cref{lem-s-tri},
   we can construct the symmetric tree on $s$ via \cref{sym-prep} using
   $\sigma \leq n-1$ gates if $n \leq 2$ and
   $\sigma \leq \bktD{2 \log_2 n} - 2$ additional gates otherwise.
   For $n \leq 2$, this shows the statement as $\rho(n) = n$ in this case.
   For $n \geq 3$, as $\rho(n) = \bktD{2 \log_2(n-1)}$,
   we have
   \[
   \rho(n) - \sigma - 1 \geq \BktD{2 \log_2(n-1)} - \BktD{2 \log_2 n} + 2 - 1
   \overset{\substack{n \geq 3, \\ \shortcref{lem-log-plus-2}}}{\geq} 0\,.\]

   \textbf{Case 2.2:} Assume that
   \begin{equation} \label{grin-size-m large enough}
     m > \mu(d-1, 0)\,,
   \end{equation}
   i.e., that we perform
   an alternating split
   \[f(s, t) = f\Bkt{s, \prefix{t}{k}} \land \Bkt{f\Bkt{\everysecond{\prefix{t}{k}}, \suffix{t}{k}}}^*\]
   with a prefix of odd length $k$
   in \cref{Grinchuk algo end}.



   Recall that $d \geq 5$.

   By induction hypothesis, we need at most $\Phi(d-1, k, n)$
   additional gates for $\depthckt\Bkt{s, \prefix{t}{k}}$
   and at most $\Phi\Bkt{d-1, m-k, \frac{k-1}{2}}$
   additional gates for $\Bkt{\depthckt\Bkt{\everysecond{\prefix{t}{k}}, \suffix{t}{k}}}^*$.
   Note that by \cref{grin-gate-partition},
   the concatenation gate counts as an additional gate in the case of an alternating split.
   In total, we have at most
   \newcommand{\aw}{2}
   \begin{align*}
    \Phi(d-1, k, n) + \Phi\Bkt{d-1, m-k, \frac{k-1}{2}} + 1
    \ca{\aw}{=} \Bkt{\alpha + \phi(d-1)} k + \rho(n) + \Bkt{\alpha + \phi(d-1)} (m-k) + \rho\Bkt{\frac{k-1}{2}} + 1\\
    \ca{\aw}{=} \Bkt{\alpha + \phi(d-1)} m + \rho(n) + \rho\Bkt{\frac{k-1}{2}} + 1
   \end{align*}
   additional gates.
   We need to show that this is at most $\Phi(d, m, n) = \Bkt{\alpha + \phi(d)} m + \rho(n)$.
   Hence, it suffices to show
   \[\rho\Bkt{\frac{k-1}{2}} + 1 + \Bkt{\alpha + \phi(d-1)} m \leq \Bkt{\alpha + \phi(d)} m\,.\]
   Since for $d \geq 5$, by \cref{lem-phi}, we have
   $\phi(d) - \phi(d-1) = \psi(d)$,
   it remains to show that
   \begin{equation}\label{final-cond}
    \rho\Bkt{\frac{k-1}{2}} + 1 \leq \psi(d) m \,.
   \end{equation}
   Due to assumption \labelcref{grin-size-m large enough},
   we have
   $
    m > \mu(d-1, 0) = \xi \frac{2^{d-1} - 2}{d-1} + 2
   $.
   As $m$ is integral, this implies
   $
    m \geq \BktU{\xi\frac{2^{d-1} - 2}{d-1}} + 2
   $.
   Furthermore, by the choice of $k$ in \cref{algo compute k},
   $k$ is the maximum odd integer with $k \leq \mu(d-1, n)$, in other words,
   \[
    k = \flodd\Bkt{\mu(d-1, n)} = \flodd\Bkt{\xi\frac{2^{d-1} - n - 2}{d-1} + 2}
     \leq \flodd\Bkt{\xi\frac{2^{d-1} - 2}{d-1}} + 2\,.
   \]
   Using these two bounds, the fact that $\psi(d) \geq 0$ for all $d \geq5$ by
   its definition in \cref{lem-psi}
   and the fact that $\rho$ is increasing by \cref{obs-rho},
   inequality~\labelcref{final-cond} is hence implied by
    \[  1 + \rho\left(\frac{1}{2}\Bkt{\flodd\Bkt{\xi\frac{2^{d-1} - 2}{d-1}} + 1}\right)
      = \psi(d) \Bkt{\BktU{\xi\frac{2^{d-1} - 2}{d-1}} + 2}\,,\]
   which holds by definition of $\psi$ (see \cref{lem-psi}).

   Hence, the induction step also holds in case 2.2.2
   and the \lcnamecref{lem-Phi} is proven.
  \end{proof}
\end{lemma}

Now, we can finally state and prove the main theorem of this \lcnamecref{sec-grinchuk-size}.

\linebox{
\begin{theorem} \label{grinchuk-improved-size}
 Given input variables $s = \Bkt{s_0, \dotsc, s_{n-1}}$ and $t = \Bkt{t_0, \dotsc, t_{m-1}}$
 with $m \geq 1$,
\totalref{alg-grin-leftist} computes a circuit $\depthckt\Bkt{s, t}$
for the extended \aop{} $f(s, t)$ with
 \begin{align*}
  \size(\depthckt\Bkt{s, t}) &\leq 3.67 m + n + \rho(n) - 2\,,
 \end{align*}
  where
  $\rho(n) = \begin{cases}
            n & \text{if $n \in \set{0, 1, 2}$} \\
            \BktD{2 \log_2(n-1)}     & \text{if $n \geq 3$}
           \end{cases}$
  as in \cref{def-rho}.
\end{theorem}
}           
\begin{proof}
 We partition the gates of $\depthckt\Bkt{s, t}$ as in \cref{grin-gate-partition}.
 \begin{enumerate}[align=left,labelwidth=\parindent]
  \item There are at most $m + n - 2$ gates contained in the leftist trees $S\!_0$ and $S\!_1$ by \cref{lem-leftist-gates}.
  \item[(ii) - (v)] Let $d := \dmin(n, m)$.
   By \cref{lem-Phi}, the number of additional gates is at most
 \begin{equation*}
  \Phi(d, m, n)
  =  \Bkt{\alpha + \phi(d)} m + \rho(n) 
  \overset{\substack{\shortcref{lem-phi}}}{\leq} \alpha m + \rho(n) 
  \overset{\substack{\alpha = 2.67}}{=} 2.67 m + \rho(n) \,.
 \end{equation*}
 \end{enumerate}
 Adding up all these gates yields the claimed size bound.
\end{proof}

\begin{proposition} \label{grin-runtime}
 Given symmetric inputs $s = \Bkt{s_0, \dotsc, s_{n-1}}$ and alternating inputs $t = \Bkt{t_0, \dotsc, t_{m-1}}$
 with $m \geq 1$,
 \totalref{alg-grin-leftist} has running time $\mathcal O((m+n) (\log_2 n + \log_2 m))$.
 \begin{proof}
  Constructing the leftist trees $S\!_0$ and $S\!_1$ takes time $\mathcal O(m + n)$,
  and computing the data from \cref{runtime-det-B}
  takes time $\mathcal O((m+n) \log_2(m + n))$.
  It remains to bound the running time of \totalref{grinchuk-alg}.

  Recall that in each recursive call of \cref{grinchuk-alg},
  $t$ is a consecutive set of the inputs by \cref{obs-t-consecutive}
  and $s$ is triangular by \cref{lem-s-tri},
  i.e., $s$ consists of two consecutive input sets by \cref{def-triangular}.
  Hence, we can pass on $s$ and $t$ during the algorithm via a constant number of indices.

  Note that in each recursive call of \cref{grinchuk-alg},
  we build at least one gate.
  Hence, by \cref{grinchuk-improved-size}, there are at most $\mathcal O(m + n)$ recursive calls.

  In a single recursive call, the running time is dominated by \cref{algo compute d,algo compute k}
  and the construction of symmetric trees using \cref{sym-prep}.
  Using binary search, \cref{algo compute k} can be executed in time $\mathcal O(\log_2 m)$, and,
  as by \cref{Grinchuk-own-thm}, we have $d \in \mathcal O(\log_2(m + n))$,
  \cref{algo compute d} can be executed in time $\mathcal O(\log_2\log_2(m + n))$.
  Note that each symmetric tree computed has $s$ as inputs,
  plus potentially $t_0$ and $t_1$.
  As $s$ and $s \tupleconcat (t_0)$ are both triangular by \cref{lem-s-tri},
  by \cref{sym-prep},
  computing a single symmetric tree requires time at most $\mathcal O(\log_2 n)$
  using the precomputed data from \cref{runtime-det-B}.

  In total, this means that \cref{grinchuk-alg} runs in time $\mathcal O((m + n)(\log_2 m + \log_2 n))$.
 \end{proof}
\end{proposition}

For the special case of \aop{}s, plugging together
\cref{cor-aop-depth,grinchuk-improved-size,grin-runtime} yields the following result.

\linebox{
\begin{qedtheorem} \label{depth-opt-size}
 Given input variables $t = \Bkt{t_0, \dotsc, t_{m-1}}$ with $m \geq 2$,
 \totalref{alg-grin-leftist} computes a circuit $C$
 for the \aop{} $g(t)$ with
 \[
 \depth(C) \leq \log_2 m + \log_2 \log_2 m + 0.65
 \qquad \text{and} \qquad
 \size(C)  \leq 3.67 m - 2
 \]
 in running time $\mathcal O(m \log_2 m)$.
\end{qedtheorem}
}

\section{Faster Linear-Size Adder Circuits} \label{sec-adder}

We now present the fastest known linear-size adder circuits.
In \cref{sec: step 1}, we construct adder circuits
with a good depth and sub-quadratic size,
based on our \aop{} optimization algorithm from \cref{sec-aop-opt}.
In \cref{sec: linearization framework},
we develop a general linearizing framework,
allowing us to linearize our circuits from \cref{sec: step 1}
in \cref{sec: step 2}.

\subsection{Fast Adder Circuits with Sub-Quadratic Size} \label{sec: step 1}

Our idea for constructing fast adder circuits with a sub-quadratic size
is to apply the \aop{} optimization algorithm from \cref{sec-aop-opt}
(\cref{alg-grin-leftist})
only for the computation of some carry bits.
Hence, we use the generic procedure described in \cref{alg-step-1}
as a core routine for constructing our first family of adders.
Here, we want to compute an $n$-bit adder circuit
and assume that we already know how to construct
three types of circuits for each $k < n$:
$k$-bit adder circuits $A_k$ on $k$ input pairs,
\andprefixckt{}s $S\!_k$ on $k$ inputs,
and \aop{} circuits $AOP_k$ on $k$ inputs.
Later, we will, e.g., use our \aop{} circuits from \cref{depth-opt-size}
as $AOP_k$,
but for now, \cref{alg-step-1} is formulated in a general way.

\begin{figure}[h]
\begin{algorithm}[H]
  \caption{$2$-part adder construction framework} 
  \label{alg-step-1}
  \DontPrintSemicolon
  \KwIn{$n \in \N_{\geq 2}$ and $n$ input pairs $p_0, g_0, \dotsc, p_{n-1}, g_{n-1}$,
  adder circuits $(A_{k})_{k < n}$,
  \andprefixckt{}s $(S\!_{k})_{k < n}$,
  \aop{} circuits $(AOP_k)_{k < n}$.}
  \KwOut{An adder circuit $C_n$ on $p_0, g_0, \dotsc, p_{n-1}, g_{n-1}$.}

  \BlankLine
  $k_l \gets \BktD{ \frac{n}{2} }$, $k_r \gets \BktU{\frac{n}{2}}$.\; \label{step-1-def-k}
  $P_r \gets \Bkt{p_0, g_0, \dotsc, p_{k_r - 1}, g_{k_r - 1}}$,
  $P_l \gets \Bkt{p_{k_r}, g_{k_r}, \dotsc, p_{n - 1}, g_{n - 1}}$. \label{step-1-def-parts}\;
  Compute an adder circuit $A_{k_l}$ on $P_l$.\;
  Compute an adder circuit $A_{k_r}$ on $P_r$.\;
  Compute an \andprefixckt{} $S\!_{k_l}$ on the inputs $p_i$ with $i > k_r$ of $P_{l}$.\;
  Compute an \aop{} circuit $AOP_{k_r}$ on the inputs of $P_{r}$.\;
  \For{$i \gets 1$ \KwTo $k_r$}
  {
     $\cktout_i\Bkt{C_n} \gets \cktout_i\Bkt{A_{k_r}}$.\; \label{i < k_r}
  }
  \For{$i \gets 1$ \KwTo $k_l$}
  {
     $\cktout_{k_r + i}\Bkt{C_n} \gets \cktout_{i}\Bkt{A_{k_l}} \lor \Bkt{\cktout_i\Bkt{S\!_{k_l}} \land AOP_{k_r}}$.\; \label{i > k_r}
  }
  \KwRet{$C_n$.}
\end{algorithm}
\end{figure}

\cref{alg-step-1} works as follows:
In \cref{step-1-def-parts},
we partition the $n$ input pairs $p_0, g_0, \dotsc, p_{n-1}, g_{n-1}$
into two parts
$P_r$ and $P_l$ of roughly equal sizes
$k_l$ and $k_r$ as defined in \cref{step-1-def-k}.
Note that we have
$ k_l + k_r = n\,, $
so this is indeed a partition of all input pairs.
On the parts $P_r$ and $P_l$,
we construct an adder circuit $A_{k_l}$ and $A_{k_l}$, respectively.
Hence, for the part $P_r$, we can read off all the carry bits
for $C_{n}$ from $A_{k_r}$.
Moreover, we construct an \aop{} circuit $AOP_{k_r}$ on part $P_r$
and an \andprefixckt{} $S\!_{k_l}$ on $P_l$.
Using these two circuits and $A_{k_l}$,
for the part $P_l$, the carry bits can be computed by applying the alternating split
presented in \totalref{alternating-split-propagate}, see also \cref{img-split-step-1-1}.
Thus, \cref{alg-step-1} correctly computes all the carry bits
and hence an adder circuit on $n$ input pairs.
In \cref{img-split-step-1-2}, we depict all circuits used to construct
the adder circuit in \cref{alg-step-1}.
The following \lcnamecref{lem-two-framework} estimates the depth and size of this circuit.

\begin{figure}[h]
  \begin{center}
  \begin{subfigure}{0.47\textwidth}
  \adjustbox{max width=\textwidth}{
  \centering{\begin{tikzpicture}

\node[input] (x0)  at (-1.5, 6.2){$g_5$};
\node[input] (x1)  at (-0.5, 6.2){$p_5$};
\node[input] (x2)  at (0.5,  6.2){$g_4$};
\node[input] (x3)  at (1.5,  6.2){$p_4$};
\node[input] (x4)  at (2.5,  6.2){$g_3$};
\draw[thick, ->] (x4) -- ($(x4) - (0,.9)$);
\node[input] (x5)  at (3.5,  6.2){$p_3$};
\node[input] (x6)  at (4.5,  6.2){$g_2$};
\node[input] (x7)  at (5.5,  6.2){$p_2$};
\node[input] (x8)  at (6.5,  6.2){$g_1$};
\node[input] (x9)  at (7.5,  6.2){$p_1$};
\node[input] (x10) at (8.5,  6.2){$g_0$};
\node[input] (x11) at (9.5,  6.2){$p_0$};

\draw[thick, decorate,decoration={brace, amplitude=5pt, raise=10pt}] (x0.west) -- (x5.east) node [midway, above, sloped, yshift=15pt, scale=1.6] {$P_l$};
\draw[thick, decorate,decoration={brace, amplitude=5pt, raise=10pt}] (x6.west) -- (x11.east) node [midway, above, sloped, yshift=15pt, scale=1.6] {$P_{r}$};

\node[or-gate] at (7,4) (or1){};
\node[and-gate] at (6,3) (and1){};

\node[or-gate] at (5,2) (or4){};

\node[and-gate] at (8,5) (and7){};

\node[and-gate] at (2,5) (and6){};
\node[or-gate] at (1,4) (or9){};

\node[and-gate] at (0,3) (and10){};
\node[or-gate] at (-1,2) (or11){};

\node[sym-and-gate] at (3,5) (and12){};
\draw[thick] (x5) -- (and12.input 1);
\draw[thick] (x3) -- (and12.input 2);

\node[sym-and-gate] at (2,4) (and13){};
\draw[thick] (and12.output) -- (and13.input 1);
\draw[thick] (x1) -- (and13.input 2);

\draw[thick] (or1.output) -- (and1.input 1);
\draw[thick] (x8) -- (or1.input 2);
\draw[thick] (x7) -- (and1.input 2);
\draw[thick] (x9) -- (and7.input 2);

\draw[thick] (x10) -- (and7.input 1);

\draw[thick] (x6) -- (or4.input 2);
\draw[thick] (and1.output) -- (or4.input 1);

\draw[thick] (x4) -- (and6.input 1);

\draw[thick] (x3) -- (and6.input 2);

\draw[thick] (x2) -- (or9.input 2);
\draw[thick] (and6.output) -- (or9.input 1);
\draw[thick] (and7.output) -- (or1.input 1);

\draw[thick] (or9.output) -- (and10.input 1);
\draw[thick] (x1) -- (and10.input 2);
\draw[thick] (and10.output) -- (or11.input 1);
\draw[thick] (x0) -- (or11.input 2);

\draw[thick, ->] (x10.south) -- ($(x10.south) - (-0.4, 0.5)$);

\draw[thick, ->] (or9.output) -- ($(or9.output) - (0, 0.5)$);
\draw[thick, ->] (or11.output) -- ($(or11.output) - (0, 0.5)$);

\draw[thick, ->] (and13.output) -- ($(and13.output) - (0, 0.5)$);
\draw[thick, ->] (and12.output) -- ($(and12.output) - (0, 0.5)$);

\node[draw, circle, anchor=north] at ($(x10.south) - (-0.4, 0.5)$) (c0) {$c_1$};
\node[draw, circle, anchor=north] at ($(or1.output) - (0, 0.5)$) (c1) {$c_2$};
\node[draw, circle, anchor=north] at ($(or4.output) - (0, 0.5)$) (c2) {$c_3$};
\draw[thick, ->] (or1.output) -- ($(or1.output) - (0, 0.5)$);
\draw[thick, ->] (or4.output) -- ($(or4.output) - (0, 0.5)$);

\node[concat-and-gate] at (4,1) (and15){};
\draw[thick] (or4.output) -- (and15.input 1);
\draw[thick] (x5) -- (and15.input 2);

\node[concat-or-gate] at (3,0) (or16){};
\draw[thick] (and15.output) -- (or16.input 1);
\draw[thick] (x4) -- (or16.input 2);
\draw[thick, ->] (or16.output) -- ($(or16.output) - (0, 0.5)$);
\node[draw, circle, anchor=north] at ($(or16.output) - (0, 0.5)$) (c3) {$c_4$};

\node[concat-and-gate] at (2,1) (and17){};
\draw[thick] (or4.output) -- (and17.input 1);
\draw[thick] (and12.output) -- (and17.input 2);
\node[concat-or-gate] at (1,0) (or18){};
\draw[thick] (and17.output) -- (or18.input 1);
\draw[thick] (or9.output) -- (or18.input 2);
\draw[thick, ->] (or18.output) -- ($(or18.output) - (0, 0.5)$);
\node[draw, circle, anchor=north] at ($(or18.output) - (0, 0.5)$) (c4) {$c_5$};

\node[concat-and-gate] at (0,1) (and19){};
\draw[thick] (or4.output) -- (and19.input 1);
\draw[thick] (and13.output) -- (and19.input 2);
\node[concat-or-gate] at (-1,0) (or20){};
\draw[thick] (and19.output) -- (or20.input 1);
\draw[thick] (or11.output) -- (or20.input 2);
\draw[thick, ->] (or20.output) -- ($(or20.output) - (0, 0.5)$);
\node[draw, circle, anchor=north] at ($(or20.output) - (0, 0.5)$) (c5) {$c_6$};

\end{tikzpicture}}
  }
  \caption{Example for the computation of the carry bit $c_6$
  in \cref{i > k_r} of \cref{alg-step-1}.
  We show ripple-carry adders in red and green as $A_{k_l}$ and $A_{k_r}$,
  an \andprefixckt{} in yellow and concatenation gates in blue.}
  \label{img-split-step-1-1}
 \end{subfigure}
 \hfill
 \begin{subfigure}{0.47\textwidth}
 \begin{center}
   \adjustbox{max width=0.9\textwidth}{
  \centering{\begin{tikzpicture}

 \phantom{\node[or-gate] at (0,1.5) (or1){};}

\def\partwidth{4}
\def\firstlevel{0}
\def\secondlevel{-1}
\def\thirdlevel{-2.4}
\def\blockheight{0.7}

\draw (0, \firstlevel) rectangle +(\partwidth,\blockheight) node[pos=.5] (x0) {$P_{l}$};
\draw (4, \firstlevel) rectangle +(\partwidth,\blockheight) node[pos=.5] (x1) {$P_{r}$};
\def\blockwidth{3.8}
\def\blockoffset{0.1}
\def\arrowlength{.4}
\def\arrowdist{.2}
\def\symtreeoffset{0.5}
\draw [fill=cyan] (\blockoffset, \secondlevel) rectangle +(\blockwidth,0.7) node[pos=.5] (A_l) {$A_{k_l}$};
\foreach \i in {0,...,3}
{
 \draw[thick, ->] ($(\blockoffset,\secondlevel) + (\i + \arrowdist, 0)$) -- ($(\blockoffset,\secondlevel) + (\i + \arrowdist, -\arrowlength)$);
}
\draw [fill=cyan] (\partwidth+\blockoffset, \secondlevel) rectangle +(\blockwidth,\blockheight) node[pos=.5] (A_r) {$A_{k_r}$};
\foreach \i in {0,...,3}
{
 \draw[thick, ->] ($(\partwidth+\blockoffset, \secondlevel) + (\i + \arrowdist, 0)$) -- ($(4+\blockoffset, \secondlevel) + (\i + \arrowdist, -\arrowlength)$);
}
\draw [fill=mid_yellow] (\blockoffset, \thirdlevel) rectangle +(\blockwidth,\blockheight) node[pos=.5] (S) {$S_{k_l}$};
\foreach \i in {0,...,3}
{
 \draw[thick, ->] ($(\blockoffset + \symtreeoffset, \thirdlevel) + (\i + \arrowdist, 0)$) -- ($(\blockoffset+\symtreeoffset, \thirdlevel) + (\i + \arrowdist, -\arrowlength)$);
}
\draw [fill=mid_green] (\partwidth+\blockoffset, \thirdlevel) rectangle +(\blockwidth,\blockheight) node[pos=.5] (AOP) {$AOP_{k_r}$};
\draw[thick, ->] ($(\partwidth+\blockoffset, \thirdlevel) + (\arrowdist, 0)$) -- ($(4+\blockoffset, \thirdlevel) + (\arrowdist, -\arrowlength)$);

\end{tikzpicture}}
  }
  \end{center}
  \vspace{1.68cm}
   \caption{All components constructed in
        \cref{alg-step-1}.
        On $P_l$, we see the adder circuit $A_{k_l}$ in blue
        and the \andprefixckt{} $S\!_{k_l}$ in yellow;
        and on $P_r$, we see the adder circuit $A_{k_r}$ in blue
        and the \aop{} circuit $AOP_{k_r}$ in green.}
  \label{img-split-step-1-2}
  \end{subfigure}
  \end{center}
 \caption{Illustration of \cref{alg-step-1}.}
 \label{img-split-step-1}
\end{figure}

\begin{lemma} \label{lem-two-framework}
 Given $n \in \N$ with $n \geq 2$, adder circuits $(A_{k})_{k < n}$,
  \andprefixckt{}s $(S\!_{k})_{k < n}$,
  and \aop{} circuits $(AOP_k)_{k < n}$,
 \cref{alg-step-1} computes an adder circuit $C_n$ with
 \begin{align*}
  d(C_n) &\leq \max \left\{d\Bkt{A_{k_r}}, d\Bkt{A_{k_l}} + 1, d\Bkt{AOP_{k_r}} + 2, d\Bkt{S\!_{k_l}} + 2\right\} \quad \text{and}\\
  s(C_n) &\leq s\Bkt{A_{k_r}} + s\Bkt{A_{k_l}} + s\Bkt{AOP_{k_r}} + s\Bkt{S\!_{k_l}} + 2k_l\,,
 \end{align*}
where $k_l :=  \BktD{ \frac{n}{2} }$ and $k_r := \BktU{\frac{n}{2}}$.
 \begin{proof}
  The size of $C_n$ arises from adding up the sizes of all sub-circuits,
  plus $2 k_l$ since for each $i \in \set{k_r + 1, \dotsc, n}$,
  the $i$-th carry bit is computed using two additional gates in \cref{i > k_r}
  (in \cref{img-split-step-1-1}, for $c_6$, these are the two blue gates).

  For the depth estimation, we consider two different cases.
  If $i \leq k_r$, the depth of the $i$-th carry bit is simply the depth of $A_{k_r}$, see \cref{i < k_r}.
  If $i > k_r$, the $i$-th carry bit is computed in \cref{i > k_r} with a depth of at most
  $\max \BktC{d\Bkt{A_{k_l}} + 1, d\Bkt{AOP_{k_r}} + 2, d\Bkt{S\!_{k_l}} + 2}$.
  From this, the depth bound follows.
 \end{proof}
\end{lemma}

Note that in \cref{alg-step-1},
we actually compute the carry bit $\cktout_{k_r}(C_n)$ twice --
once by the \aop{}, once by the last output of $A_{k_r}$.
We do this as it does not harm the overall analysis and simplifies notations.

We now derive an adder family $\Bkt{A^{1}_n}_{n \in \N}$ with a good depth and
with a size in the order of $\mathcal O(n \log_2 n)$.
For small $n$, we use existing adder circuits.
For large $n$, we use \cref{alg-step-1} in a recursive fashion:
We apply \cref{alg-step-1} with the \aop{} circuit from \cref{depth-opt-size}
as $AOP_{k_r}$ and the circuit $S\!^f_{k_l}$ by \citet{LF80} with $f = 2$
as $S\!_{k_l}$,
see \cref{Ladner Fischer};
the adder circuits $A_{k_l}$ and $A_{k_r}$ are computed recursively
using \cref{alg-step-1}.
In order to bound the depth and size of the arising circuits,
we need two numerical inequalities.
\parbox{\textwidth}{
\begin{lemma} \label{step1-num}
For $4 \leq n \leq 17$, the following statements are fulfilled.
\begin{multicols}{2}
\begin{enumerate}
\item We have $0.441 \log_2 n \leq \log_2 \log_2 n - 0.024$. \label{num-441}
\item We have $1.5 \Bkt{n^2 - n} \leq 6.2 n \log_2 n$. \label{num-sq}
\end{enumerate}
\end{multicols}
\end{lemma}
}
\begin{proof}
See \cref{app-proofs}, \cpageref{proof-step1-num}.
\end{proof}

The following \lcnamecref{theorem::step_1_red_version} describes our concrete strategy
for computing the adder family $\Bkt{A^1_n}_{n \in \N}$ and
analyzes the resulting depth and size.

\linebox{
\begin{theorem}\label{theorem::step_1_red_version}
 Given $n \in \N$ with $n \geq 3$ and $c := 6.2$,
 we can construct
 an adder circuit $A^1_n$ on $n$ input pairs
 with
 \begin{align*}
  d\Bkt{A^1_n} \leq \log_2 n + \log_2 \log_2 n + 2.65
  \qquad \text{and} \qquad
  s\Bkt{A^1_n} \leq c n \log_2 n
 \end{align*}
  in running time $\mathcal O\Bkt{n \log_2^2 n}$.
\end{theorem}
}
 \begin{proof}
 First, by induction on $n$, we will see the depth and size bounds.

 \textbf{Case 1:} Assume that $n = 3$.

 In this case, we choose $A^1_3$ to be the ripple carry adder.
 Its depth is $2n - 2 = 4 < \log_2 n + \log_2 \log_2 n + 2.65$ and its size
 is $2n - 2 = 4 < c n \log_2 n$ (see \cref{rem-ripple}).

 \textbf{Case 2:} Assume that $4 \leq n \leq 17$.

 We apply the \aop{} optimization algorithm by \citet{Held-Spirkl-AOPs}
 (see~\cref{eq::sophie_aop_depth}) to compute
 each carry bit separately
 and obtain a circuit $A^1_n$ with
 {
  \setlength{\belowdisplayskip}{0pt}
  \setlength{\belowdisplayshortskip}{0pt}
 \newcommand{\aw}{6}
 \begin{alignat*}{5}
  d\Bkt{A^1_n} & \overset{\text{\cite{Held-Spirkl-AOPs}}}{\leq} 1.441 \log_2 n + 2.674
           & \ca{\aw}{\overset{\substack{n \leq 17, \\ \shortcref{step1-num}}}{\leq}} \log_2 n + \log_2 \log_2 n + 2.65
  \quad \text{ and } \\
 s\Bkt{A^1_n} &\overset{\text{\cite{Held-Spirkl-AOPs}}}{\leq}
 \sum_{i = 1}^n (3i - 3) = 1.5(n^2 - n)
   & \ca{\aw}{\overset{\substack{n \leq 17, c = 6.2, \\ \shortcref{step1-num}}}{\leq}} c n \log_2 n\,.
 \end{alignat*}
 }

  \textbf{Case 3:} Assume that $n \geq 18$.

  We may assume inductively that for all $i < n$, an adder $A^1_i$ with the stated depth and size can be constructed.
  Hence, we may apply \cref{alg-step-1} using the following sub-circuits:

  \begin{itemize}
   \item We use $A_{k_l} := A_{k_l}^1$ and $A_{k_r} := A^1_{k_r}$
   as adder circuits on $P_l$ and $P_r$, respectively.
   We have
   { 
    \newcommand{\aw}{3}
    \begin{alignat}{5}
     d\Bkt{A_{k_l}} \ca{\aw}{\overset{\indhyp}{\leq}}         \log_2(k_l) + \log_2 \log_2(k_l) + 2.65
                 \overset{k_l \leq \frac{n}{2}}{\leq}  \log_2 n + \log_2 \log_2 n + 1.65
     \ \ \text{ and} \ \
     & s\Bkt{A_{k_l}} \ca{\aw}{\overset{\indhyp}{\leq}}         c k_l \log_2 (k_l)\,, \label{d s A kl red}\\
     d\Bkt{A_{k_r}} \ca{\aw}{\overset{\indhyp}{\leq}}         \log_2(k_r) + \log_2 \log_2(k_r) + 2.65
                 \overset{k_r \leq n}{\leq}            \log_2 n + \log_2 \log_2 n + 2.65
     \ \ \text{ and} \ \
     & s\Bkt{A_{k_r}} \ca{\aw}{\overset{\indhyp}{\leq}}         c k_r \log_2 (k_r)\label{d s A kr red}\,.
    \end{alignat}
   }
   \item We choose $S\!_{k_l} := S\!_{k_l}^2$
        to be the Ladner-Fischer \andprefixckt{} \cite{LF80} for $f=2$,
        as denoted in \cref{Ladner Fischer}.
        Note that the choice $f = 2$ fulfills the requirement
        $f \leq \lceil \log_2 k_l \rceil$ as $n \geq 6$.
        We have
        \begin{equation}
         d(S\!_{k_l}) \leq \BktU{ \log_2 k_l} + 2
         \overset{k_l = \BktD{\frac{n}{2}}}{\leq}
         \log_2 n + 2
         \overset{n \geq 6}{\leq} \log_2 n + \log_2 \log_2 n + 0.65
         \ \ \text{ and } \ \
         s(S\!_{k_l}) \leq 2 \left(1 + 2^{-2}\right) k_l = 2.5 k_l
         \,. \label{d s LF red}
        \end{equation}
   \item As \aop{} circuit $AOP_{k_r}$, we use the circuit
         computed by \cref{alg-grin-leftist}.
         Since $k_r \leq \frac{n + 1}{2}$ and $AOP_{k_r}$ has $2 k_r - 1 \leq n$ inputs,
         by \cref{depth-opt-size}, we have
    \begin{equation}
    d(AOP_{k_r}) \leq \log_2 n + \log_2 \log_2 n + 0.65
    \quad \text{ and } \quad
    s(AOP_{k_r}) \leq 3.67 n - 2\,. \label{Grinchuk-own-depth-size-red}
    \end{equation}
 \end{itemize}

 For the depth of $A^1_{n}$, these observations together with \cref{lem-two-framework} imply
 { 
        \newcommand{\aw}{5}
 \begin{equation*}
  d\Bkt{A^1_{n}} \overset{\text{Lem. }\labelcref{lem-two-framework}}{\leq}
                \max \left\{d\Bkt{A_{k_r}}, d\Bkt{A_{k_l}} + 1, d\Bkt{AOP_{k_r}} + 2, d\Bkt{S\!_{k_l}} + 2\right\}
             \overset{\substack{\labelcref{d s A kl red}, \labelcref{d s A kr red}, \\
                                         \labelcref{Grinchuk-own-depth-size-red}, \labelcref{d s LF red}}}
                                        {\leq}
                \log_2 n + \log_2 \log_2 n + 2.65\,.
 \end{equation*}
 }

 It remains to compute the size of $A^1_{n}$.
 As $n \geq 18$, we have
  \begin{equation} \label{new log(n/2) red}
    \log_2 k_r \overset{k_r = \BktU{\frac{n}{2}}}{\leq} \log_2 \left( \frac{n+1}{2} \right)
      = \log_2 n - 1 + \log_2 \left(1 + \frac{1}{n}\right)
      \overset{n \geq 18}{\leq} \log_2 n - 0.921\,.
  \end{equation}
  Moreover, for $0 \leq \alpha \leq 1$, we have
  \begin{equation}
   \alpha k_r + k_l = \begin{cases}
                        \Bkt{\alpha +1} \frac{n}{2} & \text{if $n$ even} \\
                        \alpha \frac{n+1}{2} + \frac{n-1}{2} & \text{if $n$ odd}
                      \end{cases}
                    \overset{\alpha \leq 1}{\geq} n \frac{\alpha+1}{2} + \frac{\alpha-1}{2}\,. \label{a k_r + k_l}
  \end{equation}

  Based on these two inequalities, we can bound the total size of the recursively computed adder circuits by
  { 
  \setlength{\belowdisplayskip}{0pt}
  \setlength{\belowdisplayshortskip}{0pt}
  \newcommand{\aw}{6}
  \begin{align}
  \begin{split}
      s(A_{k_r}) + s(A_{k_l})
  \overset{\labelcref{d s A kr red}, \labelcref{d s A kl red}}{\leq}
     c k_r \log_2 (k_r) + c k_l \log_2 (k_l)
  \ca{\aw}{\overset{\labelcref{new log(n/2) red}, k_l \leq \frac{n}{2}}{\leq}}
     c \Bkt{k_r (\log_2 n - 0.921) + k_l (\log_2 n - 1)}  \\
  \ca{\aw}{\overset{k_r + k_l = n}{=}} c \Bkt{n \log_2 n - 0.921 k_r - k_l}  \\
  \ca{\aw}{\overset{\labelcref{a k_r + k_l}}{\leq}}
     c \Bkt{n \log_2 n - n \frac{0.921+1}{2} - \frac{0.921-1}{2}} \\
  \ca{\aw}{=} c \Bkt{n\big( \log_2 n - 0.9605\big) + 0.0395}\,. \label{summed ind size red}
  \end{split}
  \end{align}
  }
  In total, the size of $A^1_n$ is hence at most
  { 
  \newcommand{\aw}{10}
  \setlength{\belowdisplayskip}{0pt}
  \setlength{\belowdisplayshortskip}{0pt}
  \begin{align*}
   s(A^1_n) \ca{\aw}{\overset{\text{Lem. } \labelcref{lem-two-framework}}{\leq}}
      s(A_{k_r}) + s(A_{k_l}) + s(AOP_{k_r}) + s(S\!_{k_l}) + 2k_l \\
  \ca{\aw}{\overset{\labelcref{summed ind size red}, \labelcref{Grinchuk-own-depth-size-red}, \labelcref{d s LF red}}{\leq}}
      c \Big(n\big( \log_2 n - 0.9605\big) + 0.0395\Big)
      + 3.67 n + 2.5 k_l + 2 k_l \\
  \ca{\aw}{\overset{k_l \leq \frac{n}{2}}{\leq}}
      cn \log_2 n - 0.9605 cn + 0.0395 c + 5.92 n \\
  \ca{\aw}{=}
       c n \left( \log_2 n -0.9605 + \frac{0.0395}{n} + \frac{5.92}{c} \right) \\
  \ca{\aw}{\overset{\substack{n \geq 18, \\ c \geq 6.2}}{\leq}}
     c n \Bkt{\log_2 n - 0.9609 + 0.003 + 0.955}\\
  \ca{\aw}{<} c n \log_2 n\,.
  \end{align*}
  }

Finally, for proving the running time bound,
it suffices to consider the case $n \geq 18$.
We show that there is a constant $\alpha$
such that we can compute our circuit in $\alpha n \log^2_2 n$ elementary steps.

In one call of \cref{alg-step-1},
we apply two recursive calls to instances with size of at most $\frac{n+1}{2}$ each,
so we may assume inductively that each of them needs at most
$\alpha \frac{n+1}{2} \log_2^2 \left( \frac{n+1}{2} \right)$ steps.
By \cref{depth-opt-size}, there is $\beta \in \N$
such that the \aop{} $AOP_{k_r}$ with at most $n$ inputs can be computed with
$\beta n \log_2 n$ steps.
The rest of one call of \cref{alg-step-1} (including the computation of
$S\!_{k_l} = S\!^2_{k_l}$, see \cref{LF-runtime}) takes linear time
$\gamma n$ for some $\gamma \in \N$.

For $n \geq 21$, we have
\begin{equation}
\alpha n \log_2^2 \left( \frac{n+1}{2} \right)
\stackrel{n \geq 18}{\leq} \alpha n \log_2^2 \left( \frac{n}{1.8} \right)
<  \alpha n \left(\log_2 n - 0.84 \right)^2
 =  \alpha n \log_2^2 n - 1.68 \alpha n  \log_2 n
  + 0.7056 \alpha n \,, \label{bound-runtime-A}
\end{equation}
so in total, the number of steps for computing $A^1_n$ can be bounded by
{ 
        \newcommand{\aw}{3}
\begin{align*}
2 \alpha \frac{n+1}{2} \log_2^2 \left( \frac{n+1}{2} \right)
   + \beta n \log_2 n
   + \gamma n
\ca{\aw}{=} \alpha n \log_2^2 \left( \frac{n+1}{2} \right)
  + \alpha \log^2_2 \left( \frac{n+1}{2} \right)
  + \beta n \log_2 n
   + \gamma n\\
\ca{\aw}{\overset{\labelcref{bound-runtime-A}}{\leq}}  \alpha n \log_2^2 n - 1.68 \alpha n  \log_2 n
  + 0.7056 \alpha n
  + \alpha \log^2_2 \left( \frac{n+1}{2} \right)
  + \beta n \log_2 n
   + \gamma n\,.
\end{align*}
}
This is at most $\alpha n \log^2 n$ if $\alpha$ is chosen sufficiently large compared to
$\beta$ and $\gamma$.
\end{proof}

\subsection{An Adder Linearization Framework} \label{sec: linearization framework}

In this \lcnamecref{sec: linearization framework},
we develop a framework that linearizes the size of a given adder family $\Bkt{B_n}_{n \in \N}$ with sub-quadratic size
in a way that depth does not increase too much.
The idea of this linearization goes back to \citet{Ofman63} and \citet{Khrapchenko-construction},
but we perform some crucial changes in order to obtain a best possible depth.
See \cref{remark-linearization-comp} for a comparison of the two linearization frameworks.

\cref{alg-framework} is a method for solving the \pradderopt{} on $n$ input pairs
that is an extension of \cref{alg-step-1}.
It depends on oracles solving the \pradderopt{}, the \praopdepthopt{} and the \prmultiand{}.
In the following, we assume w.l.o.g.\ for the family $(A_k)_{k \in \N}$ of adder circuits 
used in the algorithm that $d(A_k) \leq d(A_l)$ for $k \leq l$.
In \cref{general linearization}, we present our linearization which uses this algorithm.

\begin{figure}[t]
\LinesNumbered
 \begin{algorithm}[H]
   \caption{$l$-part adder construction framework}
   \label{alg-framework}
   \DontPrintSemicolon
    \KwIn{$n \in \N$, $n \geq 2$,
          $n$ input pairs $p_0, g_0, \dotsc, p_{n-1}, g_{n-1}$,
          a family of adder circuits $(A_k)_{k \in \N}$,
          a family of \andprefixckt{}s $(S\!_k)_{k \in \N}$,
          a family of \aop{} circuits $(AOP_k)_{k \in \N}$.}
    \KwOut{An adder circuit $C_n$ on $p_0, g_0, \dotsc, p_{n-1}, g_{n-1}$.}
    \BlankLine
      Choose $k \in \N_{> 0}$ and $l \gets \lceil n / k \rceil$.\;
      \label{alg-frame-partition-start}
      \For{$j \gets 0$ \KwTo $l-2$}
      {
        $P^{(j)} \gets \bkt{p_{j k}, g_{jk}, \dotsc, p_{(j+1)k - 1}, g_{(j+1)k - 1}}$.\;
      }
      $P^{(l-1)} \gets \bkt{p_{(l-1) k}, g_{(l-1)k}, \dotsc, p_{n-1}, g_{n-1}}$.\;
      \For{$j \gets 0$ \KwTo $l-1$}
      {
        $n_j \gets |P^{(j)}|$,
        $N_j \gets n_0 + \dotsc + n_{j-1}$.\; \label{alg-frame-partition-end}
      }
      Construct an adder circuit $\ckt{A}{0}{n_0}$ on $\inpart{0}$.\;
      \For{$i \gets 1$ \KwTo $n_0$}{
        Let $\cktout_i\Bkt{C_n} \gets \cktout_i\Bkt{\ckt{A}{0}{n_0}}\label{final 0}$.\; \label{alg-frame-P0}
      }
      \For{$j \gets 1$ \KwTo $l-1$}{
        Construct an adder circuit $\ckt{A}{j}{n_j}$ on $\inpart{j}$.\; \label{alg-frame-A}
        Construct an \andprefixckt{} $\ckt{S}{j}{n_j}$ on $\inpart{j}$.\; \label{alg-frame-S}
        Construct an \aop{} circuit $\ckt{AOP}{j}{N_j}$ on the $N_j$ input pairs in
         $\inpart{j-1}, \dotsc, \inpart{0}$.\; \label{alg:aop_comp}
        Let $\cktout_{N_j}\Bkt{C_n} \gets \ckt{AOP}{j}{}$.\;
        \For{$i \gets 1$ \KwTo $n_j$}{
          Let $\cktout_{N_{j} + i}\Bkt{C_n} \gets \cktout_{i}\Bkt{\ckt{A}{j}{n_j}}
             \lor \left( \cktout_{i}\Bkt{\ckt{S}{j}{n_j}} \land \ckt{AOP}{j}{N_j} \right)$. \label{final general}\;
        }
      }
      \KwRet{$C_n$.}\;
 \end{algorithm}
\end{figure}

\begin{figure}[!h]
  \begin{center}
  \adjustbox{max width=0.9\textwidth}{
  \centering{\begin{tikzpicture}

\def\partwidth{2}
\def\firstlevel{0}
\def\secondlevel{-1}
\def\thirdlevel{-2.4}
\def\blockheight{0.7}

\draw (0, \firstlevel) rectangle +(\partwidth,\blockheight)              node[pos=.5] {$\inpart{l-1}$};
\draw (\partwidth, \firstlevel) rectangle +(\partwidth,\blockheight)     node[pos=.5] {$\inpart{l-2}$};
\draw (2*\partwidth, \firstlevel) rectangle +(\partwidth,\blockheight)   node[pos=.5] {$\dotsc$};
\draw (3*\partwidth, \firstlevel) rectangle +(\partwidth,\blockheight)   node[pos=.5] {$\inpart{j+1}$};
\draw (4*\partwidth, \firstlevel) rectangle +(2*\partwidth,\blockheight) node[pos=.5] {$\dotsc, g^{(j)}_{i}, p^{(j)}_{i}, \dotsc$};
\draw (6*\partwidth, \firstlevel) rectangle +(\partwidth,\blockheight)   node[pos=.5] {$\inpart{j-1}$};
\draw (7*\partwidth, \firstlevel) rectangle +(\partwidth,\blockheight)   node[pos=.5] {$\dotsc$};
\draw (8*\partwidth, \firstlevel) rectangle +(\partwidth,\blockheight)   node[pos=.5] {$\inpart{1}$};
\draw (9*\partwidth, \firstlevel) rectangle +(\partwidth,\blockheight)   node[pos=.5] {$\inpart{0}$};

\def\leftpartj{4*\partwidth}
\def\blockwidth{3.8}
\def\blockoffset{0.1}
\def\arrowlength{.4}
\def\arrowdist{.2}
\def\symtreeoffset{0.4}
\draw [fill=cyan] (\leftpartj+\blockoffset, \secondlevel) rectangle +(\blockwidth,0.7) node[pos=.5] {$\ckt{A}{j}{n_j}$};
\foreach \i in {0,...,3}
{
 \draw[thick, ->] ($(\leftpartj+\blockoffset,\secondlevel) + (\i + \arrowdist, 0)$) -- ($(\leftpartj+\blockoffset,\secondlevel) + (\i + \arrowdist, -\arrowlength)$);
}
\draw [fill=mid_yellow] (\leftpartj+\blockoffset, \thirdlevel) rectangle +(\blockwidth,0.7) node[pos=.5] {$\ckt{S}{j}{n_j}$};
\foreach \i in {0,...,3}
{
 \draw[thick, ->] ($(\leftpartj+\blockoffset+\symtreeoffset,\thirdlevel) + (\i + \arrowdist, 0)$) -- ($(\leftpartj+\blockoffset+ \symtreeoffset,\thirdlevel) + (\i + \arrowdist, -\arrowlength)$);
}
\draw [fill=mid_green] (\leftpartj+2*\partwidth+\blockoffset, \thirdlevel) rectangle +(4*\partwidth-2*\blockoffset,0.7) node[pos=.5] {$\ckt{AOP}{j}{N_j}$};
\draw[thick, ->] ($(\leftpartj+2*\partwidth +\blockoffset,\thirdlevel) + (\arrowdist, 0)$) -- ($(\leftpartj+2*\partwidth +\blockoffset,\thirdlevel) + (\arrowdist, -\arrowlength)$);

\end{tikzpicture}}
  }
  \end{center}
 \caption{Illustration of the circuits constructed for part $P^{(j)}$
       in \crefrange{alg-frame-A}{alg:aop_comp}
       of \cref{alg-framework}.}
 \label{fig-adder-framework}
\end{figure}

\cref{alg-framework} constructs an adder circuit $C_n$ as follows:
Given $k \in \N_{> 0}$ and $l = \BktU{\frac{n}{k}}$,
in \crefrange{alg-frame-partition-start}{alg-frame-partition-end},
we partition the inputs into $l$ consecutive parts $\inpart{0}, \dotsc, \inpart{l-1}$,
where each part has $n_j \leq k$ input pairs.
On each part $\inpart{j}$, we compute an adder circuit $\ckt{A}{j}{n_j}$.
In part $\inpart{0}$, we can directly read off the carry bits of $C_n$ from $\ckt{A}{0}{n_j}$
in \cref{alg-frame-P0}.
For computing the carry bits of $C_n$ in part $\inpart{j}$ for $j > 0$,
as in \cref{alg-step-1}, we use the alternating split from \cref{alternating-split-propagate}
in \cref{final general}.
For this, we need an adder circuit $\ckt{A}{j}{n_j}$
and an \andprefixckt{} $\ckt{S}{j}{n_j}$ on $\inpart{j}$,
and an \aop{} circuit $\ckt{AOP}{j}{N_j}$ on the input parts $\inpart{j-1}, \dotsc, \inpart{0}$.
These are constructed in \crefrange{alg-frame-A}{alg:aop_comp}.
Hence, all carry bits are computed correctly and $C_n$ is an adder circuit on $n$ input pairs.
\cref{fig-adder-framework} illustrates the circuits computed for part $\inpart{j}$.

Note that \cref{alg-step-1} is a special case of \cref{alg-framework}
where $k = \lceil \frac{n}{2}\rceil$.

It is easy to read off the depth and size of $C_n$ from the construction:

\begin{observation} \label{thm-framework}
The circuit $C_n$ computed by \cref{alg-framework} fulfills
{
\newcommand{\aw}{2}
\begin{align*}
d(C_n) \ca{\aw}{\leq} \max \BktC{ d(A_{n_0}),\, \max_{j \in \bktCfixed{normal}{1, \dotsc, l-1}} \BktC{d\bktRfixed{normal}{A_{n_j}} + 1,
              \max \BktC{d\bktRfixed{normal}{S\!_{n_j}}, d\bktRfixed{normal}{AOP_{N_j}}}+ 2 }} \\
       \ca{\aw}{\overset{n_j \leq k}{\leq}} \max \BktC{ d\Bkt{A_{k}} + 1, d\Bkt{S\!_{k}} + 2,
              \max_{j \in \{1, \dotsc, l-1\}} \BktC{d\bktRfixed{normal}{AOP_{N_j}} + 2 }}\,, \\
s(C_n) \ca{\aw}{\leq} s\Bkt{A_{n_0}} + \sum_{j = 1}^{l-1} \Bkt{s\bktRfixed{normal}{A_{n_j}} + s\bktRfixed{normal}{S\!_{n_j}} + s\bktRfixed{normal}{AOP_{N_j}} + 2(n_j-1)}\\
       \ca{\aw}{<}    s\bktRfixed{normal}{A_{n_0}} + \sum_{j = 1}^{l-1} \Bkt{s\bktRfixed{normal}{A_{n_j}} + s\bktRfixed{normal}{S\!_{n_j}} + s\bktRfixed{normal}{AOP_{N_j}} }
           + 2n \,.
\end{align*}
}
\end{observation}

\cref{alg-framework} will be the main ingredient of our linearization framework.
However, the circuit to be linearized is not $A_k$,
but an adder circuit $B_l$ with $l$ input pairs which we introduce now.
For this, we compute the \aop{}s $AOP_{N_j}$
in \cref{alg:aop_comp} in a special way.
Here, given $j \in \set{0, \dotsc, l-1}$,
we write $a(P^{(j)}) := \sym\DBkt{p_{jk}, p_{jk+1}, \dotsc, p_{jk + n_j - 1}}$
for the conjunction of all ``$p$-inputs`` of part $P^{(j)}$,
and $h^*(P^{(j)}) := g^*(g_{jk + n_j - 1}, p_{jk + n_j - 1}, \dotsc, g_{jk+1}, p_{jk+1}, g_{jk})$
for the \aop{} on all inputs of $P^{(j)}$ but $p_{jk}$.

\begin{figure}[!h]
  \begin{center}
  \adjustbox{max width=0.8\textwidth}{
  \centering{\begin{tikzpicture}

\node[input] (x-12)  at (-13.5, 6.2){$g_{11}$};
\node[input] (x-11)  at (-12.5, 6.2){$p_{11}$};
\node[input] (x-10)  at (-11.5,  6.2){$g_{10}$};
\node[input] (x-9)  at (-10.5,  6.2){$p_{10}$};
\node[input] (x-8)  at (-9.5,  6.2){$g_9$};
\node[input] (x-7)  at (-8.5,  6.2){$p_9$};
\draw[thick, ->] (x-7) -- ($(x-7) - (0,.7)$);
\node[input] (x-6)  at (-7.5, 6.2){$g_8$};
\node[input] (x-5)  at (-6.5, 6.2){$p_8$};
\node[input] (x-4)  at (-5.5,  6.2){$g_7$};
\node[input] (x-3)  at (-4.5,  6.2){$p_7$};
\node[input] (x-2)  at (-3.5,  6.2){$g_6$};
\node[input] (x-1)  at (-2.5,  6.2){$p_6$};
\draw[thick, ->] (x-1) -- ($(x-1) - (0,.7)$);
\node[input] (x0)  at (-1.5, 6.2){$g_5$};
\node[input] (x1)  at (-0.5, 6.2){$p_5$};
\node[input] (x2)  at (0.5,  6.2){$g_4$};
\node[input] (x3)  at (1.5,  6.2){$p_4$};
\node[input] (x4)  at (2.5,  6.2){$g_3$};
\node[input] (x5)  at (3.5,  6.2){$p_3$};
\draw[thick, ->] (x5) -- ($(x5) - (0,.7)$);
\node[input] (x6)  at (4.5,  6.2){$g_2$};
\node[input] (x7)  at (5.5,  6.2){$p_2$};
\node[input] (x8)  at (6.5,  6.2){$g_1$};
\node[input] (x9)  at (7.5,  6.2){$p_1$};
\node[input] (x10) at (8.5,  6.2){$g_0$};
\node[input] (x11) at (9.5,  6.2){$p_0$};

\draw[thick, decorate,decoration={brace, amplitude=5pt, raise=10pt}] (x-12.west) -- (x-7.east) node [midway, above, sloped, yshift=15pt, scale=1.6] {$P^{(3)}$};
\draw[thick, decorate,decoration={brace, amplitude=5pt, raise=10pt}] (x-6.west) -- (x-1.east) node [midway, above, sloped, yshift=15pt, scale=1.6] {$P^{(2)}$};
\draw[thick, decorate,decoration={brace, amplitude=5pt, raise=10pt}] (x0.west) -- (x5.east) node [midway, above, sloped, yshift=15pt, scale=1.6] {$P^{(1)}$};
\draw[thick, decorate,decoration={brace, amplitude=5pt, raise=10pt}] (x6.west) -- (x11.east) node [midway, above, sloped, yshift=15pt, scale=1.6] {$P^{(0)}$};

\node[or-gate] at (7,4) (or1){};

\node[and-gate] at (6,3) (and1){};

\node[or-gate] at (5,2) (or4){};
\draw[thick, ->] (or4.output) -- ($(or4.output) - (0,.5)$);

\node[and-gate] at (8,5) (and7){};

\node[and-gate] at (2,5) (and6){};

\node[or-gate] at (1,4) (or9){};

\node[and-gate] at (0,3) (and10){};

\node[or-gate] at (-1,2) (or11){};
\draw[thick, ->] (or11.output) -- ($(or11.output) - (0,.5)$);

\node[sym-and-gate] at (3,5) (and12){};
\draw[thick, ->] (and12.output) -- ($(and12.output) - (0,.5)$);
\draw[thick] (x5) -- (and12.input 1);
\draw[thick] (x3) -- (and12.input 2);

\node[sym-and-gate] at (2,4) (and13){};
\draw[thick, ->] (and13.output) -- ($(and13.output) - (0,.5)$);
\draw[thick] (and12.output) -- (and13.input 1);
\draw[thick] (x1) -- (and13.input 2);

\node[concat-and-gate, fill=blue] at (3.5,1) (and14){};
\draw[thick] (or4.output) -- (and14.input 1);
\draw[thick] (and13.output) -- (and14.input 2);

\node[concat-or-gate] at (2,0) (or15){};
\draw[thick, ->] (or15.output) -- ($(or15.output) - (0,.5)$);
\draw[thick] (and14.output) -- (or15.input 1);
\draw[thick] (or11.output) -- (or15.input 2);

\draw[thick] (or1.output) -- (and1.input 1);
\draw[thick] (x8) -- (or1.input 2);
\draw[thick] (x7) -- (and1.input 2);
\draw[thick] (x9) -- (and7.input 2);

\draw[thick] (x10) -- (and7.input 1);

\draw[thick] (x6) -- (or4.input 2);
\draw[thick] (and1.output) -- (or4.input 1);

\draw[thick] (x4) -- (and6.input 1);

\draw[thick] (x3) -- (and6.input 2);

\draw[thick] (x2) -- (or9.input 2);
\draw[thick] (and6.output) -- (or9.input 1);
\draw[thick] (and7.output) -- (or1.input 1);

\draw[thick] (or9.output) -- (and10.input 1);
\draw[thick] (x1) -- (and10.input 2);
\draw[thick] (and10.output) -- (or11.input 1);
\draw[thick] (x0) -- (or11.input 2);

\node[and-gate] at (-4,5) (and16){};
\draw[thick] (x-2) -- (and16.input 1);
\draw[thick] (x-3) -- (and16.input 2);

\node[or-gate] at (-5,4) (or17){};
\draw[thick] (and16.output) -- (or17.input 1);
\draw[thick] (x-4) -- (or17.input 2);

\node[and-gate] at (-6,3) (and18){};
\draw[thick] (or17.output) -- (and18.input 1);
\draw[thick] (x-5) -- (and18.input 2);

\node[or-gate] at (-7,2) (or19){};
\draw[thick, ->] (or19.output) -- ($(or19.output) - (0,.5)$);
\draw[thick] (and18.output) -- (or19.input 1);
\draw[thick] (x-6) -- (or19.input 2);

\node[sym-and-gate] at (-3,5) (and20){};
\draw[thick, ->] (and20.output) -- ($(and20.output) - (0,.5)$);
\draw[thick] (x-1) -- (and20.input 1);
\draw[thick] (x-3) -- (and20.input 2);

\node[sym-and-gate] at (-4,4) (and21){};
\draw[thick, ->] (and21.output) -- ($(and21.output) - (0,.5)$);
\draw[thick] (and20.output) -- (and21.input 1);
\draw[thick] (x-5) -- (and21.input 2);

\node[concat-and-gate, fill=blue] at (0.5,-1) (and22){};
\draw[thick] (or15.output) -- (and22.input 1);
\draw[thick] (and21.output) -- (and22.input 2);

\node[concat-or-gate] at (-1,-2) (or23){};
\draw[thick, ->] (or23.output) -- ($(or23.output) - (0,.5)$);
\draw[thick] (and22.output) -- (or23.input 1);
\draw[thick] (or19.output) -- (or23.input 2);

\node[and-gate] at (-10,5) (and24){};
\draw[thick] (x-8) -- (and24.input 1);
\draw[thick] (x-9) -- (and24.input 2);

\node[or-gate] at (-11,4) (or25){};
\draw[thick] (and24.output) -- (or25.input 1);
\draw[thick] (x-10) -- (or25.input 2);

\node[and-gate] at (-12,3) (and26){};
\draw[thick] (or25.output) -- (and26.input 1);
\draw[thick] (x-11) -- (and26.input 2);

\node[or-gate] at (-13,2) (or27){};
\draw[thick, ->] (or27.output) -- ($(or27.output) - (0,.5)$);
\draw[thick] (and26.output) -- (or27.input 1);
\draw[thick] (x-12) -- (or27.input 2);

\node[sym-and-gate] at (-9,5) (and28){};
\draw[thick, ->] (and28.output) -- ($(and28.output) - (0,.5)$);
\draw[thick] (x-7) -- (and28.input 1);
\draw[thick] (x-9) -- (and28.input 2);

\node[sym-and-gate] at (-10.2,3.9) (and29){};
\draw[thick, ->] (and29.output) -- ($(and29.output) - (0,.5)$);
\draw[thick] (and28.output) -- (and29.input 1);
\draw[thick] (x-11) -- (and29.input 2);

\node[concat-and-gate, fill=blue] at (-2.5,-3) (and30){};
\draw[thick] (or23.output) -- (and30.input 1);
\draw[thick] (and29.output) -- (and30.input 2);

\node[concat-or-gate] at (-4,-4) (or31){};
\draw[thick, ->] (or31.output) -- ($(or31.output) - (0,.5)$);
\draw[thick] (and30.output) -- (or31.input 1);
\draw[thick] (or27.output) -- (or31.input 2);

\end{tikzpicture}}
  }
  \end{center}
 \caption{Circuit arising from applying \cref{alternating split iteratively}
       to $l = 4$ input parts with $k = 3$ input pairs each.}
 \label{fig-linearization-circuit}
\end{figure}
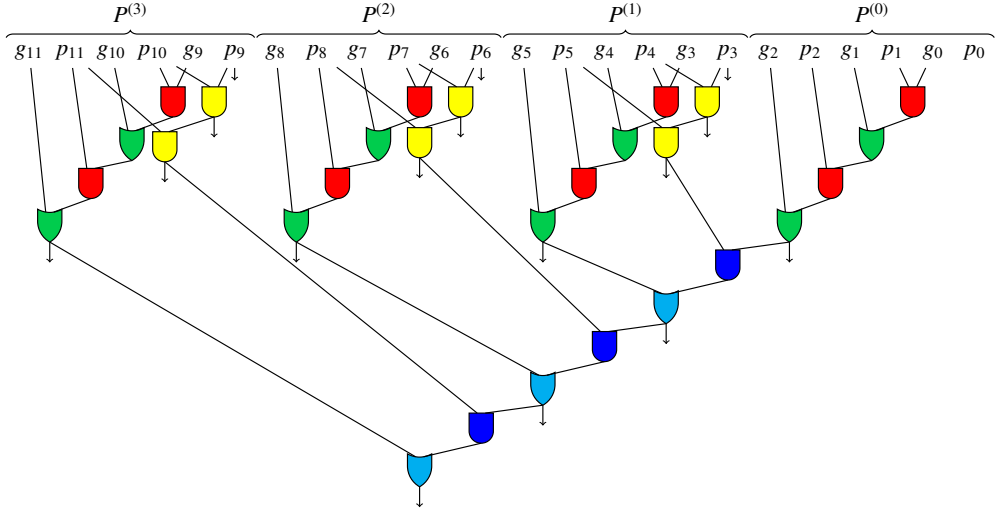

\begin{corollary}\label{alternating split iteratively}
Let $n$ pairs of inputs $g_0, p_1, g_1 \dotsc, p_{n-1}, g_{n-1}$
and a partition of the input pairs into $l$ parts $P^{(l-1)}, \dotsc, P_0$ be given.
We have
\begin{equation*}
  g^*\Bkt{g_{n-1}, p_{n-1},\dotsc, g_1, p_1, g_0}
  = g^*\Bkt{h^*(P^{(l-1)}), a(P^{(l-1)}), \dotsc,
       h^*(P^{(1)}), a(P^{(1)}), h^*(P^{(0)})}\,.
\end{equation*}
\begin{proof}
Applying \cref{alternating-split-propagate} iteratively yields
{ 
   \newcommand{\aw}{2}
\begin{align*}
 g^*(g_{n-1}, p_{n-1},\dotsc, g_0)
 \ca{\aw}{=} h^*(P^{(l-1)}) \lor \Bkt{a(P^{(l-1)}) \land g^*\Bkt{P^{(l-2)} \tupleconcat \dotsc \tupleconcat P^{(0)}}} \\
 \ca{\aw}{=} h^*(P^{(l-1)}) \lor \Bkt{ a(P^{(l-1)}) \land \Bkt{h^*(P^{(l-2)}) \lor \Bkt{a(P^{(l-2)}) \land \Bkt{ \dotsc \lor (a(P_1) \land h^*(P_0))}}}} \\
 \ca{\aw}{=} g^*\Bkt{h^*(P^{(l-1)}), a(P^{(l-1)}), \dotsc, h^*\Bkt{P^{(1)})), a(P^{(1)})), h^*(P^{(0)})}}\,. \tag*{\qedhere}
\end{align*}
}
\end{proof}
\end{corollary}

\cref{fig-linearization-circuit} illustrates this \lcnamecref{alternating split iteratively} on $l = 4$ input parts:
In yellow, we show the \AND{} trees on every part $P^{(j)}$,
in red and green the \aop{}s on every part $P^{(j)}$,
and in dark and light blue the \aop{} that has the outputs of the former mentioned circuits as inputs.
In order to compute all the \aop{}s $AOP_{N_j}$ for $j = 1, \dotsc, l - 1$,
we need the output of the blue \aop{} after each \OR{} gate.
Hence, the blue part can be realized by an adder circuit
as in the following \lcnamecref{lem-aop-by-adder}.

\begin{qedcorollary}\label{lem-aop-by-adder}
 In addition to the input of \cref{alg-framework},
 let a family $(B_k)_{k \in \N}$ of adder circuits be given.
 For each $j \in \set{0, \dotsc, l-1}$,
 compute an \aop{} circuit $AOP_{n_j}$ on input part $P^{(j)}$.
 Compute an adder $B_l$ on $l$ input pairs
 \[\ckt{AOP}{0}{n_0}, \cktout_{n_1}\Bkt{\ckt{S}{1}{n_1}}, \ckt{AOP}{1}{n_1}, \dotsc,
  \cktout_{n_{l-1}}\Bkt{\ckt{S}{l-1}{n_{l-1}}}, \ckt{AOP}{l-1}{n_{l-1}}\,.\]
 For all $j \in \{1, \dotsc, l-1\}$, we have
 $\ckt{AOP}{j}{N_j} = \cktout_{j}\Bkt{B_l}$.
\end{qedcorollary}

\begin{figure}[!h]
  \begin{center}
  \adjustbox{max width=0.7\textwidth}{
  \centering{\begin{tikzpicture}

\def\partwidth{4}
\def\firstlevel{0}
\def\secondlevel{-1}
\def\thirdlevel{-2.4}
\def\fourthlevel{-3.8}
\def\blockheight{0.7}
\def\maxj{4}
\def\blockwidth{3.8}
\def\blockoffset{0.1}
\def\arrowlength{.4}
\def\arrowdist{.2}
\def\symtreeoffset{0.4}
\def\dummy{0.}

\draw [fill=blue] (\blockoffset, \fourthlevel) rectangle +(4*\partwidth-2*\blockoffset,\blockheight) node[pos=.5, color=white] {$B_l$};
\foreach \j in {0,...,2}
{
\pgfmathsetmacro\index{\maxj-\j-1}
 \draw [fill=mid_yellow] (\partwidth*\j+\blockoffset, \secondlevel) rectangle +(\blockwidth,\blockheight) node[pos=.5] {$S_{n_{\pgfmathprintnumber[precision=0]{\index}}}^{(\pgfmathprintnumber[precision=0]{\index})}$};
 \foreach \i in {0,...,3}
 {
  \node (s\j\i) at ($(\partwidth*\j+\blockoffset+\symtreeoffset,\secondlevel) + (\i + \arrowdist, 0)$) {};
  \draw[thick, ->]  (s\j\i.center) -- ($(\partwidth*\j+\symtreeoffset+\blockoffset,\secondlevel) + (\i + \arrowdist, -\arrowlength)$);
 }
 \draw[thick,->] (s\j0.center) -- (s\j0|-\dummy,\fourthlevel+\blockheight);
}
\foreach \j in {0,...,3}
{
 \pgfmathsetmacro\index{\maxj-\j-1}
 \draw (\partwidth*\j, \firstlevel) rectangle + (\partwidth, \blockheight) node[pos=.5] {$P^{(\pgfmathprintnumber[precision=0]{\index})}$};
 \draw [fill=mid_green] (\partwidth*\j+\blockoffset, \thirdlevel) rectangle +(\blockwidth,\blockheight) node[pos=.5] {$AOP_{n_{\pgfmathprintnumber[precision=0]{\index}}}^{(\pgfmathprintnumber[precision=0]{\index})}$};
 \foreach \i in {0,...,3}
 {
   \node (a\j\i) at ($(\partwidth*\j+\blockoffset,\thirdlevel) + (\i + \arrowdist, 0)$) {};
 }
 \draw[thick,->] (a\j0.center) -- (a\j0|-\dummy,\fourthlevel+\blockheight);
 \draw[thick, ->]($(\partwidth*\j+\blockoffset,\fourthlevel) + (\arrowdist, 0)$) -- ($(\partwidth*\j+\blockoffset,\fourthlevel) + (\arrowdist, -\arrowlength)$);
}

\end{tikzpicture}}
  }
  \end{center}
 \caption{Illustration of the construction of the \aop{}s $\ckt{AOP}{j}{N_j}$
       for each $j \in \set{0, \dotsc, l-1}$ as in \cref{lem-aop-by-adder}.}
 \label{fig-aop-constr}
\end{figure}

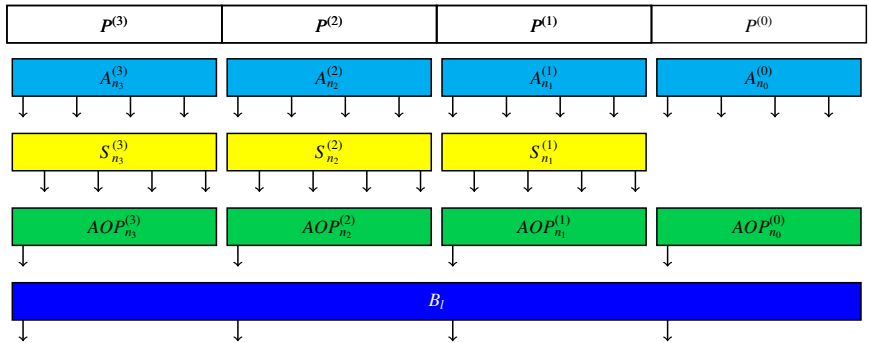
\begin{figure}[!h]
  \begin{center}
  \adjustbox{max width=0.7\textwidth}{
  \centering{\begin{tikzpicture}

\def\partwidth{4}
\def\firstlevel{0}
\def\secondlevel{-1}
\def\thirdlevel{-2.4}
\def\fourthlevel{-3.8}
\def\fifthlevel{-5.2}
\def\blockheight{0.7}
\def\maxj{4}
\def\blockwidth{3.8}
\def\blockoffset{0.1}
\def\arrowlength{.4}
\def\arrowdist{.2}
\def\symtreeoffset{0.4}
\def\dummy{0.}

\foreach \j in {0,...,2}
{
 \pgfmathsetmacro\index{\maxj-\j-1}
 \draw (\partwidth*\j, \firstlevel) rectangle + (\partwidth, \blockheight) node[pos=.5] {$P^{(\pgfmathprintnumber[precision=0]{\index})}$};
 \draw [fill=mid_yellow] (\partwidth*\j+\blockoffset, \thirdlevel) rectangle +(\blockwidth,\blockheight) node[pos=.5] {$S_{n_{\pgfmathprintnumber[precision=0]{\index}}}^{(\pgfmathprintnumber[precision=0]{\index})}$};
 \foreach \i in {0,...,3}
 {
  \node (s\j\i) at ($(\partwidth*\j+\blockoffset+\symtreeoffset,\thirdlevel) + (\i + \arrowdist, 0)$) {};
  \draw[thick, ->]  (s\j\i.center) -- ($(\partwidth*\j+\symtreeoffset+\blockoffset,\thirdlevel) + (\i + \arrowdist, -\arrowlength)$);
 }
}

\foreach \j in {0,...,3}
{
 \pgfmathsetmacro\index{\maxj-\j-1}
 \draw (\partwidth*\j, \firstlevel) rectangle + (\partwidth, \blockheight) node[pos=.5] {$P^{(\pgfmathprintnumber[precision=0]{\index})}$};
 \draw [fill=cyan] (\partwidth*\j+\blockoffset, \secondlevel) rectangle +(\blockwidth,\blockheight) node[pos=.5] {$A_{n_{\pgfmathprintnumber[precision=0]{\index}}}^{(\pgfmathprintnumber[precision=0]{\index})}$};
 \draw [fill=mid_green] (\partwidth*\j+\blockoffset, \fourthlevel) rectangle +(\blockwidth,\blockheight) node[pos=.5] {$AOP_{n_{\pgfmathprintnumber[precision=0]{\index}}}^{(\pgfmathprintnumber[precision=0]{\index})}$};
 \foreach \i in {0,...,3}
 {
  \node (a\j\i) at ($(\partwidth*\j+\blockoffset,\secondlevel) + (\i + \arrowdist, 0)$) {};
  \draw[thick, ->] (a\j\i.center) -- ($(\partwidth*\j+\blockoffset,\secondlevel) + (\i + \arrowdist, -\arrowlength)$);
 }
 \draw[thick, ->]($(\partwidth*\j+\blockoffset,\fourthlevel) + (\arrowdist, 0)$) -- ($(\partwidth*\j+\blockoffset,\fourthlevel) + (\arrowdist, -\arrowlength)$);
 \draw[thick, ->]($(\partwidth*\j+\blockoffset,\fifthlevel) + (\arrowdist, 0)$) -- ($(\partwidth*\j+\blockoffset,\fifthlevel) + (\arrowdist, -\arrowlength)$);
}
\draw [fill=blue] (\blockoffset, \fifthlevel) rectangle +(4*\partwidth-2*\blockoffset,\blockheight) node[pos=.5, color=white] {$B_l$};

\end{tikzpicture}}
  }
  \end{center}
 \caption{Illustration of the adder linearization framework from \cref{general linearization}.}
 \label{fig-linearization}
\end{figure}

Our linearization framework is \cref{alg-framework}
with this \lcnamecref{lem-aop-by-adder} applied
for the computation of the \aop{}s $AOP_{N_j}^{(j)}$
in \cref{alg:aop_comp} of \cref{alg-framework}.
\cref{fig-linearization} shows all the circuits used in
the linearization framework.
These are the circuits used in the adder construction framework from \cref{fig-adder-framework},
where the \aop{}s $\ckt{AOP}{j}{N_j}$ for each $j \in \set{0, \dotsc, l-1}$
are computed using \cref{lem-aop-by-adder} as depicted in \cref{fig-linearization-circuit,fig-aop-constr}.
Here, we need another adder $B_l$ and \aop{}s $\ckt{AOP}{j}{n_j}$.
After analyzing the depth and size of the resulting family of adders,
we shall explain why this is actually a linearization framework for the adder family $\Bkt{B_l}_{l \in \N}$.

\linebox{
\begin{theorem} \label{general linearization}
 Let $n \in \N_{\geq 2}$,
 two families of adder circuits $(A_k)_{k \in \N}$ and $(B_l)_{l \in \N}$,
 a family of \andprefixckt{}s $(S\!_k)_{k \in \N}$,
 and
 a family of \aop{} circuits $(AOP_k)_{k \in \N}$ be given.
 Using \cref{lem-aop-by-adder} for the computation of the \aop{}s in \cref{alg:aop_comp},
 \cref{alg-framework} computes an adder circuit $C_n$ on $n$ input pairs
 with
 \begin{align*}
  d(C_n)  &\leq \max \BktC{ d\bkt{A_{k}} + 1,
              d(B_l) + \max \left\{ d(AOP_{k}), d(S\!_{k}) \right\} + 2} \quad \text{and} \\
  s(C_n) & \leq s(A_{n_0}) + \sum_{j = 1}^{l-1} \Big(s(A_{n_j}) + s(AOP_{n_j})  + s(S\!_{n_j})\Big) + s(B_l) + 2n.
 \end{align*}
\end{theorem}
}
 \begin{proof}
   By construction, the depth of $AOP_{N_j}$ can be bounded by
   \begin{equation}
    d(AOP_{N_j}) \overset{\shortcref{lem-aop-by-adder}}{\leq} d(B_l) + \max \left\{ d(AOP_{n_0}),
        \max_{j = 1, \dotsc, l - 1} \left\{ d(AOP_{n_j}), d(S\!_{n_j}) \right\} \right\}
        \overset{n_j \leq k}{\leq} d(B_l) + \max \left\{ d(AOP_{k}), d(S\!_{k}) \right\}\,. \label{aop-adder-depth}
   \end{equation}
   Hence, we obtain a total depth of
   {
   \newcommand{\aw}{5}
   \begin{align*}
    d(C_n) \ca{\aw}{\overset{\text{Obs. }\labelcref{thm-framework}}{\leq}}
      \max \BktC{ d\bkt{A_{k}} + 1, d\bkt{I_k} + 2,
              \max_{j \in \{1, \dotsc, l-1\}} \BktC{d\Bkt{AOP_{N_j}} + 2 }}\\
      \ca{\aw}{=} \max \BktC{ d\bkt{A_{k}} + 1,
              d(B_l) + \max \left\{ d(AOP_{k}), d(S\!_{k}) \right\} + 2}\,.
   \end{align*}
   }

   The size bound follows directly from combining \cref{thm-framework}
   and \cref{lem-aop-by-adder}.
 \end{proof}

The preceding \lcnamecref{general linearization}
gives a construction method for a family of adder circuits $\Bkt{C_n}_{n \in \N}$ based on
other families of adder circuits $\Bkt{B_l}_{l \in \N}$ and $\Bkt{A_k}_{k \in \N}$,
\aop{} circuits and \AND{}-prefix circuits.
We will see in \cref{cor-linearization} that this is actually a linearization
of the family $\Bkt{B_l}_{l \in \N}$ once the sizes of the \aop{} circuits,
\AND{}-prefix circuits, and $\Bkt{A_k}_{k \in \N}$ are linear.
For this, we will use the adder circuits from the following \lcnamecref{halved-adder} as
family $\Bkt{A_k}_{k \in \N}$.

\begin{proposition} \label{halved-adder}
For each $n \in \N$,
there is an adder circuit $A_n$ with $s(A_n) \leq 3.5 n$ and $d(A_n) \leq n
 + 2$.
\begin{proof}
 For $n \leq 1$, we can construct an adder circuit with depth and size $0$.

 For $n \geq 2$, we apply \cref{alg-step-1} to the following circuit families:
 \begin{itemize}
  \item We use the ripple-carry adder (see \cref{rem-ripple})
  to compute the adder circuits $A_{k_l}$ and $A_{k_r}$.
  From $A_{k_r}$, we can also read off the \aop{} $AOP_{k_r}$.
  \item For $S\!_{k_l}$, we use an \AND{}-path 
       $S\!_{k_l} = p_{n-1} \land (p_{n-2} \land ( \dotsc \land (p_{k_r + 1} \land p_{k_r+1})))$
  as \andprefixckt{} on $P_l$.
 \end{itemize}
 Denote the resulting circuit by $A_n$.
 The depth of $A_n$ is at most
{
\newcommand{\aw}{5}
\begin{align*}
d(A_n) \ca{\aw}{\overset{\shortcref{lem-two-framework}}{\leq}} \max\{d(A_{k_l}) + 1, d(A_{k_r}) + 2, d(S\!_{k_l}) + 2\}
       \overset{\shortcref{rem-ripple}}{=} \max\{2k_l - 1, 2k_r, d(S\!_{k_l}) + 2\}\\
       \ca{\aw} = \max\{2k_l - 1, 2k_r, k_l + 1\}
       = 2 k_r
       \leq n + 2\,.
\intertext{As $A_{k_r}$ and $AOP_{k_r}$ have identical gates, the size of $A_n$ is}
s(A_n) \ca{\aw}{\overset{\shortcref{lem-two-framework}}{\leq}} s(A_{k_r}) + s(A_{k_l}) + s(S\!_{k_l}) + 2 k_l
       \overset{\shortcref{rem-ripple}}{\leq} 2k_r - 2 + 2k_l - 2 + s(S\!_{k_l}) + 2 k_l\\
       \ca{\aw}{\leq} 2k_r - 2 + 2k_l - 2 + k_l - 1 + 2 k_l
       < 2 k_r + 5 k_l
       \leq 3.5 n\,.  \tag*{\qedhere}
 \end{align*}
}
\end{proof}
\end{proposition}

\linebox{
\begin{theorem} \label{cor-linearization}
 Let an adder family $\Bkt{B_l}_{l \in \N}$
 with depth $d(B_l) = \log_2 l + \delta(l)$
 and size $s(B_l) \leq l \sigma(l)$
 be given,
 where $\sigma, \delta \colon \N \to \R_{\geq 0}$ are
 monotonely increasing functions with $\sigma(l)\leq l$ for all $l \in \N$.
 Then, there are constants $\beta$ and $\gamma$ such that for any 
 $n \in \N$, there is an adder circuit $C_n$ with $\size(C_n) \leq \beta n$
 and \[d(C_n) \leq \max \BktC{\sigma(n), d(B_n) + \log_2 \log_2 (\sigma(n))}
   + \gamma\,.\]
\end{theorem}
}
 \begin{proof}
  We apply the linearization from \cref{general linearization} to
  the adder family
  $\Bkt{B_l}_{l \in \N}$, using the following other circuits:
  For the adder family $\Bkt{A_k}_{k \in \N}$, we use
  the circuits from \cref{halved-adder};
  for the \andprefixckt{} family $\Bkt{S\!_k}_{k \in \N}$, we use
  the Ladner-Fischer circuits $\Bkt{S\!_{k}^0}_{k \in \N}$ from \cref{Ladner Fischer} with $f = 0$;
  and for the \aop s $AOP_k$, we use the circuit from \cref{depth-opt-size}
  (with $m \leq 2k$ inputs since we have $k$ input pairs).
  Thus, we have
  \begin{alignat}{5}
   d(A_k)   &\leq k + 2\,,
     && s(A_k) &&\leq 3.5 k\,,\nonumber\\
   d(AOP_k) &\leq \log_2 (2k) + \log_2 \log_2 (2k) + 0.65\,, \quad
     && s(AOP_k) &&\leq 7.34 k - 2\,, \nonumber\\
   d(S\!_k)   &\leq \lceil \log_2 k \rceil\,,
     && s(S\!_k)   &&\leq 2 \left(1 + \frac{1}{2^0}\right) k= 4 k\,. \label{lin-collected}
  \end{alignat}
  We choose $k := \BktU{\sigma(n)}$,
  and $l :=\BktU{\frac{n}{k}}$.
  This implies
  $s(B_l) \in \mathcal O\Bkt{l \sigma(l)} = \mathcal O\Bkt{\frac{n}{\sigma(n)} \sigma(n)} = \mathcal O(n)$.
  As also the sizes of $A_k$, $AOP_k$, and $S\!_k$ are linear in $k$, we obtain
  \begin{equation*}
   s(C_n)
   \overset{\shortcref{general linearization}}{\leq}
     s(A_{n_0}) + \sum_{j = 1}^{l-1} \Big(s(A_{n_j}) + s(AOP_{n_j})  + s(S\!_{n_j})\Big) + s(B_l) + 2n
   \overset{\labelcref{lin-collected}}{\in} \mathcal O(n)\,.
  \end{equation*}
  For analyzing the depth of $C_n$,
  note that $d(AOP_k) \geq d(S\!_k)$.
  We conclude
  {
  \newcommand{\aw}{6}
  \begin{align*}
   d(C_n) \ca{\aw}{\overset{\shortcref{general linearization}}{\leq}}
     \max \BktC{ d\Bkt{A_{k}} + 1,
              d(B_l) + \max \left\{ d(AOP_{k}), d(S\!_{k}) \right\} + 2} \\
   \ca{\aw}{\overset{\labelcref{lin-collected}}{\leq}} \max \BktC{k + 3,  \log_2 l + \delta(l) + \log_2 (2k) + \log_2 \log_2 (2k) + 2.65} \\
   \ca{\aw}{=} \max \BktC{\sigma(n),  \log_2 n - \log_2 k + \delta(l) + \log_2 k + \log_2 \log_2 k} + \const \\
   \ca{\aw}{\leq} \max \BktC{\sigma(n), d(B_n) + \log_2 \log_2 (\sigma(n))} + \const\,. \tag*{\qedhere}
  \end{align*}
  }
 \end{proof}

In particular, if the size of $B_l$ is in the order of $\mathcal O(n \log_2 n)$,
as for the adder from \cref{theorem::step_1_red_version},
we can linearize the adder family $\Bkt{B_l}_{l \in \N}$ with a depth increase of only $\log_2 \log_2 \log_2 n$
(up to an additive constant).
When the size of $B_l$ is even smaller, the depth increase will also become smaller.

\begin{remark} \label{remark-linearization-comp}
Our linearization framework presented in \cref{general linearization,cor-linearization}
is closely related to the linearization by \citet{Ofman63}
and \citet{Khrapchenko-construction},
see \citet{gashkov07} for a concise description.
In \cref{depth-opt-size},
we present the first family of \aop{} circuits with asymptotically optimum depth of $\log_2 n + \log_2 \log_2 n + \const$ and,
at the same time, a linear size.
Hence, we use these circuits for construction of the \aop{}s $AOP_{n_j}$ in \cref{general linearization}.

Instead, \citet{Ofman63} and \citet{Khrapchenko-construction} use the last carry bit of the adder circuits $A_k$ to compute $AOP_{n_j}$.
This way, the depth of $A_k$ is critical for the depth of $C_n$
-- very different to our approach,
where in \cref{cor-linearization}, we may use a circuit with
a high depth of $k + 3$.
The best possible result for \citet{Ofman63} and \citet{Khrapchenko-construction}
is achieved by applying their linearization iteratively,
which leads to a much higher depth increase in the order of $\mathcal O\Bkt{\sqrt[4]{\log_2 n}}$
and linear-size adder circuits with a depth of $\log_2 n + \sqrt{2 \log_2 n} + \mathcal O\Bkt{\sqrt[4]{\log_2 n}}$.
\end{remark}

Another linearization technique,
which is also based on \citet{Khrapchenko-construction},
is described concisely in \citet{HSAdders}.
Given an adder family $(B_l)_{l \in \N}$ and $\tau \leq \log_2 n - 1$,
this linearization constructs adder circuits with
depth at most $d(B_{n / 2^\tau}) + 2 \tau$
and size at most $s(B_{n / 2^\tau}) + 5n$.
If we used this linearization with our
family of adder circuits from \cref{theorem::step_1_red_version}
with a size of $\mathcal O(n \log_2 n)$ as $B_l$,
we would need to choose $\tau$ in the range of $\log_2 \log_2 n$.
This would lead to a much larger depth increase in the order of $\Theta(\log_2 \log_2 n)$.

Note that in \cref{cor-linearization}, we omit the additive constant in the depth
and the multiplicative constant in the size of the linearized adder $C_n$.
In order to tune these constants,
we will perform a more careful analysis in the next \lcnamecref{sec: step 2}.
Moreover, we use slightly different sub-circuits than in \cref{cor-linearization}.

\subsection{Linear-Size Adder Circuits} \label{sec: step 2}

Using the linearization framework presented in \cref{general linearization},
we now linearize the size of the adder circuits from \cref{theorem::step_1_red_version}.
We discuss two results:
First, in \cref{n log log logn red}, we present a linear-size adder with the best
depth we can obtain with our approach.
Secondly, in \cref{better-size-n log log logn red},
we show how to decrease the size of this adder substantially,
allowing the depth to increase by a small additive constant.

Again, in both cases,
we will use other adder circuits when the number of inputs is small.
For bounding their depth, we need a technical \lcnamecref{step-3-num-a}.

\begin{lemma} \label{step-3-num-a}
For $4 \leq n \leq 1024$, we have
$
2 \lceil \log_2 n \rceil
             \leq \log_2 n + \log_2 \log_2 n + \log_2 \log_2 \log_2 n + 6.3
$.
\begin{proof}
See \cref{app-proofs}, \cpageref{proof-step-3-num-a}.
\end{proof}
\end{lemma}

For bounding the depth and size of our adder circuits,
we need another two numerical lemmas.

\begin{lemma} \label{step2-main-depth-bound-a}
Given $k, l, n \in \N_{\geq 1}$ with $n \geq 1024$, $k = \bktU{\log_2 n}$ and $l = \BktU{\frac{n}{k}}$,
we have
\[
\log_2 l + \log_2 \log_2 l + \log_2 k + \log_2 \log_2 (2k) + 6.3
\leq \log_2 n + \log_2 \log_2 n + \log_2 \log_2 \log_2 n + 6.6 \,.
\]
\begin{proof}
See \cref{app-proofs}, \cpageref{proof-step2-main-depth-bound-a}.
\end{proof}

\end{lemma}
\begin{lemma} \label{step2-bound-rest-size-a}
Given $k, l, n \in \N_{\geq 1}$ with $n \geq 1024$, $k = \bktU{\log_2 n}$ and $l = \BktU{\frac{n}{k}}$,
we have
\[
(6.2 \log_2 l - 2) l \leq 6.2 n\,.
\]
\begin{proof}
See \cref{app-proofs}, \cpageref{proof-step2-bound-rest-size-a}.
\end{proof}
\end{lemma}

Using the previous lemmas, we now prove the
main theorem of this \lcnamecref{sec-adder}.

\linebox{
\begin{theorem}\label{n log log logn red}
 Given $n \in \N_{\geq 4}$,
 we can construct
 an adder circuit $A^2_n$ on $n$ input pairs
 in running time $\mathcal O(n \log_2 n)$ with
 \begin{align*}
  d\left(A^2_n\right) \leq \log_2 n + \log_2 \log_2 n + \log_2 \log_2 \log_2 n + 6.6
  \qquad \text{and} \qquad
 s\left(A^2_n\right) \leq 21.6 n\,.
 \end{align*}
\end{theorem}
}

  \begin{proof}
  If $4 \leq n \leq 1024$, we use the adder $L^f_n$ (with $f = 0$)
  by \citet{LF80}
  with size at most $12 n$ and depth at most $2 \lceil \log_2 n \rceil$,
  see \cref{Ladner Fischer}.
  Its size is apparently small enough;
  its depth is sufficiently small by \cref{step-3-num-a}.

  Hence, assume that $n \geq 1024$.
  We apply the linearization from \cref{general linearization} to
  the adder family
  $\Bkt{B_l}_{l \in \N} :=\Bkt{A^1_l}_{l \in \N}$
  from \cref{theorem::step_1_red_version}
  with
  $k = \bktU{\log_2 n}$ and thus $l = \BktU{\frac{n}{k}}\,.$
  For the adder family $\Bkt{A_k}_{k \in \N}$, we use
  the circuits from \cref{halved-adder}.
  For the \andprefixckt{} family $\Bkt{S\!_k}_{k \in \N}$, we use
  the Ladner-Fischer \citep{LF80}
  circuits $\Bkt{S\!_{k}^2}_{k \in \N}$ from \cref{Ladner Fischer} with $f = 2$,
  which is viable as $\bktU{\log_2 k} \geq \log_2 \log_2 n \overset{n \geq 1024}{>} 2$.
  This yields
  $s(S\!_k) \leq 2 \left(1 + \frac{1}{2^2}\right) k = 2.5 k$.
  For the \aop{}s $AOP_k$, we use the circuits from \cref{depth-opt-size}
  (with at most $2k$ inputs).
  Thus, we have
  \begin{alignat}{5}
   & d(A_k)   &&\leq k + 2\,,
     && s(A_k) &&\leq 3.5 k\,, \nonumber \\
   & d(B_l)   &&\leq \log_2 l + \log_2 \log_2 l + 2.65\,,
     && s(B_l) &&\leq 6.2 l \log_2 l\,, \nonumber \\
   & d(AOP_k) &&\leq \log_2 (2k) + \log_2 \log_2 (2k) + 0.65\,, \quad
     && s(AOP_k) &&\leq 7.34k - 2\,,\nonumber \\
   & d(S\!_k)   &&\leq \lceil \log_2 k \rceil + 2\,,
     && s(S\!_k)   &&\leq 2.5 k\,.
  \label{step-2-prerequisits-red}
  \end{alignat}
  We denote the adder circuit resulting from \cref{general linearization}
  applied with this setting by $A^2_n$.

  We now bound the depth of $A^2_n$.
Since $n \geq 1024$ and thus $k \geq 10$, we have
\begin{equation} \label{size-bound-S}
\log_2 (2k) + \log_2 \log_2 (2k) + 0.65
 \overset{k \geq 10}{\geq} \log_2 k + 2.11 + 1.65
 \geq \lceil \log_2 k \rceil + 2
 \geq d(S\!_k)\,.
\end{equation}
Furthermore, we have
\begin{equation} \label{eq::maximum_estimation_in_proof_of_step_2}
\log_2 l + \log_2 \log_2 l + \log_2 (2k) + \log_2 \log_2 (2k) + 5.3
\overset{l = \BktU{\frac{n}{k}}}{>}  \log_2 n + 6.3
\overset{k = \bktU{\log_2 n}}{\geq} k + 5.3 \\
\overset{\labelcref{step-2-prerequisits-red}}{\geq} d(A_k) + 1\,.
\end{equation}
Therefore, the depth of our adder circuit can be bounded by
{
\newcommand{\aw}{7}
\begin{align*}
d(A^2_n) \ca{\aw}{\overset{\shortcref{general linearization}}{\leq}}
\max \BktC{ d(A_{k}) + 1, d(B_l)
               + \max \big\{ d(AOP_{k}), d(S\!_{k}) \big\}  + 2}\\
    \ca{\aw}{\overset{\labelcref{step-2-prerequisits-red}, \labelcref{size-bound-S}, \labelcref{eq::maximum_estimation_in_proof_of_step_2}}{\leq}}
         \log_2 l + \log_2 \log_2 l + 2.65 + \log_2 (2k) + \log_2 \log_2 (2k) + 0.65 + 2\\
     \ca{\aw}{=} \log_2 l + \log_2 \log_2 l + \log_2 k + \log_2 \log_2 (2k) + 6.3 \\
     \ca{\aw}{\overset{\shortcref{step2-main-depth-bound-a}}{\leq}}
       \log_2 n + \log_2 \log_2 n + \log_2 \log_2 \log_2 n + 6.6\,.
\end{align*}
}

  The size of $A^2_n$ can be bounded by
 {
 \setlength{\belowdisplayskip}{0pt}
 \setlength{\belowdisplayshortskip}{0pt}
  \newcommand{\aw}{5}
  \begin{align*}
   s(A^2_n) \overset{\text{Thm. } \labelcref{general linearization}}{<}
      \sum_{j=0}^{l-1} \Bkt{s(A_{n_j}) + s(S\!_{n_j}) + s(AOP_{n_j})}
        + s(B_l) + 2n
   \ca{\aw}{\overset{\labelcref{step-2-prerequisits-red}}{\leq}}
              \sum_{j=0}^{l-1} (3.5 n_j + 2.5 n_j + 7.34 n_j - 2)
      + 6.2 l \log_2 l + 2n\\
   \ca{\aw}{=} 15.34 n + (6.2 \log_2 l - 2) l \\
   \ca{\aw}{\overset{\shortcref{step2-bound-rest-size-a}}{\leq}} 15.34 n + 6.2 n \\
   \ca{\aw}{<} 21.6 n\,.
   \end{align*}
   }

For deducing the running time, it suffices to consider the case
$n \geq 1024$.
The adder circuit $B_l$
on $l = \lceil \frac{n}{k} \rceil$ inputs is constructed
using \cref{theorem::step_1_red_version} in time
$\mathcal O(l \log_2^2 l) = \mathcal O(n \log_2 n)$.
For the construction of the $l$ \aop{}s $AOP_k$,
we apply the algorithm from \cref{depth-opt-size} with a running time of
$\mathcal O(k \log_2 k)$ for each \aop{},
which yields a
running time of $\mathcal O(l k \log_2 k) = \mathcal O(n \log_2 \log_2 n)$
for constructing all these \aop{}s.
The remaining part of the algorithm (including the computation of the
\andprefixckt{} family $S$ (see \cref{LF-runtime})
and adder family $A$ (see \cref{halved-adder}) can be done in linear time.
Thus, we get an overall running time of $\mathcal O(n \log_2 n)$.
 \end{proof}

These circuits are the fastest linear-size adder circuits known so far.
With a depth increase by a small additive constant of $1$,
we can reduce the size significantly as follows.

We again need some technical lemmas.

\begin{lemma} \label{step-3-num-b}
For $4 \leq n \leq 2048$, we have
$
2 \lceil \log_2 n \rceil
             \leq \log_2 n + \log_2 \log_2 n + \log_2 \log_2 \log_2 n + 7
$.
\begin{proof}
See \cref{app-proofs}, \cpageref{proof-step-3-num-b}.
\end{proof}
\end{lemma}

\begin{lemma} \label{step2-main-depth-bound-b}
Given $k, l, n \in \N_{\geq 1}$ with $n \geq 2048$, $k = \BktU{\log_2^2 n}$ and $l = \BktU{\frac{n}{k}}$,
we have
\begin{equation*}
\log_2 l + \log_2 \log_2 l + \log_2 k + \log_2 \log_2 (2k) + 6.3
\leq \log_2 n + \log_2 \log_2 n + \log_2 \log_2 \log_2 n + 7.6\,.
\end{equation*}
\begin{proof}
See \cref{app-proofs}, \cpageref{proof-step2-main-depth-bound-b}.
\end{proof}

\end{lemma}
\begin{lemma} \label{step2-bound-rest-size-b}
Given $k, l, n \in \N_{\geq 1}$ with $n \geq 2048$, $k = \BktU{\log_2^2 n}$ and $l = \BktU{\frac{n}{k}}$,
we have
\[
   (6.2 \log_2 l - 2) l \leq 0.57 n\,.
\]
\begin{proof}
See \cref{app-proofs}, \cpageref{proof-step2-bound-rest-size-b}.
\end{proof}
\end{lemma}

\linebox{
\begin{theorem} \label{better-size-n log log logn red}
Let $n \in \N$ with $n \geq 4$ be given.
 In running time $\mathcal O(n \log_2 \log_2 n)$,
 we can construct
 an adder circuit $A^3_n$ on $n$ input pairs with
 \begin{align*}
  d\left(A^3_n\right) \leq \log_2 n + \log_2 \log_2 n + \log_2 \log_2 \log_2 n + 7.6
  \qquad \text{and} \qquad
 s\left(A^3_n\right) \leq 16.7 n \,.
 \end{align*}
\end{theorem}
}
 \begin{proof}
  Consider again the proof of \cref{n log log logn red}.
  We proceed similarly, but choose $k$
  and the adders $A_k$ differently.

  If $4 \leq n \leq 2048$, we again let $A^3_n$ be the adder circuit $L^f_n$
  (with $f = 0$) by \citet{LF80}
  with size at most $12 n$ and depth at most $2 \lceil \log_2 n \rceil$, see \cref{Ladner Fischer},
  which fulfills the stated bounds by \cref{step-3-num-b}.

  Thus, assume that $n \geq 2048$.
  We again apply the linearization from \cref{general linearization}
  with $B_l$ and $AOP_k$ chosen as in \cref{n log log logn red},
  but with $S\!_k$ and $A_k$ both computed by the Ladner-Fischer parallel
  prefix circuit $L^f_k$
  from \citep{LF80} with $f = 3$, see also \cref{Ladner Fischer}.
  This way, we obtain
  \begin{equation}
   d(A_k) \leq 2 \Bkt{\bktU{\log_2 k} + 3}\,, \qquad
   d(S\!_k) \leq \bktU{\log_2 k} + 3\,, \qquad
   s(A_k) + s(S\!_k) \leq 6.75 k\,, \label{better-size-pre}
  \end{equation}
  and the bounds regarding $B_l$ and $AOP_k$ from \cref{step-2-prerequisits-red} still hold.
  This time, we choose $k = \BktU{\log_2^2 n}$ and $l = \BktU{\frac{n}{k}}$.
  The choice of $f$ is valid as
  $\log_2 k \geq \log_2 (\log_2^2 n) = 2 \log_2 \log_2 n \overset{n \geq 2048}{>} 6.9 > 3$.
  Denote the arising adder circuit by $A_n^3$.

  Since $n \geq 2048$ and thus $k \geq 121$, we have
  \begin{equation} \label{size-bound-S2}
  \log_2 (2k) + \log_2 \log_2 (2k) + 0.65
   \overset{k \geq 121}{\geq} \log_2 k + 2.98 + 1.65
   \geq \lceil \log_2 k \rceil + 3
   \geq d(S\!_k)\,.
  \end{equation}
  Note that we have $l = \BktU{\frac{n}{k}} \geq \frac{2048}{12^2} \geq 14$
  and thus
  \begin{equation} \label{eq::bound_l}
      \log_2 \log_2 (l) \geq \log_2 \log_2 (14) \geq 1.92\,.
  \end{equation}
  Moreover, we have $k \geq 121$, so
  \begin{equation} \label{eq::bound_k}
      \log_2 \log_2 (2k) \geq  \log_2 \log_2 (242) \geq 2.98\,.
  \end{equation}
  For bounding the depth of $A_k$, we calculate
  \begin{equation} \label{better-size-bound-lf-depth}
    2 \BktU{\log_2 \Bkt{1 + \log_2^2 n}}
    \overset{n \geq 2048}{\leq} 2 \BktU{\log_2 \Bkt{1.0083 \log_2^2 n}}
    \overset{}{<} 2 \BktU{0.012 + 2 \log_2 \log_2 n}
    \overset{}{<} 4 \log_2 \log_2 n + 2.024
   \end{equation}
  and thus, using in the third inequality that $x \mapsto \log_2 x$
  grows faster than $x \mapsto 4 \log_2 \log_2 x$,
  \begin{equation}
    \label{better-size-bound-A}
    d(A_k)
    \overset{\labelcref{better-size-pre}, k = \BktU{\log_2^2 n}}{\leq}
       2 \BktU{\log_2 \BktU{\log_2^2 n}} + 6
    \quad{\overset{\labelcref{better-size-bound-lf-depth}}{<}}\quad 4 \log_2 \log_2 n + 8.024 
    \quad{\overset{n \geq 2048}{\leq}}\quad \log_2 n + 10.9\\
   \end{equation}

Thus we have 
  \begin{equation}
    \label{better-size-bound-B}
   d(A_k) +1 \quad{\overset{\labelcref{better-size-bound-A}}{\leq}}\quad \log_2 n + 11.9 
    \quad{\overset{n \geq 2048}{\leq}}\quad \log_2 n + \log_2 \log_2 n + \log_2 \log_2 \log_2 n + 7.6
  \end{equation}

Moreover, we have
  {
  \newcommand{\aw}{7}
  \begin{align} \label{better-size-bound-C}
   \begin{split}
     d(B_l)
               + \max \big\{ d(AOP_{k}), d(S\!_{k}) \big\}  + 2
    \ca{\aw}{\overset{\labelcref{step-2-prerequisits-red}, \labelcref{size-bound-S2}}{\leq}}
         \log_2 l + \log_2 \log_2 l + 2.65 + \log_2 (2k) + \log_2 \log_2 (2k) + 0.65+ 2\\
     \ca{\aw}{=} \log_2 l + \log_2 \log_2 l + \log_2 k + \log_2 \log_2 (2k) + 6.3 \\
     \ca{\aw}{\overset{\shortcref{step2-main-depth-bound-b}}{\leq}}
       \log_2 n + \log_2 \log_2 n + \log_2 \log_2 \log_2 n + 7.6\,.
   \end{split}
  \end{align}
  }

  Hence, the depth of $A^3_n$ can be bounded by
  {
   \newcommand{\aw}{7}
   \begin{align}
   \begin{split}
    d(A^3_n) \ca{\aw}{\overset{\shortcref{general linearization}}{\leq}}
\max \BktC{ d(A_{k}) + 1, d(B_l)
               + \max \big\{ d(AOP_{k}), d(S\!_{k}) \big\}  + 2}\\
\ca{\aw}{\overset{\labelcref{better-size-bound-B},\labelcref{better-size-bound-C}}{\leq}} 
   \log_2 n + \log_2 \log_2 n + \log_2 \log_2 \log_2 n + 7.6\,.
   \end{split}
  \end{align}
  }

 For the size of $A^3_n$, we obtain
 { 
  \newcommand{\aw}{6}
  \begin{align*}
   s(A^3_n) \overset{\text{Thm. } \labelcref{general linearization}}{<}
      \sum_{j=0}^{l-1} \Bkt{s(A_{n_j}) + s(S\!_{n_j})
      + s(AOP_{n_j})} + s(B_l) + 2n
   \ca{\aw}{\overset{\substack{\labelcref{better-size-pre}, \\ \labelcref{step-2-prerequisits-red}}}{<}}
              \sum_{j=0}^{l-1} (6.75 n_j + 7.34 n_j - 2)
      + 6.2 l \log_2 l + 2n\\
   \ca{\aw}{=} 16.09 n + (6.2 \log_2 l - 2) l \\
   \ca{\aw}{\overset{\shortcref{step2-bound-rest-size-b}}{\leq}} 16.09 n + 0.57 n \\
   \ca{\aw}{<} 16.7 n\,.
   \end{align*}
   }
For proving the running time, it again suffices to consider the case
$n \geq 2048$.
For $k = \BktU{\log_2^2 n}$, constructing the adder circuits $B_l$
using \cref{theorem::step_1_red_version}
takes time
$\mathcal O(l \log_2^2 l) = \mathcal O(n)$.
For the construction of the $l$ \aop{}s $AOP_k$ using \cref{depth-opt-size},
we need a running time of $\mathcal O(l k \log_2 k) = \mathcal O(n \log_2 \log_2 n)$.
Again, the remaining part of the algorithm (including the computation of $A_k$ and $S\!_k$
(see \cref{LF-runtime}) can be done in linear time.
Thus, we get an overall running time of $\mathcal O(n \log_2 \log_2 n)$.
  \end{proof}

\section*{Conclusions}
We presented the fastest circuits for binary addition of $n$-bit numbers
among all circuits of sub-quadratic size. 
With a size of $O(n \log_2 n)$,
we reach a depth of $\log_2 n + \log_2 \log_2 n + \const$ which differs from
the lower bound only by an additive constant. 
Moreover, we presented circuits with linear size and a depth of
$\log_2 n + \log_2 \log_2 n + \log_2 \log_2 \log_2 n + \const$, which is still
significantly better than the best previously known depths of circuits of sub-quadratic size. 
In addition, our approach allows, up to a certain degree, to trade off depth for size
(compare \cref{n log log logn red} and  \cref{better-size-n log log logn red} 
for two possible specifications).
The construction is based on new \aop{} circuits, which are the fastest
 linear-size circuits for this problem.

It is an interesting open question if there are adder circuits with linear size
reaching the lower bounds on depth of $\log_2 n + \log_2 \log_2 n + \const$
at least up to a constant additive term. 
Another topic for future research would be the question if our conjecture that
\cref{grinchuk-alg} always finds a depth-optimum solution is true.

\FloatBarrier 

\begin{appendix}

\section{Missing Proofs} \label{app-proofs}

In this \lcnamecref{app-proofs},
we collect all proofs that were skipped in
earlier sections,
sometimes with additional helper lemmas.

\phantomsection
\begin{proof}[Proof of \cref{small-values-of-d}] \label{proof-small-values-of-d}
Given $n, m \in \N$ with $m \geq 1$
and $d = \dmin(n, m)$, we have $m \leq \mu(d, n)$
by \cref{def-dmin}.
Hence,  we have $m \leq \mu(d, n) \leq \xi \frac{2^d - n - 2}{d} + 2$
and thus $n \leq 2^d - \frac{d}{\xi}(m-2) - 2$, which leads to the following results:
\begin{itemize}
   \item If $d = 1$, we have $n \leq 2 - \frac{1}{\xi}(m-2) - 2 = - \frac{1}{\xi}(m-2)$ which is true if and
     only if $m \in \{1,2\}$ and $n=0$.
   \item If $d = 2$, we have  $n \leq 2^2 - \frac{2}{\xi}(m-2) - 2 = 2 - \frac{2}{\xi}(m-2)$. For $m=1$,
     this is true if and only if $n \leq \left \lfloor 2 + \frac{2}{\xi} \right\rfloor = 3$. For $m=2$,
     the condition true if and only if $n \leq 2$. For $m=3$, the condition is true if and only if 
     $n \leq \left \lfloor 2 - \frac{2}{\xi} \right\rfloor = 0$. For $m \geq 4$, the condition cannot be true.
   \item If $d = 3$, we have  $n \leq 2^3 - \frac{3}{\xi}(m-2) - 2 = 6 - \frac{3}{\xi}(m-2)$. 
     For $m=1$, this is true if and only if $n \leq \left \lfloor 6 + \frac{3}{\xi} \right\rfloor = 7$.
     For $m=2$, the condition is true if and only if $n \leq 6$.
     For $m=3$, the condition is true if and only if $n \leq \left \lfloor 6 - \frac{3}{\xi} \right\rfloor = 4$.
     For $m=4$, the condition is true if and only if $n \leq \left \lfloor 6 - 2 \cdot \frac{3}{\xi} \right\rfloor = 2$.
     For $m=5$, the condition is true if and only if $n \leq \left \lfloor 6 - 3 \cdot \frac{3}{\xi} \right\rfloor = 1$.
     For $m \geq 6$, the condition cannot be true.
   \item If $d = 4$, we have  $n \leq 2^4 - \frac{4}{\xi}(m-2) - 2 = 14 - \frac{4}{\xi}(m-2)$.
     For $m=1$, this is true if and only if $n \leq \left \lfloor 14 + \frac{4}{\xi} \right\rfloor = 16$.
     For $m=2$, the condition is true if and only if  $n \leq 14$.
     For $m=3$, the condition is true if and only if $n \leq \left \lfloor 14 - \frac{4}{\xi} \right\rfloor = 11$.
     For $m=4$, the condition is true if and only if $n \leq \left \lfloor 14 - 2\cdot\frac{4}{\xi} \right\rfloor = 9$.
     For $m=5$, the condition is true if and only if $n \leq \left \lfloor 14 - 3\cdot\frac{4}{\xi} \right\rfloor = 7$.
     For $m=6$, the condition is true if and only if $n \leq \left \lfloor 14 - 4\cdot\frac{4}{\xi} \right\rfloor = 5$.
     For $m=7$, the condition is true if and only if $n \leq \left \lfloor 14 - 5\cdot\frac{4}{\xi} \right\rfloor = 3$.
     For $m=8$, the condition is true if and only if $n \leq \left \lfloor 14 - 6\cdot\frac{4}{\xi} \right\rfloor = 1$.
     For $m \geq 9$, the condition cannot be true.
   \item $d = 5$, we have  $n \leq 2^5 - \frac{5}{\xi}(m-2) - 2 = 30 - \frac{5}{\xi}(m-2)$ which is true for
     $m = 9$ and $n = 12$ and thus for all cases considered in the table.
\end{itemize}
From these statements, together with the minimum choice of $d$, the table follows.
\end{proof}

\begin{lemma}\label{upper bound mu}
Given $d, n \in \N$ with $d \geq 2$, we have
$
\mu(d, n) < 2^d
$.
\begin{proof}
We have 
\[\mu(d, n) = \xi \frac{2^d - n - 2}{d} + 2 \leq \xi \frac{2^d - 2}{d} + 2
            < 2^d\,,\]
where the last inequality follows from 
$2\left(1 - \frac{\xi}{d} \right) < 2^d\left(1 - \frac{\xi}{d} \right)$
(for $\xi < 2 \leq d$).
\end{proof}
\end{lemma}

\spacingbetweenappproofs

\phantomsection
\begin{proof}[Proof of \cref{m-k-small-enough}] \label{proof-m-k-small-enough}
  Note that $k$ is well-defined by condition~\labelcref{cond-mu-large}.
  Since $k$ is odd by condition~\labelcref{cond-k-odd},
  we have $\frac{k-1}{2} \in \N$.
  Moreover, since $d \geq 2$ by condition~\labelcref{cond-d-large},
  we have
  \[\frac{k-1}{2} \overset{\labelcref{cond-k-odd}}{\leq} \frac{\mu(d, n)-1}{2}
   \overset{\shortcref{upper bound mu}}{<} \frac{2^d-1}{2} < 2^d\,.\]
  It remains to show \labelcref{to-show},
  or, equivalently,
  $\mu \Bkt{d, \frac{k-1}{2}} + k \geq m$. Thus, since 
  $m \leq \mu(d+1,n)$, it is sufficient to show that
  \begin{equation}\label{eq::goal_lemma_3.8}
    \mu \Bkt{d, \frac{k-1}{2}} + k - \mu(d+1,n) \geq 0
  \end{equation}
  We have
  {
  \newcommand{\aw}{6}
  {\allowdisplaybreaks
  \begin{align*}
   \mu \Bkt{d, \frac{k-1}{2}} + k - \mu(d+1,n) 
   \ca{\aw}{\overset{\shortcref{def-mu}}{=}} 
        \xi \frac{2^d - \frac{k-1}{2} - 2}{d} + 2 + k - \mu(d+1,n)\\
   \ca{\aw}{=} 
        k \Bkt{1 - \frac{\xi}{2d}}  + \xi\frac{2^d-2}{d} + 2 + \frac{\xi}{2d} - \mu(d+1,n)\\
   \ca{\aw}{\overset{\substack{\xi<2,\\ k \geq 1}}{>}}
        k \Bkt{1 - \frac{1}{d}}  + \xi\frac{2^d-2}{d} + 2 + \frac{\xi}{2d} - \mu(d+1,n)\\
   \ca{\aw}{\overset{\substack{d \geq 1, \\ k + 2 > \mu(d, n)}}{>}} 
       (\mu(d, n) - 2) \Bkt{1 - \frac{1}{d}}  + \xi\frac{2^d-2}{d} + 2 + \frac{\xi}{2d} - \mu(d+1,n)\\
   \ca{\aw}{\overset{\shortcref{def-mu}}{=}} 
        \xi\frac{2^{d} - n -2}{d} \Bkt{1 - \frac{1}{d}} + \xi\frac{2^d-2}{d} + 2 + \frac{\xi}{2d} - \xi\frac{2^{d+1} - n -2}{d+1} - 2\\
   \ca{\aw}{=} 
        \xi\frac{2^{d} -2}{d} \Bkt{1 - \frac{1}{d}} + \xi\frac{2^d-2}{d} + \frac{\xi}{2d} - \xi\frac{2^{d+1} -2}{d+1} 
    + n \xi \Bkt{\frac{1}{d+1} - \frac{1}{d} + \frac{1}{d^2}}\\
   \ca{\aw}{=} 
        \xi\frac{2^{d} -2}{d} \Bkt{1 - \frac{1}{d}} + \xi\frac{2^d-2}{d} + \frac{\xi}{2d} - \xi\frac{2^{d+1} -2}{d+1} 
       + n \xi \frac{d^2 - d(d+1) + d+1}{d^2(d+1)}\\
   \ca{\aw}{\overset{n\geq 0}{\geq}} 
      \xi\frac{2^{d} -2}{d} \Bkt{1 - \frac{1}{d}} + \xi\frac{2^d-2}{d} + \frac{\xi}{2d} - \xi\frac{2^{d+1} -2}{d+1}\\
    \ca{\aw}{=} 
      \xi \Bkt{\frac{2^{d} -2}{d} \frac{d - 1}{d} + \frac{2^d-2}{d} + \frac{1}{2d} - \frac{2^{d+1} -2}{d+1}}\\
    \ca{\aw}{=} 
      \frac{\xi}{d^2(d+1)} \Bkt{(2^{d} -2)(d - 1)(d+1)+ (2^d-2)d(d+1) + \frac{1}{2}d(d+1) - (2^{d+1} -2)d^2}\\
    \ca{\aw}{=} 
      \frac{\xi}{d^2(d+1)} \Bkt{(2^{d} -2)(d^2 - 1)+ (2^d-2)(d^2+d) + \frac{1}{2}(d^2+d) - (2^{d+1} -2)d^2}\\
    \ca{\aw}{=} 
      \frac{\xi}{d^2(d+1)} \Bkt{2^{d}d^2 - 2^{d} -2d^2 + 2 + 2^dd^2 + 2^dd -2d^2 -2d + \frac{1}{2}(d^2+d) - 2^{d+1}d^2 + 2 d^2}\\
    \ca{\aw}{=} 
      \frac{\xi}{d^2(d+1)} \Bkt{- 2^{d} -2d^2 + 2 + 2^dd -2d + \frac{1}{2}(d^2+d)}\\
    \ca{\aw}{=} 
      \frac{\xi}{d^2(d+1)} \Bkt{- 2^{d} -\frac{3d^2}{2} + 2 + 2^dd - \frac{3d}{2}}\\
    \ca{\aw} {\overset{d\geq 3}{\geq}} 
      \frac{\xi}{d^2(d+1)} \Bkt{2^{d+1} -\frac{3d^2}{2} + 2 - \frac{3d}{2}}
  \end{align*}
  } 
  }
  However, for the function
  $x \mapsto - 2^{x+1} -\frac{3x^2}{2} + 2 - \frac{3x}{2}$ its first derivative
  is $\ln(2) 2^{x+1} - 3x - \frac{3}{2}$ and its second derivative is
  $(\ln(2))^2 2^{x+1} -3$. Thus, the second derivative is obviously
  positive for $x \geq 3$. Since the value of the first derivative is also positive
  for $x = 3$, this means that the first derivative if positive for $x \geq 3$.
  This implies for $d \geq 3$
  \begin{equation*}
     2^{d+1} -\frac{3d^2}{2} + 2 - \frac{3d}{2} 
         \geq 2^{4} -\frac{3\cdot3^2}{2} + 2 - \frac{3\cdot3}{2} = 0
  \end{equation*}
  which proves \cref{eq::goal_lemma_3.8} and thus the \lcnamecref{m-k-small-enough}.
\end{proof}

\spacingbetweenappproofs

\phantomsection
\begin{proof}[Proof of \cref{lem-theta-0-m}] \label{proof-lem-theta-0-m}
 We first prove the statement for $n = 0$ and then for $n > 0$.

   \textbf{Case 1:} Assume that $n = 0$.

   Note that $\vartheta(0, m) = 2^{c} \xi m \log_2 m - (m - 2) \Bkt{\log_2 m + \log_2 \log_2 m + c} - 2 \xi $.

   Writing
   $\lambda(m) := \Bkt{2^c\xi - 1} \log_2 m - \log_2 \log_2 m - c$
   and
   $\kappa(m) :=2\Bkt{\log_2 m + \log_2 \log_2 m + c} -2\xi$,
   we have
   \begin{equation}
    \vartheta(0, m) = m \lambda(m) + \kappa(m)\,. \label{rewrite-theta}
   \end{equation}
   We will examine these functions in a series of claims.


   \begin{claim} \label{claim-lambda}
    We have $\lambda(m) > 0 $ for $m \in \N_{\geq 3}$.
    \begin{proof_of_claim}
   The derivative of $\lambda(m)$ is
   \begin{equation}
    \frac{d}{dm} \lambda(m) =\frac{2^c\xi - 1}{\ln (2) m} - \frac{1}{\ln(2) m \ln m}
                            = \frac{1}{\ln(2) m} \Bkt{2^c\xi - 1 - \frac{1}{\ln m }}\,, \label{lambda-derived}
   \end{equation}
   so (because $\ln (2) m > 0$)
   its only minimum is at $m=e^{\frac{1}{2^c\xi-1}} \approx 5.809$.
   Since $\lambda \Bkt{e^{\frac{1}{2^c\xi-1}}} > 0.44 > 0$, this proves the claim.
   \end{proof_of_claim}
   \end{claim}

   Based on the computations in the proof of \cref{claim-lambda}, we can compute the derivative 
   of the function $m \lambda (m)$:
   \begin{equation}
    \frac{d}{dm} \Bkt{m \lambda(m)} =\lambda(m) + m \frac{d}{dm} \lambda(m)
                                    \overset{\labelcref{lambda-derived}}{=} \lambda(m) + \frac{1}{\ln(2)} \Bkt{2^c\xi - 1 - \frac{1}{\ln m }} \label{derivative-m-lambda-m}
   \end{equation}
   Now, we can calculate the derivatives of $\kappa(m)$ and $\vartheta(0, m)$:
   {
   \newcommand{\aw}{6}
   \begin{align}
    \frac{d}{dm} \kappa(m) \ca{\aw}{=} 2\Bkt{\frac{1}{\ln(2) m} + \frac{1}{\ln(2) m \ln m}} \label{derivative-kappa} \\
    \frac{d}{dm} \vartheta(0, m) \ca{\aw}{\overset{\labelcref{rewrite-theta}}{=}} \frac{d}{dm}(m \lambda (m)) + \frac{d}{dm} \kappa(m)
                                 \overset{\substack{\labelcref{derivative-m-lambda-m}, \\ \labelcref{derivative-kappa}}}{=}
                                    \lambda(m) + \frac{2^c\xi - 1}{\ln(2)}
                                    + \frac{- m + 2 \ln m + 2}{\ln(2) m \ln m} \nonumber
   \end{align}
   }

   Writing
   \[\nu(m) :=  \frac{2^c\xi - 1}{\ln(2)} + \frac{- m + 2 \ln m + 2}{\ln(2) m \ln m}\,,\]
   we thus have
   \[\frac{d}{dm} \vartheta(0, m) = \lambda(m) + \nu(m)\,.\]

   \begin{claim} \label{claim-nu}
    We have $\nu(m) > 0$  for $m \in \N_{\geq 3}$.
    \begin{proof_of_claim}
   We compute the derivative of $\nu(m)$:
   {
   \newcommand{\aw}{2}
   \begin{align*}
    \frac{d}{dm} \nu(m)
    \ca{\aw}{=} \frac{  \Bkt{-1 + \frac{2}{ m}} \ln(2) m \ln m
                      - \Bkt{-m + 2 \ln m + 2} \ln(2) \Bkt{\ln m + 1}}{\ln^2(2) m^2 \ln^2 m} \\
    \ca{\aw}{=} \frac{  -m \ln m + 2 \ln m + m \ln m - 2 \ln^2 m - 2 \ln m + m - 2 \ln m - 2}{\ln(2) m^2 \ln^2 m} \\
    \ca{\aw}{=} \frac{  m - 2 \ln ^2 m - 2 \ln m - 2}{\ln(2) m^2 \ln^2 m}
   \end{align*}
   }
   Note that $\nu(m)$ is monotonely increasing if and only if
   its numerator $\widetilde{\nu}(m) :=  m - 2 \ln ^2 m - 2 \ln m - 2$
   fulfills $\widetilde{\nu}(m) \geq 0$.
   In order to see when this is the case, we compute the derivative of $\widetilde{\nu}(m)$:
   \begin{align*}
    \frac{d}{dm} \widetilde{\nu}(m) &= 1 - \frac{4\ln m}{m} - \frac{2}{m}
   \end{align*}
   Note that each summand of $\frac{d}{dm} \widetilde{\nu}(m)$ is monotonely increasing
   (for $-\frac{4 \ln m}{m}$, this follows from
   $m \geq 3 > e$ and \cref{d / ln d incr}),
   hence so is $\frac{d}{dm} \widetilde{\nu}(m)$.
   From this, by plugging in $m = 11, 12$ in $\frac{d}{dm} \widetilde{\nu}(m)$, we deduce
   $\frac{d}{dm} \widetilde{\nu}(m) \begin{cases}
                                       < 0 & \text{ for } 3 \leq m \leq 11\\
                                       > 0 & \text{ for } m \geq 12
                                      \end{cases}
   $.
   In other words, the function $\widetilde \nu (m)$ is decreasing for $3 \leq m \leq 11$
   and increasing for $m \geq 12$.
   Evaluating $\widetilde \nu(m)$ for $m \in \{3, 33, 34\}$, we see that for $m \in \N_{\geq 3}$,
   we have $\widetilde \nu(m) \geq 0$ if and only if $m \geq 34$.
   Hence, for $m \geq 3$, the function $\nu(m)$ is monotonely decreasing
   in the range $3 \leq m \leq 33$
   and monotonely increasing for $m \geq 34$.
   Since $\nu(33) > 0.5 > 0$ and $\nu(34) > 0.5 > 0$, this proves the claim.
   \end{proof_of_claim}
   \end{claim}

   We have  
   \[
      \frac{d}{dm} \vartheta(0, m) = \lambda(m) + \nu(m)\,.
   \]
   so by \cref{claim-lambda} and \cref{claim-nu}, 
   the function $\vartheta(0, m)$ is monotonely increasing for $m \geq 3$.
   Therefore, for $m \geq 3$, we have 
   \[
      \vartheta(0, m) \geq  \vartheta(0, 3) > 1.5 > 0\,.
   \]
   This proves this \lcnamecref{lem-theta-0-m} for $n = 0$.

  \textbf{Case 2:} Assume that $n \geq 1$.

  Here, we show that for all $n \geq 0$ and $m \geq 3$, we have
   \begin{toshowequation} \label{derivative >= 0}
    \frac{d}{dn} \vartheta(n, m) \geq 0\,,
   \end{toshowequation}
   implying that for $n \geq 1$, we have $\vartheta(n, m) \geq \vartheta(0, m) \geq 0$.
   Writing $\kappa(n, m) := \frac{d}{dn} \vartheta(n, m)$, we compute
   \[
    \kappa(n, m) = 2^c \xi \log_2 m - \frac{m-2}{\ln (2) (m + n)} - \xi\,.
   \]
   Since $\kappa(n, m) \geq \kappa(0, m)$ for all $n, m \in \N$,
   it suffices to show $\kappa(0, m) \geq 0$ for all $m \geq 3$ in order to prove \cref{derivative >= 0}.
   For small values of $m$, we have $\kappa(0, 3) = 2^c \xi \log_2 3 - \frac{1}{3 \ln (2)} - \xi > 0.005 > 0$,
   and $\kappa(0, 4) = 2^c \xi \log_2 4 - \frac{1}{2 \ln (2)} - \xi > 0.4 > 0$.
   And for $m \geq 5$, we have
   \begin{equation*}
    \kappa(0, m) = 2^c \xi \log_2 m + \frac{1}{\ln (2)} \Bkt{\frac{2}{m} - 1} - \xi
                 \overset{m > 0}{\geq} 2^c\xi \log_2 m - \frac{1}{\ln (2)} - \xi
                 \overset{m \geq 5}{>} 0\,.
   \end{equation*}
   This proves \cref{derivative >= 0} and thus the \lcnamecref{lem-theta-0-m} in the final case that $n \geq 1$.
\end{proof}

\spacingbetweenappproofs

\phantomsection
 \begin{proof}[Proof of \cref{lem-tri-easy}]\label{lem-tri-easy-proof}
\begin{enumerate}
 \item For the empty set, the boundary tree sequence is empty, hence the first statement.

 \item If $K = \cktins(S)$, then the boundary tree sequence
  consists of exactly the trees of the leftist circuit~$S$.
  As their depths are strictly decreasing by definition of a leftist circuit,
  $K = \set{x_0, \dotsc, x_{n-1}}$ is triangular.

  \item Now assume that $K \subseteq \cktins(S)$ is triangular with $x_i$ being its right-most vertex.
  Let $B := B(K, S)$ and
  let $T_0, \dotsc, T_{|B| - 1}$ be the boundary tree sequence of $K$ with respect to $S$.
  Then, $T_{|B|-1}$ is the unique tree in the sequence which contains~$x_i$.
  Let $d$ denote the depth of $T_{|B|-1}$.

  If $d = 0$, then $K \backslash \set{x_i}$ is certainly triangular.

  Thus, assume that $d \geq 1$.
  As $x_i$ is the right-most vertex of $T_{|B|-1}$ and $T_{|B|-1}$ is a full binary tree,
  deleting $x_i$ and all its successors from $T_{|B|-1}$
  results in a sequence $T'_0, \dotsc, T'_{d-1}$ of full binary trees,
  ordered from left to right.
  The boundary tree sequence of $K \backslash \set{x_i}$ is given by
  $T_0, \dotsc, T_{|B|-2}, T'_0, \dotsc, T'_{d-1}$.
  As $x_i$ is the right-most vertex of $T_{|B|-1}$, the inputs of these trees
  are consecutive, and if the inputs of $T_{|B|-2}$ and $T_{|B|-1}$
  are consecutive, the inputs of $T_{|B|-2}$ and $T'_0$ are also consecutive.
  As $K$ is triangular and $\depth(T'_j) = d - j - 1$ for all $j \in \set{0, \dotsc, d-1}$,
  \cref{triangular-card} is fulfilled for this sequence.
  Hence, $K \backslash \set{x_i}$ is triangular.

  The proof works analogously if $x_i$ is the left-most vertex of $K$.
  This proves the third statement.

  \item The fourth statement follows from the second and the third statement
  as we can obtain any consecutive set $K$ from $\set{x_0, \dotsc, x_{n-1}}$
  by successively deleting the right-most or left-most vertex. \qedhere
  \end{enumerate}
\end{proof}

\spacingbetweenappproofs

\phantomsection
\begin{proof}[Proof of \cref{lem-triangular}]\label{proof-lem-triangular}
     We use the notation from \cref{det-triangular-subset}.
   We start the proof with showing $2$ claims.

  \begin{claim} \label{claim-tri-power}
   We have $n - k \leq 2^{d - |B| + 1} -2$.
   \begin{proof_of_claim_no_qed}
   If $K$ is built in \cref{tri-ret-K-d-1},
   we have $|B| = 1$.
   This implies
   $ n - k \overset{n < 2^d,\, k \geq 1}{\leq} 2^d -2 \overset{|B| = 1}{=} 2^{d-|B| + 1} - 2$.

   Otherwise, $K$ is constructed in \cref{tri-ret-K-d-2}.
   By the choice of $D$ in \cref{tri-choose-D} and the proof of \cref{lem-triangular-remains},
   in this case, $K$ contains exactly one tree for each depth $i \in \set{D+1, \dotsc, d-2}$
   and exactly $2$ trees of depth $D$.
   This implies $|B| = (d - 2) - D + 1 + 1 = d - D$.
   Hence, we have
   \setlength{\belowdisplayskip}{0pt}
   \setlength{\belowdisplayshortskip}{0pt}
   \[  n - k \leq 2 \sum_{i = 0}^{D-1} 2^{i} = 2 \Bkt{2^{D} - 1}
     = 2 \Bkt{2^{d - |B|} - 1}
     = 2^{d - |B| + 1} - 2\,. \tag*{\claimqed}
   \]
   \end{proof_of_claim_no_qed}
   \end{claim}
   \begin{claim} \label{tri-log-bound-first-step}
    If $n - k \geq 1$, we have $|B| + 2 \log_2 (n-k) \leq 2 \log_2 n$.
   \begin{proof_of_claim}
   Choose $\lambda \in (1, 2)$ with $n = \lambda 2^{d-1}$.
   From \cref{claim-tri-power} and $k = 2^{d-1}$, we obtain
   $ n = k + n - k \leq 2^{d-1} + 2^{d - |B| + 1} - 2$.
   This implies
   \begin{equation} \label{tri-bound-lambda}
    \lambda
     = \frac{n}{2^{d-1}} \leq \frac{2^{d-1} + 2^{d - |B| + 1} - 2}{2^{d-1}} \leq 1 + 2^{2 - |B|}\,.
   \end{equation}
   In order to prove this \lcnamecref{tri-log-bound-first-step},
   we will show that
   \begin{toshowequation} \label{tri-to-show}
     2 \log_2 n - 2 \log_2(n-k) - |B| \geq 0\,.
   \end{toshowequation}
   We have
   {
   \newcommand{\aw}{5}
   \begin{align}
    \begin{split}
    2 \log_2 n - 2 \log_2(n-k) - |B|
    \ca{\aw}{\overset{\text{Def. } \lambda, k}{=}} 2 \log_2 (\lambda 2^{d-1}) - 2 \log_2((\lambda-1) 2^{d-1}) - |B|  \\
    \ca{\aw}{=} 2 \log_2 \lambda + 2 (d-1) - 2 \log_2(\lambda - 1) - 2 (d-1) - |B| \\
    \ca{\aw}{=} 2 \Bkt{\log_2 \lambda - \log_2(\lambda - 1)} - |B|\,. \label{tri-bound-greater-0}
    \end{split}
   \end{align}
   }
   The function $x \mapsto \log_2 x - \log_2 (x-1)$ is strictly monotonely decreasing for $x > 1$
   as its derivative $\frac{1}{\ln(2)x} - \frac{1}{\ln(2) (x-1)} = -\frac{1}{\ln(2) x (x-1)}$
   is negative for all $x > 1$.
   Hence, we obtain
   {
   \newcommand{\aw}{5}
   \begin{align*}
     2 \log_2 n - 2 \log_2(n-k) - |B|
     \ca{\aw}{\overset{\labelcref{tri-bound-greater-0}}{=}}
       2 \Bkt{\log_2 \lambda - \log_2(\lambda - 1)} - |B| \\
     \ca{\aw}{\overset{\labelcref{tri-bound-lambda}}{\geq}}
      2 \Bkt{\log_2 \Bkt{1 + 2^{2 - |B|}} - \log_2\Bkt{1 + 2^{2 - |B|} - 1}} - |B| \\
     \ca{\aw}{=} 2 \log_2 \Bkt{1 + 2^{2 - |B|}} + |B| - 4\,.
   \end{align*}
   }
   The first summand is always positive.
   Thus, the entire term is apparently non-negative for all $|B| \geq 4$,
   but also for $|B| \in \set{1, 2, 3}$, as can be seen by explicit calculation.
   In any case, \cref{tri-to-show}
   and the \lcnamecref{tri-log-bound-first-step} follow.
   \end{proof_of_claim}
   \end{claim}

   For proving the first statement of this \lcnamecref{lem-triangular},
   assume $n - k \geq 2$.
   Recall the above remark that the function
   $x \mapsto \log_2 x - \log_2 (x-1)$ is strictly monotonely decreasing for $x > 1$.
   Together with \cref{tri-log-bound-first-step},
   this implies
   \[
   2 \log_2 (n-1) - 2 \log_2(n-k-1) - |B|
   \overset{\cref{tri-log-bound-first-step}}{\geq} 2 \log_2 (n-1) - 2 \log_2 n + 2 \log_2 (n-k) - 2 \log_2(n-k-1)
   \geq 0\,.
   \]
   From this, as $B$ and $0$ are integral, we obtain
   $\bktD{2 \log_2 (n-1)} - \bktD{2 \log_2(n-k-1)} - |B| \geq 0$,
   i.e., the first statement.

   For proving the second statement, we distinguish two cases.

   \textbf{Case 1:} Assume that $n - k \leq 2$.

   By \cref{claim-tri-power},
   we have $n - k \leq 2^{d - |B| + 1} - 2$.
   This implies $\log_2(n-k + 2) \leq d - |B| +1$ and thus
   \begin{equation} \label{bound |B| case 1}
     |B| \leq d + 1 - \log_2(n-k + 2) \overset{k = 2^{d-1}}{=} \log_2 k - \log_2(n-k + 2) + 2\,.
   \end{equation}
   From this, we obtain
   {
   \newcommand{\aw}{5}
   \begin{align}
    \begin{split}
    \rho(n) - \rho(n-k) - |B|
    \ca{\aw}{\overset{\substack{n \geq 16, \\ n-k \leq 2}}{=}}
      \bktD{2 \log_2(n-1)} - (n-k) - |B| \\
    \ca{\aw}{\overset{\labelcref{bound |B| case 1}}{\geq}}
      \bktD{2 \log_2 (n-1)} - (n-k) - \Bkt{\log_2 k - \log_2(n-k + 2) + 2} \\
    \ca{\aw}{\overset{n \geq 16}{\geq}} \log_2 (n-1) + \log_2(15) - 1 - (n - k) - \log_2 k + \log_2(n-k + 2) - 2 \\
    \ca{\aw}{>} \log_2 (n-1) - (n - k) - \log_2 k + \log_2(n-k + 2) + 0.9\,. \label{rho-B-unfinished}
    \end{split}
  \intertext{
   For $k = n$, this reduces to
   $\log_2 (n-1) - \log_2 n + 1.9 \geq 0$,
   which is fulfilled by $n \geq 16$ and \cref{lem-log-plus-2},
   so the second statement is proven.
   For $1 \leq n - k \leq 2$ and thus $n - 2 \leq k \leq n - 1$, we prove it by
  }
   \begin{split}
    \rho(n) - \rho(n-k) - |B|
    \ca{\aw}{\overset{\labelcref{rho-B-unfinished}}{\geq}} \log_2 (n-1) - (n - k) - \log_2 k + \log_2(n-k + 2) + 0.9 \\
    \ca{\aw}{\overset{n - 2 \leq k \leq n - 1}{\geq}} \log_2 (n-1) - 2 - \log_2 (n-1) + \log_2(3) + 0.9 \\
    \ca{\aw}{>} 0\,. \nonumber
   \end{split}
   \end{align}
   }
   \indent \textbf{Case 2:} Assume that $n-k \geq 3$.

   Here, by definition of $\rho$
   (see \cref{def-rho}),
   we have $\rho(n) = \bktD{2 \log_2(n-1)}$ since $n \geq 16 > 3$
   and $\rho(n-k) = \bktD{2 \log_2(n-k-1)}$ as $n -k \geq 3$.
   Hence, in this case, the second statement coincides with the first statement.

   This finishes the proof of the second statement
   and hence of this \lcnamecref{lem-triangular}.
\end{proof}

\spacingbetweenappproofs


\phantomsection
\begin{proof}[Proof of \cref{size-small-values}] \label{proof-size-small-values}
 For the values $\dmin(n, m)$, see \totalref{table-d-m-n}.

 \textbf{Case 1:} Assume that $1 \leq m \leq 2$.

 In this case, we compute $\depthckt(s, t)$ as an optimum symmetric tree on $m + n \geq 1$ inputs in \cref{algo m 2} of \totalref{grinchuk-alg}.
 By \cref{lem-s-tri}, the set $s \tupleconcat (t_0)$ is triangular with respect to $S\!_0$.
 Setting $K = s \tupleconcat (t_0)$ and $L = (t_1, \dotsc, t_{m-1})$,
 we have $|K| = n + 1 \geq 3$ and $|L| = m - 1$.
 By \cref{sym-prep},
 for the construction of $\depthckt(s, t)$,
 we need at most $m + n - 1$ gates if $n \leq 1$,
 and $2 \log_2(n+1) + m - 1 - 2 = m + 2 \log_2(n+1) - 3$ additional gates otherwise.

 This bounds the number of additional gates for $m \in \{1, 2\}$, i.e., the first two rows of \totalref{table-add-gates-m-n}.
 In particular, by \totalref{table-d-m-n}, this covers all cases of $m$ and $n$ with $\dmin(n, m) = 1$.

\textbf{Case 2:} Assume that $m \geq 3$.

Let $d := \dmin(n, m)$ as in \cref{algo compute d} of \cref{grinchuk-alg}.
We traverse the remaining cases of $m$ and $n$ in order of increasing $d$.
Thus, we may assume that the number of additional gates needed in the circuit $\depthckt(m', n')$ as shown in \cref{table-add-gates-m-n}
with $\dmin(m', n') < d$ has already been verified.

\textbf{Case 2.1:} Assume that $d = 2$.

Then, by \cref{table-d-m-n}, we have $m=3$ (since $m \geq 3$) and $n=0$.
Thus we need $2 = m + n - 1$ additional gates.

\textbf{Case 2.2:} Assume that $d = 3$.

By \cref{table-d-m-n}, this implies $m \leq 5.$

In this case, we construct $\depthckt(s, t)$ in \cref{algo m 3} or \cref{algo base end} of \cref{grinchuk-alg}.

If $m = 3$ then (again by \cref{table-d-m-n}) we have $n \leq 4$.
 By \cref{lem-s-tri}, $s \tupleconcat (t_0)$ is triangular with respect to $S\!_0$.
 Hence, we can compute a delay-optimum circuit for
 $\sym\DBkt{s_0, \dotsc, s_{n-1}, t_0, t_1 \lor t_2}$
 via \cref{sym-prep} in \cref{algo m 3},
 which is a depth optimum-circuit for $f(s, t)$.
 By \cref{sym-prep}, this has $m + n - 1$ gates if $n \leq 1$
 and $2 \log_2(n+1) + m - 1 - 2 = m + 2 \log_2(n+1) - 3$ additional gates otherwise.

If $m = 4$ and thus $n \leq 2$ by \cref{table-d-m-n},
we construct a circuit using in total $m + n - 1$ gates in \cref{algo base end}
(see \cref{img-proof_small_1}). However, if $n=2$, then the gate with inputs $s_0$
and $s_1$ in \cref{img-proof_small_1} is contained in the precomputed leftist trees,
so for $m=4$ and $n=2$, we need only $m+n-2 = 4 = m+2\log_2 n - 2$ additional gates.

If $m = 5$ and thus $n \leq 1$ by \cref{table-d-m-n},
we construct a circuit using in total $m + n$ gates in \cref{algo base end}
(see \cref{img-proof_small_2}). However, in this case, the gate with inputs $t_1$
and $t_3$ in \cref{img-proof_small_2} is contained in the precomputed leftist trees,
so we only need $m + n - 1$ additional gates.

\textbf{Case 2.3:} Assume that $d = 4$.

Here, \cref{table-d-m-n} yields $m \leq 8$.

\textbf{Case 2.3.1:} Assume that $n < 2^{d-1}$ and $m \leq \mu(d-1, 0)$.

 Note that we have $n < 2^{d-1} = 8$ and $3 \leq m \leq \mu(4-1, 0) = \xi \frac{8 - 2}{3} + 2 = 5.998$,
 i.e.\ $3 \leq m \leq 5$.
 We use the simple split
 \[f(s, t) = \sym(s) \land f\Bkt{(), (t_0, \dotsc, t_{m-1})}\]
 in \cref{algo simple split} of \cref{grinchuk-alg}.
 Here, as $s$ is triangular by \cref{lem-s-tri}, by \cref{sym-prep},
 for realizing $\sym(s)$, if $n \in \{1, 2\}$, we need $n - 1$ gates,
 and at most $2 \log_2 n - 2$ additional gates otherwise;
 and at most $m - 1$ additional gates for $\depthckt\Bkt{(), (t_0, \dotsc, t_{m-1})}$ 
 by \cref{table-add-gates-m-n}.
 In total, if $n \in \{1, 2\}$, we need at most 
 $n - 1 + m - 1 + 1 = m + n - 1$ additional gates,
 and otherwise at most $2 \log_2 n - 2 + m - 1 + 1 = m + 2 \log_2 n - 2$ additional gates.

\textbf{Case 2.3.2:} Assume that $n \geq 2^{d-1}$.

 As $n \geq 2^{d-1} = 8$, \cref{table-d-m-n} together with $m \geq 3$ implies $m \in\{3,4\}$.
 Then, we apply the symmetric split $f(s, t) = \sym(s') \land f(s \backslash s', t)$
 in \cref{algo sym split} of \cref{grinchuk-alg},
 where $s'$ with length $8$ is computed by \cref{det-triangular-subset}.
 By \cref{lem-triangular-remains},
 both $s'$ and $s \backslash s'$ are triangular.
 Using \cref{sym-prep} to construct a circuit for $\sym(s')$,
 we need at most $2 \log_2 8 - 2 = 4$ additional gates.

 First, assume in addition that $n \in \{8,9\}$. Then, the number of 
 additional gates for $f(s \backslash s', t)$ is $m + n-8 -1 = m + n - 9$,
 so in total we have $m + n - 9 + 4 + 1 = m + n - 4$ additional gates.
 For $n = 8$, this is $m + 4= m + 2 \log_2 n - 2$, and for $n=9$,
 this is $m + 5$.

 Now, assume that $n \in \{10,11\}$. Thus \cref{table-d-m-n} implies $m=3$.
 Since $s \backslash s'$ has length $2$ or $3$,
 the number of additional gates for $f(s \backslash s', t)$ is at most
 $m + \lfloor 2 \log_2(n-8+1)\rfloor - 3 = m + \lfloor 2 \log_2(n-7)\rfloor - 3$,
 which is $m$ for $n=10$ and $m+1$ for $n=11$. Thus, the total number of additional
 gates in those cases is $m + 4 + 1 = m + 5$ for $n=10$ and 
 $m + 1 + 4 + 1 = m + 6$ for $n=11$.

\textbf{Case 2.3.3:} Assume that $n < 2^{d-1}$ and $m > \mu(d-1, 0)$.
If $n < 2^{d-1} = 8$ and $m > \mu(d-1, 0) = 5.998$, we have $m \in \{6,7,8\}$.
Moreover, \cref{table-d-m-n} implies $n \leq 5$. 
In this case, we perform the alternating split
\[
\depthckt(s, t) = \depthckt\Bkt{s, \prefix{t}{k}} \land \Bkt{\depthckt\Bkt{\everysecond{\prefix{t}{k}}, \suffix{t}{k}}}^*\,,
\]
where
$\prefix{t}{k} = \Bkt{t_0, \dotsc, t_{k-1}}$ and $\suffix{t}{k} = t \backslash \prefix{t}{k}$
and $k \in \{1,\dots,m-1\}$ is maximum with $k$ odd and $k \leq \mu(d-1, n)$.
The $\AND{}$-gate connecting $\depthckt\Bkt{s, \prefix{t}{k}}$ and
$\Bkt{\depthckt\Bkt{\everysecond{\prefix{t}{k}}, \suffix{t}{k}}}^*$
counts as one additional gate.
Since we have 
$\mu(d-1, n) = \xi\frac{2^{d-1} -n -2 }{d-1} + 2 = \xi\frac{2^{3} -n -2 }{3} + 2$, 
we get
\[
 k =
\begin{cases} \, 5 & n\in \{0,1\}\\
 \, 3 & n \in \{2,3,4\}\\
 \, 1 & n = 5\\
\end{cases}
\]

First, assume $n\in \{0,1\}$, so $k=5$. Then, we get 
\[
\depthckt(s, t) = \depthckt\Bkt{s, \prefix{t}{k}} \land \Bkt{\depthckt\Bkt{\everysecond{\prefix{t}{k}}, \suffix{t}{k}}}^*
  = \depthckt\Bkt{s, (t_0,\dots,t_4)} \land \Bkt{\depthckt\Bkt{(t_1,t_3),(t_5,\dots,t_{m-1})}}^*\,.
\]
Since $n\in \{0,1\}$, we need $5+n-1$ additional gates for $\depthckt\Bkt{s, (t_0,\dots,t_4)}$.
For $\Bkt{\depthckt\Bkt{(t_1,t_3),(t_5,\dots,t_{m-1})}}^*$, we need 
$m-5 + \lfloor 2 \log_2(2+1) \rfloor - 3 = m - 5$ additional gates.
Hence, in total  we have $5+n-1+m-5+1 = m+n$ additional gates.

Now, assume $n\in \{2,3,4\}$, so $k=3$. Then, we get 
\[
\depthckt(s, t) = \depthckt\Bkt{s, \prefix{t}{k}} \land \Bkt{\depthckt\Bkt{\everysecond{\prefix{t}{k}}, \suffix{t}{k}}}^*
  = \depthckt\Bkt{s, (t_0,t_1,t_2)} \land \Bkt{\depthckt\Bkt{(t_1),(t_3,\dots,t_{m-1})}}^*\,.
\]
From the known entries of \cref{table-add-gates-m-n} we see that
for $\depthckt\Bkt{s, (t_0,t_1,t_2)}$ we need $3 + 2\log_2(n+1) - 3 = 2\log_2(n+1)$ additional gates
and for $\Bkt{\depthckt\Bkt{(t_1),(t_3,\dots,t_{m-1})}}^*$ we need $m - 3 + 1 - 1 = m - 3$ additional
gates. Thus, in total, we need $2\log_2(n+1) + m - 3 + 1 = m +  2\log_2(n+1) - 2$ additional gates.

Finally, assume $n = 5$, so $k=1$. Since, $d=4$, \cref{table-d-m-n} implies $m \leq 6$, so $m = 6$.
We get 
\[
\depthckt(s, t) = \depthckt\Bkt{s, \prefix{t}{k}} \land \Bkt{\depthckt\Bkt{\everysecond{\prefix{t}{k}}, \suffix{t}{k}}}^*
  = \depthckt\Bkt{s, (t_0)} \land \Bkt{\depthckt\Bkt{(),(t_1,\dots,t_{m-1})}}^*\,.
\]
For $\depthckt\Bkt{s, (t_0)}$, we need $1 + 2\log_2(n+1) - 3 = 2\log_2(n+1) - 2$ 
additional gates, and for $\Bkt{\depthckt\Bkt{(),(t_1,\dots,t_{m-1})}}^*$ we need
$m + 0 - 1 = m - 1$ additional gates, so in total $2\log_2(n+1) - 2 + m - 1 + 1 = m +  2\log_2(n+1) - 2$ 
additional gates.
\end{proof}

\spacingbetweenappproofs

\phantomsection
\begin{proof}[Proof of \cref{add-gates-upper-bound}] \label{proof-add-gates-upper-bound}
We partition all cases to consider based on the coloring of \cref{table-add-gates-m-n}.

First consider the dark red part,
i.e., the cases $m \in\{6,7\}$ and $n \geq 2$.
Here, by \cref{table-add-gates-m-n}, we need at most $m + 2 \log_2(n+1) - 2$ additional gates.
If $n = 2$, we have at most
$\bktD{m + 2 \log_2(3) - 2} = m + 1 \leq m + \rho(n)$
additional gates,
and for $n \geq 3$, we have
\[\BktD{m + 2 \log_2(n+1) - 2}
  \overset{n \geq 3, \shortcref{lem-log-plus-2}}{\leq}
  m + \bktD{2 \log_2(n-1)}  \overset{n \geq 3}{=} m + \rho(n) \,.\]
This also proves the result for the light red area (i.e.\ $1 \leq m \leq 2$
and $n \geq 2$ or $m=3$ and $n \in \{2,3,4\}$ because 
$m + 2 \log_2(3) - 3 < m + 2 \log_2(3) - 2$.

Now consider the (light and dark) blue part, i.e., the cases
$m \in \set{1, \dotsc, 8}$ and $n \in \set{0, 1}$,
This implies that $n \leq 2$, hence we need at most
$m +  n  \overset{n \leq 2}{=} m + \rho(n) $ additional gates.

For the yellow part
(i.e., $m = 3$ and $n \in \set{5, \dotsc, 8}$;
$m = 4$ and $n \in \set{2, \dotsc, 8}$;
$m = 5$ and $n \in \set{2, \dotsc, 7}$),
we need at most $\bktD{m + 2 \log_2 n - 2}$ additional gates.
For $n = 2$, we have $m + 2 \log_2 n - 2 = m < m + 2 = m + \rho(n) $.
For $3 \leq n \leq 8$, we have
\[\BktD{m + 2 \log_2 n - 2}
\overset{n \geq 3, \shortcref{lem-log-plus-2}}{\leq}
m + \BktD{2 \log_2(n-1)} - 1 \overset{n \geq 3}{=} m + \rho(n) - 1 < m + \rho(n)\,.\]

For the light green part
(i.e., $m = 3$ and and $n \in \set{9, 10}$ or $m=4$ and $n = 9$),
we need at most
$m  + 5 \overset{n \geq 9}{\leq} m + 2 \log_2(n-1) - 1 \overset{n \in \set{9, 10}}{\leq} m + \rho(n)$
additional gates.

Finally, for the dark green part (i.e.\ $m=3$ and $n=11$) we need at most
$ m + 6 \overset{n = 11}{\leq} m + 2 \log_2(n-1) = \overset{n = 11}{=} m + \rho(n)$
additional gates.

Note that this a complete enumeration of all cases by \cref{table-add-gates-m-n}.
\end{proof}

\spacingbetweenappproofs

\phantomsection
\begin{proof}[Proof of \cref{lem-S}] \label{proof-lem-S}
We prove the first statement by induction on $n$.
For $n = 1$, we have $S\!_n = 0 = 12 - \frac{4}{2}\Bkt{1+2+3}$, 
and for $n = 2$, we have
$S\!_n = 1 = 12 - \frac{4}{4} \Bkt{4+4+3}$.
The induction step for $n \geq 2$ and hence first statement is proven via
{
\newcommand{\aw}{4}
\[
S\!_{n+1} = \sum_{k = 2}^{n+1} \frac{(k-1)^2}{2^{k-2}}
        \overset{\indhyp}{=}  12 - \frac{4}{2^n} \Bkt{n^2 + 2 n+ 3} + \frac{n^2}{2^{n-1}}
        = 12 - \frac{4}{2^{n+1}} \Bkt{(n+1)^2 + 2 (n+1) + 3}\,.
\]
}

To see that the second statement is fulfilled,
note that the first statement implies
$\sum_{k \geq 2} \frac{(k-1)^2}{2^{k-2}} \leq 12$.
From this, together with the first statement, we conclude
\[
  \sum_{k \geq 19} \frac{(k-1)^2}{2^{k-2}}
  = \sum_{k \geq2} \frac{(k-1)^2}{2^{k-2}} - \sum_{k = 2}^{18} \frac{(k-1)^2}{2^{k-2}}
  \leq
    12 - \Bkt{12 - \frac{4}{2^{18}} \Bkt{18^2 + 2 \cdot 18 + 3}}
  = \frac{1452}{2^{18}}
  < 0.006\,. \qedhere
\]
\end{proof}

\spacingbetweenappproofs

\phantomsection
\begin{proof}[Proof of \cref{lem-psi}] \label{proof-lem-psi}
 We have
\begin{equation} \label{bound-for-nom}
2^{d-1} \cdot 2 (d-1) - \Bkt{2^{d-1} + \frac{d}{\xi} - 2 - \frac{1}{\xi}}
 = 2^{d-1} (d-3) + d\Bkt{2^{d-1} - \frac{1}{\xi}} + 2 + \frac{1}{\xi}
 \overset{d \geq 3}{>} 0
\end{equation}
and hence
\begin{equation} \label{bound-nominator}
\frac{1}{2}\Bkt{\flodd\Bkt{\xi\frac{2^{d-1} - 2}{d-1}} + 1}
\leq \frac{1}{2}\Bkt{\xi\frac{2^{d-1} - 2}{d-1} + 1}
= \xi\frac{2^{d-1} - 2 + \frac{d}{\xi} - \frac{1}{\xi}}{2(d-1)}
\overset{\labelcref{bound-for-nom}}{<} \xi 2^{d-1}\,.
\end{equation}
Moreover, since $\frac{1}{2}\Bkt{\flodd\Bkt{\xi\frac{2^{d-1} - 2}{d-1}} + 1}$ is 
integral, we have
\begin{equation} \label{bound-nominator-floor}
\frac{1}{2}\Bkt{\flodd\Bkt{\xi\frac{2^{d-1} - 2}{d-1}} + 1}
\leq
\lfloor \xi 2^{d-1} \rfloor\,.
\end{equation}
Furthermore, we have
\begin{equation} \label{bound-denominator}
 \BktU{\xi\frac{2^{d-1} - 2}{d-1}} + 2 \geq \xi\frac{2^{d-1} - 2}{d-1} + 2 
  = \xi\frac{2^{d-1} - 2 + \frac{2d}{\xi} - \frac{2}{\xi}}{d-1}
\overset{d \geq 5 > 1 + \xi}{\geq} \xi\frac{2^{d-1}}{d-1}\,.
\end{equation}
As both the nominator and denominator of $\psi(d)$ are positive for $d \geq 5$,
we conclude from these inequalities that
{
\newcommand{\aw}{6}
\begin{align*}
\psi(d) \ca{\aw}{=}
   \frac{1 + \rho\Bkt{\frac{1}{2}\Bkt{\flodd\bktRfixed{big}{\xi\frac{2^{d-1} - 2}{d-1}} + 1}} }{\BktU{\xi\frac{2^{d-1} - 2}{d-1}} + 2}
 \overset{\substack{\labelcref{bound-nominator-floor}\,,\\ \shortcref{obs-rho}}}{\leq}
   \frac{1 + \rho \Bkt{\lfloor\xi 2^{d-1}\rfloor} }{\BktU{\xi\frac{2^{d-1} - 2}{d-1}} + 2}
 \overset{\substack{\shortcref{def-rho} \\ \labelcref{bound-denominator}}}{\leq}
   \frac{1 + \BktD{2 \log_2\Bkt{\lfloor\xi 2^{d-1}\rfloor - 1}} }{\xi\frac{2^{d-1}}{d-1}}\\
  \ca{\aw}{\leq}
  \frac{1 + 2 \log_2\Bkt{\xi 2^{d-1} } }{\xi\frac{2^{d-1}}{d-1}}
 \,\,=\,\,
   \frac{d-1}{\xi 2^{d-1}} + \frac{2(d-1)^2}{\xi 2^{d-1}} + \frac{2(d-1)\log_2(\xi)}{\xi 2^{d-1}}
 \,\,\leq\,\,
   \frac{2(d-1)^2}{2^{d-1}}\,,
\end{align*}
}
where the last inequality is equivalent to
$d-1 + 2(d-1) \log_2(\xi) \leq 2 (\xi - 1) (d-1)^2$ and $1 + 2 \log_2(\xi) \leq 2 (\xi - 1) (d-1)$
which is obviously true for $d\geq 5$.
\end{proof}

\spacingbetweenappproofs


\spacingbetweenappproofs

\phantomsection
\begin{proof}[Proof of \cref{step1-num}] \label{proof-step1-num}
To prove the first statement
consider $\frac{d}{dx} \log_2(\log_2(x)) - 0.441 \log_2(x) -0.024
= \frac{1 - 0.441 \ln(x)}{\ln(2)\ln(x)x}$ which is positive for
$x \leq e^{\frac{1}{0.441}}\approx 9.656$ and negative for $x > e^{\frac{1}{0.441}}$.
Hence the statement follows from $\log_2(\log_2(4)) - 0.441 \log_2(4) -0.024 > 0$
and $\log_2(\log_2(17)) - 0.441 \log_2(17) -0.024 > 0$.

 To prove the second statement, we need that
 by \cref{d / log d incr},
 the function $x \mapsto \frac{2^x}{x}$ is monotonely increasing for $x \geq \frac{1}{\ln(2)}$.
 Hence, the function $x \mapsto \frac{\log_2 x}{x}$ is monotonely decreasing
 for $x \geq 2^{\frac{1}{\ln(2)}} = e$.
 Consequently, for $n \geq 3$, we have
 \[ 6.2 \log_2 n - 1.5 \Bkt{n - 1} = n \Bkt{\frac{6.2 \log_2 n  + 1.5}{n} - 1.5}
   \overset{\substack{4 \leq n \leq 17, \\ \shortcref{d / log d incr}}}{\geq} n \Bkt{\frac{6.2 \log_2 17  + 1.5}{17} - 1.5}
   > 0\,.
 \]
 Multiplying with $n$ yields the second statement.
\end{proof}

\spacingbetweenappproofs

\phantomsection
\begin{proof}[Proof of \cref{step-3-num-a}] \label{proof-step-3-num-a}
 Define the function $\nu_r \colon x \mapsto \log_2 \log_2 x + \log_2 \log_2 \log_2 x + 4.3 - \log_2 x $ for $x \geq 4$.
 For $x \leq 512$, we will prove the stronger statement that for $x\in\N_{\geq 4}$ we have
\begin{toshowequation} \label{nu-greater-0}
\nu_r(x) \geq 0\,.
\end{toshowequation}
Note that for $x \leq 19 < 2^{4.3}$ and thus $\log_2 x < 4.3$, this is clearly fulfilled.

Thus, assume that $x \geq 20$.
The derivative of $\nu_r(x)$ is
\begin{equation*}
\frac{d}{dx} \nu_r(x)
= \frac{1}{\ln^2(2) x \log_2 x} + \frac{1}{\ln^3(2) x \log_2 x \log_2 \log_2 x}
  - \frac{1}{\ln(2) x }
  = \frac{\ln(2) \log_2 \log_2 x + 1 - \ln^2(2) \log_2 x \log_2 \log_2 x}{\ln^3(2) x \log_2 x \log_2 \log_2 x}\,.
\end{equation*}
This function is negative as its denominator is always positive and
for its nominator, we have
{
\newcommand{\aw}{4}
\begin{align*}
 \ln(2) \log_2 \log_2 x + 1 - \ln^2(2) \log_2 x \log_2 \log_2 x
 \ca{\aw}{=} 1 + \ln(2) \log_2 \log_2 x (1 - \ln(2) \log_2 x) \\
 \ca{\aw}{\overset{x \geq 20}{\leq}} 1 - 1.99 \ln(2) \log_2 \log_2 x \\
 \ca{\aw}{\overset{x \geq 20}{<}}  0\,.
\end{align*}
}
Thus, for $20 \leq x \leq 512$, we have
$\nu_r(x) \geq \nu_r(512) > 0.1 > 0$.
This proves \cref{nu-greater-0}.

Now, we may assume that $512 < x \leq 1024$,
i.e., $9 < \log_2 x \leq 10$.
This implies
{
\newcommand{\aw}{4}
\begin{equation*}
\log_2 x + \log_2 \log_2 x + \log_2 \log_2 \log_2 x + 6.3
\overset{n > 512}{>} 9 + \log_2 9 + \log_2 \log_2 9 + 6.3
> 20
\overset{x \leq 1024}{\geq} 2\BktU{\log_2 x}. \tag*{\qedhere}
\end{equation*}
}
\end{proof}

\begin{lemma} \label{lem-l-func}
The following statements are fulfilled:
\begin{enumerate}
 \item For $r \in \{1, 2\}$, the function
$\mu_r(x) := \frac{x}{\log_2^r x + 1}$
is monotonely increasing in $x$ for all $x \geq 8$. \label{lem-l-func-first}
 \item For $r \in \{2, 3\}$, the function
$\nu_r(x) := \frac{\log_2^r x}{x \log_2 \log_2 x}$
is monotonely decreasing in $x$ for all $x \geq 32$. \label{lem-l-func-second}
\end{enumerate}
\begin{proof}
 Note that for all $r \in \N_{\geq 1}$, we have
 \begin{equation} \label{deriv-nu-0}
   \frac{d}{dx} \Bkt{\log_2^r x}= \frac{r (\log_2 x)^{r-1}}{\ln(2) x}\,.
  \end{equation}

  To prove the first statement for $r \in \set{1, 2}$, we compute
  \begin{align*}
   \frac{d}{dx} \mu_r(x)
   \overset{\labelcref{deriv-nu-0}}{=}
      \frac{\log_2^r x + 1 - x \frac{r (\log_2 x)^{r-1}}{\ln(2) x}}{\Bkt{\log_2^r x + 1}^2}
   = \frac{(\log_2 x)^{r-1} \Bkt{\log_2 x - \frac{r}{\ln(2)}} + 1}{\Bkt{\log_2^r x + 1}^2}\,.
  \end{align*}
  As $r \leq 2$, the term $\log_2 x - \frac{r}{\ln 2}$ is positive
  for $x \geq 8$, so $\frac{d}{dx} \mu_r(x) > 0$,
  hence the first statement.

  To prove the second statement for $r \in \set{2, 3}$, note that
  \begin{equation} \label{deriv-nu-2}
   \frac{d}{dx} \Bkt{x \log_2 \log_2 x} = \log_2 \log_2 x + x \frac{1}{\ln^2(2) x \log_2 x}\,.
  \end{equation}
  Hence, the derivative of $\nu_r$ is
  {
  \newcommand{\aw}{4}
  \begin{align*}
   \frac{d}{dx} \nu_r(x) \ca{\aw}{\overset{\substack{\labelcref{deriv-nu-0}, \\ \labelcref{deriv-nu-2}}}{=}}
     \frac{\frac{r (\log_2 x)^{r-1}}{\ln(2) x}x \log_2 \log_2 x - \log_2^r x \Bkt{\log_2 \log_2 x + x \frac{1}{\ln^2(2) x \log_2 x}}}{x^2 (\log_2 \log_2 x)^2} \\
   \ca{\aw}{=} \frac{\frac{r (\log_2 x)^{r-1} \log_2 \log_2 x}{\ln(2)} - \log_2^r x \log_2 \log_2 x -  \frac{(\log_2 x)^{r-1}}{\ln^2(2)}}{x^2 (\log_2 \log_2 x)^2}\,.
  \end{align*}
  }
  This is strictly negative for $x \geq 32$ as
  \begin{equation*}
   \frac{r (\log_2 x)^{r-1} \log_2 \log_2 x}{\ln(2)} - \log_2^r x \log_2 \log_2 x -  \frac{(\log_2 x)^{r-1}}{\ln^2(2)}
   < (\log_2  x)^{r-1} \log_2 \log_2 x \Bkt{\frac{r}{\ln(2)} - \log_2 x}
   \overset{\substack{2 \leq r \leq 3, \\ x \geq 32}}{<} 0\,.
  \end{equation*}
  Hence, for $x \geq 32$,
  we have $\frac{d}{dx} \nu_r(x) < 0$,
  which proves the second statement.
 \end{proof}
\end{lemma}

\spacingbetweenappproofs

\phantomsection
\begin{proof}[Proof of \cref{step2-main-depth-bound-a}] \label{proof-step2-main-depth-bound-a}

 We have $k = \bktU{\log_2 n}$.

  With $n \geq 1024$, we have
 \begin{equation} \label{eq::n/k_lower_bound_step_2}
  \frac{n}{k} \overset{k = \bktU{\log_2 n}}{\geq} \frac{n}{\log_2 n + 1}
  \overset{\substack{n \geq 1024, \\ \shortcref{lem-l-func}, \labelcref{lem-l-func-first}}}{\geq} 93
 \end{equation}
 and thus
 \begin{equation} \label{eq::log_n/k_lower_bound_step_2} \log_2 l
 \overset{l = \BktU{\frac{n}{k}}}{\leq}
    \log_2\Bkt{\frac{n}{k} \Bkt{1 + \frac{k}{n}}}
    \stackrel{\labelcref{eq::n/k_lower_bound_step_2}}{\leq} \log_2 n - \log_2 k + 0.016\,.
 \end{equation}

\textbf{Case 1:} Assume that $n\leq 2^{32}$.

We have
 {
 \newcommand{\aw}{6}
 \begin{align}
  \begin{split} \label{step2-bound-loglog2k-a}
   \log_2 \log_2 (2k) 
   \ca{\aw}{\overset{k = \bktU{\log_2 n}}{\leq}}
     \log_2 \Bkt{1 + \log_2 \Bkt{\log_2 n + 1}}
    \,\overset{n \geq 1024}{\leq}\, \log_2 \Bkt{1+ \log_2 \Bkt{\frac{11}{10}\log_2 n}} \\
    \ca{\aw}{<} \log_2(\log_2 \log_2 n + 1.14) 
    \,\overset{n \geq 1024}{<}\, \log_2 \Bkt{1.35 \log_2 \log_2 n} \\
    \ca{\aw}{<} \log_2 \log_2 \log_2 n + 0.44\,.
  \end{split}
 \end{align}
}

Furthermore, we have
 {
 \newcommand{\aw}{6}
 \begin{align}
  \begin{split} \label{step2-bound-loglog-l}
\log_2 \log_2 l
\ca{\aw}{\leq} \log_2 \log_2\Bkt{\frac{n}{k} + 1}
\quad\leq\quad \log_2 \log_2\Bkt{\frac{n}{\log_2n} + 1}\\
\ca{\aw}{\overset{\substack{n \geq 1024, \\ \shortcref{d / log d incr}}}{\leq}}
  \log_2 \log_2\Bkt{\frac{n}{\log_2 n}\Bkt{1 + \frac{10}{1024}}}
\quad\leq\quad
  \log_2 \log_2\Bkt{1.01\frac{n}{\log_2 n}}\\
\ca{\aw}{\leq}
  \log_2 \Bkt{0.015 + \log_2\Bkt{\frac{n}{\log_2 n}}}
\quad=\quad
  \log_2 \Bkt{0.015 + \log_2 n - \log_2 \log_2 n}\\
\ca{\aw}{\overset{\substack{n \leq 2^{32}, \\ \shortcref{d / log d incr}}}{\leq}}
\log_2 \Bkt{0.015 + \log_2 n - \log_2 n \frac{\log_2 \log_2\Bkt{2^{32}}}{\log_2\Bkt{2^{32}}}}
\quad = \quad
  \log_2 \Bkt{0.015 + \log_2 n - \frac{5}{32} \log_2 n}\\
\ca{\aw}{\overset{n>1024}{\leq}}
  \log_2 \Bkt{0.0015 \log_2 n + \log_2 n - \frac{5}{32} \log_2 n}
\quad{\leq}\quad
   \log_2 \Bkt{0.85 \log_2 n}\\
\ca{\aw}{\leq}
   \log_2 \log_2 n - 0.23
  \end{split}
 \end{align}
}

We conclude
   { 
  \newcommand{\aw}{6}
  \begin{align*}
    \ca{\aw}{} \log_2 l + \log_2 \log_2 l + \log_2 k + \log_2 \log_2 (2k) + 6.3\\
    \ca{\aw}{\overset{\labelcref{eq::log_n/k_lower_bound_step_2} }{\leq}}
            \log_2 n - \log_2 k + \log_2 \log_2 l + \log_2 k + \log_2 \log_2(2k) + 6.3 + 0.016 \\
   \ca{\aw}{\overset{\labelcref{step2-bound-loglog2k-a}}{<}}
   \log_2 n + \log_2 \log_2 l + \log_2 \log_2 \log_2 n + 6.8\\
   \ca{\aw}{\overset{\labelcref{step2-bound-loglog-l}}{<}}
   \log_2 n + \log_2 \log_2 n + \log_2 \log_2 \log_2 n + 6.6
  \end{align*}
  }

\textbf{Case 2:} Assume that $n > 2^{32}$.

We have
 {
 \newcommand{\aw}{6}
 \begin{align}
  \begin{split} \label{step2-bound-loglog2k-b}
   \log_2 \log_2 (2k) \ca{\aw}{\overset{k = \bktU{\log_2 n}}{\leq}}
     \log_2 \Bkt{1 + \log_2 \Bkt{\log_2 n + 1}}
    \,\overset{n > 2^{32}}{\leq}\, \log_2 \Bkt{1+ \log_2 \Bkt{\frac{33}{32}\log_2 n}} \\
    \ca{\aw}{<} \log_2(\log_2 \log_2 n + 1.045) 
    \,\overset{n > 2^{32}}{<}\, \log_2 \Bkt{1.21 \log_2 \log_2 n} \\
    \ca{\aw}{<} \log_2 \log_2 \log_2 n + 0.28\,.
  \end{split}
 \end{align}
}

We conclude
   { 
  \newcommand{\aw}{6}
  \begin{align*}
    \ca{\aw}{} \log_2 l + \log_2 \log_2 l + \log_2 k + \log_2 \log_2 (2k) + 6.3\\
    \ca{\aw}{\overset{\labelcref{eq::log_n/k_lower_bound_step_2} }{\leq}}
            \log_2 n - \log_2 k + \log_2 \log_2 l + \log_2 k + \log_2 \log_2(2k) + 6.3 + 0.016 \\
   \ca{\aw}{\overset{\substack{l \leq n, \\ \labelcref{step2-bound-loglog2k-b}}}{<}}
   \log_2 n + \log_2 \log_2 n + \log_2 \log_2 \log_2 n + 6.6 \tag*{\qedhere}
  \end{align*}
  }
\end{proof}

\spacingbetweenappproofs

\phantomsection
\begin{proof}[Proof of \cref{step2-bound-rest-size-a}] \label{proof-step2-bound-rest-size-a}
   We have $k = \bktU{\log_2 n}$.

   Thus, we get
   \begin{equation} \label{step2-bound-l}
   l = \BktU{\frac{n}{k}}
     \overset{k \geq \log_2 n}{\leq} \BktU{\frac{n}{\log_2 n}} \leq \frac{n}{\log_2 n} + 1
   \end{equation}
   and thus
   \begin{equation} \label{bound-log-l}
    \log_2 l
   \overset{\labelcref{step2-bound-l}}{\leq}  \log_2 \Bkt{ \frac{n}{\log_2 n} + 1 } \\
   \overset{\substack{\shortcref{d / log d incr}, \\ n \geq 1024}}{\leq}
              \log_2 \Bkt{ 1.01 \frac{n}{\log_2 n} }\\
   \leq \log_2 n - \log_2 \log_2 n + 0.015\,.
   \end{equation}
   Hence, we obtain
  { 
  \newcommand{\aw}{5}
   \begin{align*}
   (6.2 \log_2 l - 2) l
   \ca{\aw}{\overset{\substack{\labelcref{bound-log-l}, \\ \labelcref{step2-bound-l}}}{\leq}}
     \Bkt{6.2 \Bkt{\log_2 n - \log_2 \log_2 n + 0.015} - 2} \Bkt{\frac{n}{\log_2 n} + 1} \\
   \ca{\aw}{\leq}
              6.2n - \frac{6.2n}{\log_2 n} \log_2 \log_2 n + \frac{0.1 n}{\log_2 n}
                  + 6.2 \log_2 n - 6.2 \log_2 \log_2 n + 0.1
              - \frac{2 n}{\log_2 n} - 2 \\
   \ca{\aw}{<}  6.2 n - \frac{6.2n}{\log_2 n} \log_2 \log_2 n + 6.2 \log_2 n \\
   \ca{\aw}{=} 6.2 n + \frac{6.2n\log_2 \log_2 n }{\log_2 n}  \Bkt{
              - 1 + \frac{\log^2_2 n}{n \log_2 \log_2 n} }\,.
    \end{align*}
    }
    The last term can be bounded from above by $6.2n$ as
    \[
   \frac{\log^2_2 n}{n \log_2 \log_2 n}
   \overset{\substack{n \geq 1024, \\ \shortcref{lem-l-func}, \labelcref{lem-l-func-second}}}{\leq}
              \frac{\log^2_2 (1024)}{1024 \log_2 \log_2 (1024)}
   < 1\,.\]
   This shows the statement.
\end{proof}

\spacingbetweenappproofs

\phantomsection
\begin{proof}[Proof of \cref{step-3-num-b}] \label{proof-step-3-num-b}
 Define the function $\nu_r \colon x \mapsto \log_2 \log_2 x + \log_2 \log_2 \log_2 x + 5 - \log_2 x $ for $x \geq 4$.
 For $x \leq 1024$, we will prove the stronger statement that for $x\in\N_{\geq 4}$ we have
\begin{toshowequation} \label{nu-greater-0b}
\nu_r(x) \geq 0\,.
\end{toshowequation}
Note that for $x \leq 32 = 2^{5}$ and thus $\log_2 x < 5$, this is clearly fulfilled.

Thus, assume that $x \geq 32$.
The derivative of $\nu_r(x)$ is
\begin{equation*}
\frac{d}{dx} \nu_r(x)
= \frac{1}{\ln^2(2) x \log_2 x} + \frac{1}{\ln^3(2) x \log_2 x \log_2 \log_2 x}
  - \frac{1}{\ln(2) x }
  = \frac{\ln(2) \log_2 \log_2 x + 1 - \ln^2(2) \log_2 x \log_2 \log_2 x}{\ln^3(2) x \log_2 x \log_2 \log_2 x}\,.
\end{equation*}
This function is negative as its denominator is always positive and
for its nominator, we have
{
\newcommand{\aw}{4}
\begin{align*}
 \ln(2) \log_2 \log_2 x + 1 - \ln^2(2) \log_2 x \log_2 \log_2 x
 \ca{\aw}{=} 1 + \ln(2) \log_2 \log_2 x (1 - \ln(2) \log_2 x) \\
 \ca{\aw}{\overset{x \geq 32}{\leq}} 1 - 2.4 \ln(2) \log_2 \log_2 x \\
 \ca{\aw}{\overset{x \geq 32}{<}}  0\,.
\end{align*}
}
Thus, for $32 \leq x \leq 1024$, we have
$\nu_r(x) \geq \nu_r(1024) > 0.05 > 0$.
This proves \cref{nu-greater-0b}.

Now, we may assume that $1024 < x \leq 2048$,
i.e., $10 < \log_2 x \leq 11$.
This implies
{
\newcommand{\aw}{4}
\begin{equation*}
\log_2 x + \log_2 \log_2 x + \log_2 \log_2 \log_2 x + 7
\overset{n > 1024}{>} 10 + \log_2 10 + \log_2 \log_2 10 + 7
> 22
\overset{x \leq 2048}{\geq} 2\BktU{\log_2 x}. \tag*{\qedhere}
\end{equation*}
}
\end{proof}

\spacingbetweenappproofs

\phantomsection
\begin{proof}[Proof of \cref{step2-main-depth-bound-b}] \label{proof-step2-main-depth-bound-b}

 We have $k = \BktU{\log_2^2 n}$.

 With $n \geq 2048$, we have
 \begin{equation} \label{better-eq::n/k_lower_bound_step_2}
  \frac{n}{k} \overset{k = \BktU{\log_2^2 n}}{\geq} \frac{n}{\log_2^2 n + 1}
  \overset{\substack{n \geq 2048, \\ \shortcref{lem-l-func}, \labelcref{lem-l-func-first}}}{\geq} 16
 \end{equation}
 and thus
 \begin{equation} \label{better-eq::log_n/k_lower_bound_step_2} \log_2 l
 \overset{l = \BktU{\frac{n}{k}}}{\leq}
    \log_2\left(\frac{n}{k} \Bkt{1 + \frac{k}{n}}\right)
    \stackrel{\labelcref{better-eq::n/k_lower_bound_step_2}}{\leq}
    \log_2 n - \log_2 k + 0.088\,.
 \end{equation}

Furthermore, we have
 {
 \newcommand{\aw}{6}
 \begin{align}
  \begin{split} \label{better-step2-bound-loglog2k-a}
   \log_2 \log_2 (2k) \ca{\aw}{\overset{k = \BktU{\log_2^2 n}}{\leq}}
     \log_2 \Bkt{1 + \log_2 \Bkt{\log_2^2 n + 1}}
    \,\overset{n \geq 2048}{\leq}\, \log_2 \Bkt{1+ \log_2 \Bkt{\frac{122}{121}\log_2^2 n}} \\
    \ca{\aw}{<} \log_2(2\log_2 \log_2 n + 1.012) 
    \,\overset{n \geq 2048}{<}\, \log_2 \Bkt{2.293 \log_2 \log_2 n} \\
    \ca{\aw}{<} \log_2 \log_2 \log_2 n + 1.2\,.
  \end{split}
 \end{align}
 }

We conclude
   { 
  \newcommand{\aw}{6}
  \begin{align*}
    \ca{\aw}{} \log_2 l + \log_2 \log_2 l + \log_2 k + \log_2 \log_2 (2k) + 6.3\\
    \ca{\aw}{\overset{\labelcref{better-eq::log_n/k_lower_bound_step_2}}{\leq}}
            \log_2 n - \log_2 k + \log_2 \log_2 l + \log_2 k + \log_2 \log_2 (2k) + 6.3
           + 0.088 \\
   \ca{\aw}{\overset{\substack{l \leq n, \\ \labelcref{better-step2-bound-loglog2k-a}}}{<}}
   \log_2 n + \log_2 \log_2 n + \log_2 \log_2 \log_2 n +
             7.6 \,.  \tag*{\qedhere}
  \end{align*}
  }

\end{proof}

\spacingbetweenappproofs

\phantomsection
\begin{proof}[Proof of \cref{step2-bound-rest-size-b}] \label{proof-step2-bound-rest-size-b}

  We have $k = \BktU{\log_2^2 n}$.

  Thus, we get
   \begin{equation} \label{better-size-step2-bound-l}
   l = \BktU{\frac{n}{k}}
     \overset{k \geq \log_2^2 n}{\leq} \BktU{\frac{n}{\log_2^2 n}} \leq \frac{n}{\log_2^2 n} + 1\,,
   \end{equation}
   and for the derivative of $x \mapsto \frac{\log_2^2 x}{x}$ with $x \geq 2048$ we get
   \begin{equation} \label{missing-deriv}
    \frac{d}{dx} \frac{\log_2^2 x}{x}
    = \frac{x\frac{2}{\ln(2)x}\log_2{x} - \log_2^2 x}{x^2}
    = \frac{\log_2 x \Bkt{\frac{2}{\ln(2)} - \log_2 x}}{x^2}
    \overset{x \geq 2048}{<} 0
   \end{equation}
   and thus
    \begin{equation} \label{better-size-bound-log-l}
    \log_2 l
   \overset{\labelcref{better-size-step2-bound-l}}{\leq}
      \log_2 \Bkt{ \frac{n}{\log_2^2 n} + 1 } \\
   \overset{\substack{\labelcref{missing-deriv}, \\ n \geq 2048}}{\leq}
              \log_2 \Bkt{ 1.06 \frac{n}{\log_2^2 n} }\\
   \leq \log_2 n - 2 \log_2 \log_2 n + 0.09 \,.
   \end{equation}
   This implies
  { 
  \newcommand{\aw}{5}
   \begin{align*}
   (6.2 \log_2 l - 2) l
   \ca{\aw}{\overset{\substack{\labelcref{better-size-bound-log-l} \\ \labelcref{better-size-step2-bound-l}}}{\leq}}
     \Bkt{6.2 \Bkt{\log_2 n - 2 \log_2 \log_2 n + 0.09} - 2} \Bkt{\frac{n}{\log_2^2 n} + 1} \\
   \ca{\aw}{=}
              6.2\frac{n}{\log_2 n} - \frac{12.4 n}{\log_2^2 n} \log_2 \log_2 n + \frac{0.558 n}{\log_2^2 n}
                  + 6.2 \log_2 n - 12.4 \log_2 \log_2 n + 0.558
              - \frac{2 n}{\log_2^2 n} - 2 \\
   \ca{\aw}{<}  6.2 \frac{n}{\log_2 n} - \frac{12.4 n}{\log_2^2 n} \log_2 \log_2 n + 6.2 \log_2 n \\
   \ca{\aw}{\overset{n \geq 2048}{\leq}} 0.57 n + \frac{n\log_2 \log_2 n }{\log_2^2 n}  \Bkt{
              - 12.4 + 6.2 \frac{\log^3_2 n}{n \log_2 \log_2 n} }\,.
   \end{align*}
    }
    The last term can be bounded from above by $0.57n$ as
\[    6.2 \frac{\log^3_2 n}{n \log_2 \log_2 n}
   \overset{\substack{n \geq 2048, \\ \shortcref{lem-l-func}, \labelcref{lem-l-func-second}}}{\leq}
      6.2 \frac{\log^3_2 (2048)}{2048 \log_2 \log_2 (2048)}
   < 12.4\,.
\]
This proves this \lcnamecref{step2-bound-rest-size-b}.
\end{proof}

\section{Glossary of Symbols}\label{sec::glossary}
\renewcommand{\arraystretch}{1.05}
\begin{table}[H]
\begin{center}
\begin{tabular}{|l|l|}\hline
$A^1_n$ & Adder Circuit on $n$ input pairs with
         $d\Bkt{A^1_n} \leq \log_2 n + \log_2 \log_2 n + 2.65$
         and
         $s\Bkt{A^1_n} \leq 6.2 n \log_2 n$,\\
&        \cref{theorem::step_1_red_version}\\ \hline
$A^2_n$ & Adder Circuit on $n$ input pairs with
        $d\left(A^2_n\right) \leq \log_2 n + \log_2 \log_2 n + \log_2 \log_2 \log_2 n + 6.6$
        and\\ 
&       $s\left(A^2_n\right) \leq 21.6 n$, \cref{n log log logn red}\\ \hline
$A^3_n$ & Adder Circuit on $n$ input pairs with
        $d\left(A^3_n\right) \leq \log_2 n + \log_2 \log_2 n + \log_2 \log_2 \log_2 n + 7.6$
        and\\ 
&       $s\left(A^2_n\right) \leq 16.7 n$, \cref{better-size-n log log logn red}\\ \hline
$B(K, S)$ &
   Set of boundary vertices of a set $K$ of inputs in a leftist circuit $S$,
   \cref{def-boundary} \\ \hline
$d(C)$ &
   Depth of a circuit $C$, \cref{depth-page}, \cpageref{depth-page}\\ \hline
$\delay(C)$ &
   Delay of a circuit $C$ with input arrival times,
   \cref{sec-sym-share}, \cpageref{delay-page}\\ \hline
$\dmin(n, m)$ & 
   Given $n, m \in \N$ with $m \geq 1$, we define
   $\dmin(n, m) := \min \set{d \in \N_{\geq 1} \where{}{} m \leq \mu(d, n)}$,
   \cref{def-dmin}\\ \hline
$\cktedges(C)$ &
   Edges of a circuit $C$, \cref{cktedges-page}, \cpageref{cktedges-page}\\ \hline
$f(s, t)$ &
   Extended \aop{} $f(s, t) = s_0 \land \dotsc \land s_{n-1} \land g(t)$
   with symmetric inputs $s = \Bkt{s_0, \dotsc, s_{n-1}}$ and \\ 
&  alternating inputs $t = \Bkt{t_0, \dotsc, t_{m-1}}$,
   \cref{def-extended-aop}\\ \hline
$f^*(s, t)$ &
   Extended \aop{} $f^*(s, t) = s_0 \lor \dotsc \lor s_{n-1} \lor g^*(t)$
   with symmetric inputs $s = \Bkt{s_0, \dotsc, s_{n-1}}$ and \\ 
&  alternating inputs $t = \Bkt{t_0, \dotsc, t_{m-1}}$,
   \cref{def-extended-aop}\\ \hline
$\fanout(C)$ &
   Maximum fanout of any vertex of a circuit $C$, \cref{fanout-page}, \cpageref{fanout-page}\\ \hline
$\flodd(x)$ & Given $x \in \R$, we write
             $\flodd(x) = \max \set{y \in \Z \where y \text{ odd}, y \leq x}$,
             \cref{def-flodd}\\ \hline
$\cktgates(C)$ &
   Gates of a circuit $C$, \cref{cktgates-page}, \cpageref{cktgates-page} \\ \hline
$g(t)$ &
   \aop{} $g(t) = t_0 \land (t_1 \lor (t_2 \land \dotsc ))$
   with Boolean input variables $t = \Bkt{t_0, \dotsc, t_{m-1}}$,
   \cref{def-aop}\\ \hline
$g^*(t)$ &
   \aop{} $g^*(t) = t_0 \lor (t_1 \land (t_2 \lor \dotsc ))$
   with Boolean input variables $t = \Bkt{t_0, \dotsc, t_{m-1}}$,
   \cref{def-aop}\\ \hline
$\cktins(C)$ &
   Inputs of a circuit $C$, \cref{cktins-page}, \cpageref{cktins-page}\\ \hline
$\cktins_v(C)$ & 
   Inputs in the input cone of a vertex $v$ of a circuit $C$.
   \cref{inputconeinputs-page}, \cpageref{inputconeinputs-page}\\ \hline
$L^f_n$ & Circuit from \cite{LF80} solving the \pradderopt{}
  and the \prmultiand{} on $n$\\
& input pairs  at the same time.
  The depth of the adder sub-circuit is at most $2(\lceil \log_2 n \rceil + f)$,
  and the depth\\ 
& of the {\sc And-Prefix} sub-circuit is at most $\lceil \log_2 n \rceil + f$.
  $s\left(L^f_n\right) \leq 6\left(1 + 2^{-f}\right)n$; \cref{Ladner Fischer}\\ \hline
$m(d, n)$ & 
   Capacity of $d$ and $n$ for $d, n \in \N$.
   Maximum number $m$ of alternating inputs such that an extended\\
&  \aop{} with $n$ symmetric inputs and $m$
   alternating inputs can be realized with depth $d$,\\
&  \cref{def m}\\ \hline
$s(C)$ &
   Size, i.e., number of gates, of a circuit $C$, \cref{size-page}, \cpageref{size-page}\\ \hline
$S^f_n$ & Circuit from \cite{LF80} for the \prprefix{} with
   prefix depth at most $\lceil \log_2 n \rceil + f$ and
   prefix\\
&  size at most $2\left(1 + 2^{-f}\right)n$; \cref{Ladner Fischer}\\ \hline
$\widehat{t}$ &
   For $t = \Bkt{t_0, \dotsc, t_{m-1}}$ with $m$ odd, we write
   $\widehat{t} := (t_1, t_3, t_5, \dotsc, t_{m-2})$, \cref{everysecond-page}, \cpageref{everysecond-page}\\ \hline
$\cktnodes(C)$ &
   Vertices of a circuit $C$, \cref{cktnodes-page}, \cpageref{cktnodes-page}\\ \hline
$\cktnodes_v(C)$ & 
   Input cone of a vertex $v$ of a circuit $C$,
   \cref{inputcone-page}, \cpageref{inputcone-page}\\ \hline
$\mu(d, n)$ &
   Given $d, n \in \N$, we have $\mu(d, n) = \xi \frac{2^d - n - 2}{d} + 2$,
   \cref{def-mu}\\ \hline
$\rho(n)$ &
   Defined as $\rho(n) = n$ for $n \in \{0,1,2\}$ and
   $\rho(n) = \bktD{2 \log_2(n-1)}$ for $n \in \N_{\geq 3}$, \cref{def-rho}\\ \hline
$\phi(d)$ &
   For $d \in \N_{\geq 1}$, we set 
  $\phi(d) = -1.67$ if $d \leq 4$ and 
  $\phi(d) = \phi(d-1) + \psi(d)$ if $d \geq 5$, \cref{lem-phi}\\ \hline
$\psi(d)$ & 
   For $d \in \N_{\geq 5}$,  $\psi(d)$ is defined by
   $\psi(d) := \Bkt{1 + {\rho\Bkt{\frac{1}{2}\Bkt{\flodd\bktRfixed{big}{\frac{2^{d-1} - 2}{d-1}} + 1}}}}\Big/\Bkt{\BktD{\frac{2^{d-1} - 2}{d-1}} + 3}$,
   \cref{lem-psi} \\ \hline
$\cdot \tupleconcat \cdot$ &
   Concatenation of two tuples.
   For example, $(w, x) \tupleconcat (y, z) = (w, x, y, z)$.
   \cref{tupleconcat-notation}\\ \hline
\end{tabular}
\end{center}
\caption{List of symbols used in this paper.}
\label{table::notation}
\end{table}
\cref{table::notation} gives an overview of the symbols and notation used in
this paper.\newpage
\renewcommand{\arraystretch}{1.0}

\end{appendix}

\bibliographystyle{plainnat}

\samepage{
\bibliography{all_citations}
}
\end{document}